\documentclass[aps,prX,twocolumn,nofootinbib, superscriptaddress,floatfix,showpacs]{revtex4-1}
\usepackage{url}
\usepackage{graphicx,subfigure}
\usepackage{appendix}
\usepackage{slashed}
\usepackage{multirow}
\usepackage{nicefrac}
\usepackage{notes2bib}
\usepackage{gensymb}
\usepackage{adjustbox}
\usepackage{slashed}
\usepackage{upgreek}
\usepackage{titlesec}
\usepackage{multirow}
\usepackage{amsmath}
\usepackage{latexsym}
\usepackage{amssymb}
\usepackage{color}
\usepackage{hyperref}
\DeclareMathOperator{\sgn}{sgn}

\newcommand{\CFF}{$H_1^{\perp}$}

\begin{document}

\preprint{\vbox{ \hbox{Version 1.93 -- \today, BELLE DRAFT, intended for {\it Phys.Rev.D}}
                 \hbox{Authors: H. Li, A. Vossen, Committee: G. Schnell (chair), W.W. Jacobs, M. Naruki}
}}


\title{Azimuthal asymmetries of back-to-back $\pi^\pm-(\pi^0,\eta,\pi^\pm)$ pairs in $e^+e^-$ annihilation}

\noaffiliation
\affiliation{University of the Basque Country UPV/EHU, 48080 Bilbao}
\affiliation{Beihang University, Beijing 100191}
\affiliation{Brookhaven National Laboratory, Upton, New York 11973}
\affiliation{Budker Institute of Nuclear Physics SB RAS, Novosibirsk 630090}
\affiliation{Faculty of Mathematics and Physics, Charles University, 121 16 Prague}
\affiliation{Deutsches Elektronen--Synchrotron, 22607 Hamburg}
\affiliation{Duke University, Durham, North Carolina 27708}
\affiliation{Key Laboratory of Nuclear Physics and Ion-beam Application (MOE) and Institute of Modern Physics, Fudan University, Shanghai 200443}
\affiliation{SOKENDAI (The Graduate University for Advanced Studies), Hayama 240-0193}
\affiliation{Gyeongsang National University, Jinju 52828}
\affiliation{Department of Physics and Institute of Natural Sciences, Hanyang University, Seoul 04763}
\affiliation{University of Hawaii, Honolulu, Hawaii 96822}
\affiliation{High Energy Accelerator Research Organization (KEK), Tsukuba 305-0801}
\affiliation{J-PARC Branch, KEK Theory Center, High Energy Accelerator Research Organization (KEK), Tsukuba 305-0801}
\affiliation{Forschungszentrum J\"{u}lich, 52425 J\"{u}lich}
\affiliation{IKERBASQUE, Basque Foundation for Science, 48013 Bilbao}
\affiliation{Indian Institute of Science Education and Research Mohali, SAS Nagar, 140306}
\affiliation{Indian Institute of Technology Guwahati, Assam 781039}
\affiliation{Indian Institute of Technology Hyderabad, Telangana 502285}
\affiliation{Indian Institute of Technology Madras, Chennai 600036}
\affiliation{Indiana University, Bloomington, Indiana 47408}
\affiliation{Institute of High Energy Physics, Chinese Academy of Sciences, Beijing 100049}
\affiliation{Institute of High Energy Physics, Vienna 1050}
\affiliation{INFN - Sezione di Napoli, 80126 Napoli}
\affiliation{INFN - Sezione di Torino, 10125 Torino}
\affiliation{Advanced Science Research Center, Japan Atomic Energy Agency, Naka 319-1195}
\affiliation{J. Stefan Institute, 1000 Ljubljana}
\affiliation{Institut f\"ur Experimentelle Teilchenphysik, Karlsruher Institut f\"ur Technologie, 76131 Karlsruhe}
\affiliation{King Abdulaziz City for Science and Technology, Riyadh 11442}
\affiliation{Korea Institute of Science and Technology Information, Daejeon 34141}
\affiliation{Korea University, Seoul 02841}
\affiliation{Kyungpook National University, Daegu 41566}
\affiliation{LAL, Univ. Paris-Sud, CNRS/IN2P3, Universit\'{e} Paris-Saclay, Orsay 91898}
\affiliation{\'Ecole Polytechnique F\'ed\'erale de Lausanne (EPFL), Lausanne 1015}
\affiliation{P.N. Lebedev Physical Institute of the Russian Academy of Sciences, Moscow 119991}
\affiliation{Faculty of Mathematics and Physics, University of Ljubljana, 1000 Ljubljana}
\affiliation{Ludwig Maximilians University, 80539 Munich}
\affiliation{Luther College, Decorah, Iowa 52101}
\affiliation{University of Maribor, 2000 Maribor}
\affiliation{Max-Planck-Institut f\"ur Physik, 80805 M\"unchen}
\affiliation{School of Physics, University of Melbourne, Victoria 3010}
\affiliation{University of Mississippi, University, Mississippi 38677}
\affiliation{University of Miyazaki, Miyazaki 889-2192}
\affiliation{Moscow Physical Engineering Institute, Moscow 115409}
\affiliation{Moscow Institute of Physics and Technology, Moscow Region 141700}
\affiliation{Graduate School of Science, Nagoya University, Nagoya 464-8602}
\affiliation{Universit\`{a} di Napoli Federico II, 80055 Napoli}
\affiliation{Nara Women's University, Nara 630-8506}
\affiliation{National Central University, Chung-li 32054}
\affiliation{National United University, Miao Li 36003}
\affiliation{Department of Physics, National Taiwan University, Taipei 10617}
\affiliation{H. Niewodniczanski Institute of Nuclear Physics, Krakow 31-342}
\affiliation{Nippon Dental University, Niigata 951-8580}
\affiliation{Niigata University, Niigata 950-2181}
\affiliation{Novosibirsk State University, Novosibirsk 630090}
\affiliation{Pacific Northwest National Laboratory, Richland, Washington 99352}
\affiliation{Peking University, Beijing 100871}
\affiliation{University of Pittsburgh, Pittsburgh, Pennsylvania 15260}
\affiliation{Research Center for Nuclear Physics, Osaka University, Osaka 567-0047}
\affiliation{Theoretical Research Division, Nishina Center, RIKEN, Saitama 351-0198}
\affiliation{RIKEN BNL Research Center, Upton, New York 11973}
\affiliation{University of Science and Technology of China, Hefei 230026}
\affiliation{Showa Pharmaceutical University, Tokyo 194-8543}
\affiliation{Soongsil University, Seoul 06978}
\affiliation{Sungkyunkwan University, Suwon 16419}
\affiliation{School of Physics, University of Sydney, New South Wales 2006}
\affiliation{Department of Physics, Faculty of Science, University of Tabuk, Tabuk 71451}
\affiliation{Tata Institute of Fundamental Research, Mumbai 400005}
\affiliation{Department of Physics, Technische Universit\"at M\"unchen, 85748 Garching}
\affiliation{Department of Physics, Tohoku University, Sendai 980-8578}
\affiliation{Earthquake Research Institute, University of Tokyo, Tokyo 113-0032}
\affiliation{Department of Physics, University of Tokyo, Tokyo 113-0033}
\affiliation{Tokyo Institute of Technology, Tokyo 152-8550}
\affiliation{Virginia Polytechnic Institute and State University, Blacksburg, Virginia 24061}
\affiliation{Wayne State University, Detroit, Michigan 48202}
\affiliation{Yamagata University, Yamagata 990-8560}
\affiliation{Yonsei University, Seoul 03722}
  \author{H.~Li}\affiliation{Indiana University, Bloomington, Indiana 47408} 
  \author{A.~Vossen}\affiliation{Duke University, Durham, North Carolina 27708} 
  \author{H.~Aihara}\affiliation{Department of Physics, University of Tokyo, Tokyo 113-0033} 
  \author{D.~M.~Asner}\affiliation{Brookhaven National Laboratory, Upton, New York 11973} 
  \author{V.~Aulchenko}\affiliation{Budker Institute of Nuclear Physics SB RAS, Novosibirsk 630090}\affiliation{Novosibirsk State University, Novosibirsk 630090} 
  \author{T.~Aushev}\affiliation{Moscow Institute of Physics and Technology, Moscow Region 141700} 
  \author{V.~Babu}\affiliation{Deutsches Elektronen--Synchrotron, 22607 Hamburg} 
  \author{I.~Badhrees}\affiliation{Department of Physics, Faculty of Science, University of Tabuk, Tabuk 71451}\affiliation{King Abdulaziz City for Science and Technology, Riyadh 11442} 
  \author{A.~M.~Bakich}\affiliation{School of Physics, University of Sydney, New South Wales 2006} 
  \author{J.~Bennett}\affiliation{University of Mississippi, University, Mississippi 38677} 
% ...
\collaboration{The Belle Collaboration}


\begin{abstract}
This work reports the first observation of azimuthal asymmetries around the thrust axis in $e^+e^-$ annihilation of pairs of back-to-back charged pions in one hemisphere, and $\pi^0$ and $\eta$ mesons in the opposite hemisphere. These results are complemented by a new analysis of pairs of back-to-back charged pions. The $\pi^0$ and $\eta$ asymmetries  rise with the relative momentum $z$ of the detected hadrons as well as with the transverse momentum with respect to the thrust axis.
These asymmetries are sensitive to the Collins fragmentation function \CFF{} and provide complementary information to previous measurements with charged pions and kaons in the final state. In particular, the $\eta$ final states will provide additional information on the flavor structure of \CFF. This is the first measurement of the explicit transverse-momentum dependence of the Collins fragmentation function from Belle data.  
It uses a dataset of 980.4~fb$^{-1}$ collected by the Belle experiment at or near a center-of-mass energy of 10.58 GeV.
\end{abstract}
\maketitle
\section{Introduction}
A description of the three-dimensional partonic structure of the nucleon is an essential test for our understanding of quantum chromodynamics (QCD). Successful tools for the study of the nucleon have been semi-inclusive hard reactions, particularly the use of leptonic probes such as electrons and muons. 
At high enough momentum transfers, QCD factorization theorems can be applied, and the process can be described using a convolution over parton distribution functions (PDFs), fragmentation functions (FFs), and the matrix element describing the elementary hard scattering of the probe off the parton inside the nucleon.
PDFs~\cite{Aidala:2012mv} can be interpreted as the leading coefficients of the wave function of the nucleon on the light-cone in a $Q^2$ expansion, where $Q^2$ is the squared 4-momentum transfer, and have an probabilistic interpretation in the parton model as the probability of finding a parton $q$ in the nucleon carrying a momentum fraction $x$ of the parent nucleon. So-called unintegrated PDFs also carry a dependence on the transverse momentum of the struck quark.
Fragmentation functions~\cite{Metz:2016swz}, on the other hand, describe the hadronization of a quark into a final-state hadrons containing at least one detected hadron.
Fragmentation functions depend on the dimensionless variable $z$, which, in a partonic picture, can be interpreted as the momentum fraction of the struck quark carried by the detected hadron. In addition, unintegrated FFs depend on the transverse momentum $\boldsymbol{P}_{h\perp}$ of the hadron with respect to the initial quark direction.
Since FFs encode the dependence of the properties of the detected hadron with the quantum numbers of the struck quark, knowledge of them is essential for the extraction of information on the partonic structure of the nucleon from semi-inclusive hard scattering experiments. This is in particular true for the transverse spin structure of the nucleon. The large single transverse spin asymmetries of $\pi^0$ and $\eta$ mesons observed in $pp$ collisions were at odds with the expectation that they would vanish due to the suppression of spin-flip amplitudes in the hard scattering~\cite{TSSA_old_theory}. However, Collins showed~\cite{TSSA_Collins} that spin-flip amplitudes for soft components of the cross section, the PDFs and FFs, are not necessarily suppressed. 
In the collinear picture, in which the dependence of the PDFs and FFs on intrinsic transverse momenta is integrated over, the PDF that corresponds to the spin-flip amplitude is the so-called transversity PDF $h_1$~\cite{Ralston:1979ys,Artru:1989zv,Jaffe:1991ra,Cortes:1991ja}. This can be interpreted as the probability of finding a transversely polarized quark in a transversely polarized nucleon with its polarization direction along the polarization of the parent nucleon and is one of the three leading-twist PDFs needed to describe the nucleon in a collinear picture. It is a chiral-odd function, and since chiral-odd amplitudes are strongly suppressed in perturbative QCD~\cite{TSSA_old_theory}, $h_1$ has to be coupled to another chiral-odd function to construct a chiral-even observable such as a cross section.
Experimentally, the most relevant channels to access transversity are transverse single spin asymmetries in semi-inclusive deep-inelastic scattering (SIDIS) or $pp$ scattering. Here transversity couples, for instance, to the transverse polarization dependent chiral-odd Collins FF $H_1^\perp$~\cite{TSSA_Collins} or the di-hadron interference FF $H_1^\sphericalangle$~\cite{Collins:1993kq,Bianconi:1999cd}. Since both the transversity PDF as well as the transverse polarization dependent FFs are a priori unknown, an independent measurement of the FF is needed.
Such a measurement can be performed in $e^+e^-$ annihilation, where a back-to-back $q\bar{q}$ pair is created and hadronizes. The azimuthal dependence of the cross section of back-to-back production of hadrons can be described by the product of the quark and anti-quark $H_1^\perp$ together with the polarization averaged FFs. This allows access to the Collins FF without the complication of other, potentially unknown, functions that cannot be calculated in perturbative QCD. A disadvantage of $e^+e^-$  annihilation at the energies relevant for FF measurements is the small sensitivity to gluon fragmentation as well as to the flavor of the fragmenting quark. This is because the production probability of all light quarks solely depend on $e_q^2$, where $e_q$ is the electric charge of the quark, and it is assumed that $e^+e^-$ annihilation into virtual photons dominates, as in selected Belle data.

The first unambiguous observation of the Collins effect came from SIDIS off transversely polarized protons~\cite{hermes}. The behavior of the observed \(\pi^+\) and \(\pi^-\) asymmetries indicated that the Collins FF had opposite signs for favored versus disfavored fragmentation [cf.~Eq.~\eqref{eqn:FF4}], motivated also by the Sch\"afer--Teryaev sum rule for the Collins FFs~\cite{Schafer:1999kn}. 
These results spurred a wide range of both theoretical and experimental activities.
The first measurement sensitive to the Collins FF for charged pions in $e^+e^-$ annihilation was performed at Belle~\cite{Abe:2005zx,Seidl:2008xc}. It was subsequently used, together with SIDIS data, for the first extraction of transversity in a global fit~\cite{TransversityandCollinsfromSIDISandEE}. The Belle results were confirmed by BaBar~\cite{BabarCharged}. Later, BaBar also reported the transverse momentum dependence as well as the observation of a significant signal for asymmetries involving kaons~\cite{BabarKaon}. At lower energies, Collins asymmetries in $e^+e^-$ annihilation have been measured by the BESIII collaboration~\cite{BESIII}. The $Q^2$ dependence of the Collins function might provide interesting insight into the non-trivial evolution of transverse momentum dependent functions~(cf.~Ref.~\cite{Metz:2016swz} and references therein).

Here, we report the first measurement of azimuthal asymmetries in back-to-back production of hadron pairs, where one hadron is a charged pion and the other hadron a $\pi^0$ or an $\eta$. We report the fractional-energy and the transverse-momentum dependence of these asymmetries as well as of asymmetries for charged pions. These results provide additional constraints on the Collins function in global fits. The final states including $\eta$ mesons will provide sensitivity to the fragmentation of strange quarks and are also of interest since there are hints that the transverse spin asymmetries of $\pi^0$ and $\eta$ mesons in $pp$ collisions are different~\cite{Adamczyk:2012xd,Adare:2014qzo}.

This paper is structured as follows:
In \hyperref[sec:theory]{Sec.~\ref{sec:theory}} the observables are introduced, \hyperref[sec:experiment]{Sec.~\ref{sec:experiment}} briefly describes the Belle detector. Section~\ref{sec:analysis} details the analysis steps, Sec.~\hyperref[sec:results]{\ref{sec:results}} reports the result, and \hyperref[sec:summary]{Sec.~\ref{sec:summary}} provides the summary and conclusion. Data tables are provided in two Appendices. 
In the following we set $c=1$.

\section{Formalism}
\label{sec:theory}
The probability of a transversely polarized quark $q^\uparrow$ to fragment into an unpolarized hadron $h$ is given by~\cite{Bacchetta:2004jz}
\begin{equation}
D_{hq^\uparrow}=D^{q/h}_1(z,\boldsymbol{P}_{h\perp}^2)+H^{\bot q/h}_1(z,\boldsymbol{P}_{h\perp}^2)\frac{(\boldsymbol{\hat{k}}\times \boldsymbol{P}_{h\perp})\cdot \boldsymbol{S}_{\perp}}{zM_h},
\label{eqn:FF1}
\end{equation}
where $\boldsymbol{S}_{\perp}$ is the transverse polarization of the quark, $\boldsymbol{\hat{k}}$ a unit vector with the direction of the quark momentum  $\boldsymbol{k}$, $M_h$ is the hadron mass, and $D_1^{q/h}$ is the polarization-averaged fragmentation function. Here, the fragmenting quark of flavor $q$, as well as the identified hadron $h$ in the final state, has been added to the notation of the FFs in order to indicate the dependence of FFs on the final hadron to describe the cross section of back-to-back production discussed below. 
Equation~\eqref{eqn:FF1} describes an azimuthal modulation of the hadron momenta around the quark axis, with the strength of the modulation given by the Collins FF \CFF.
As described in the introduction, a measurement of the effect given by Eq.~\eqref{eqn:FF1} in single inclusive hadron production in $e^+e^-$ annihilation, i.e., in the process $e^+e^-\rightarrow h +X$, is not possible due to the chiral oddness of \CFF. Instead, the process $e^+e^-\rightarrow h_1 h_2 \mid_{\text{back-to-back}} +X$ is considered, where two back-to-back hadrons are detected. In this case, the Collins effect can be probed because it appears in a product of two chiral-odd quantities: the quark and antiquark Collins FF. The specific azimuthal modulation is in turn 
sensitive to the correlation of the transverse polarizations of the produced quark and anti-quark.

\begin{figure*} 
\includegraphics[width=0.8\textwidth]{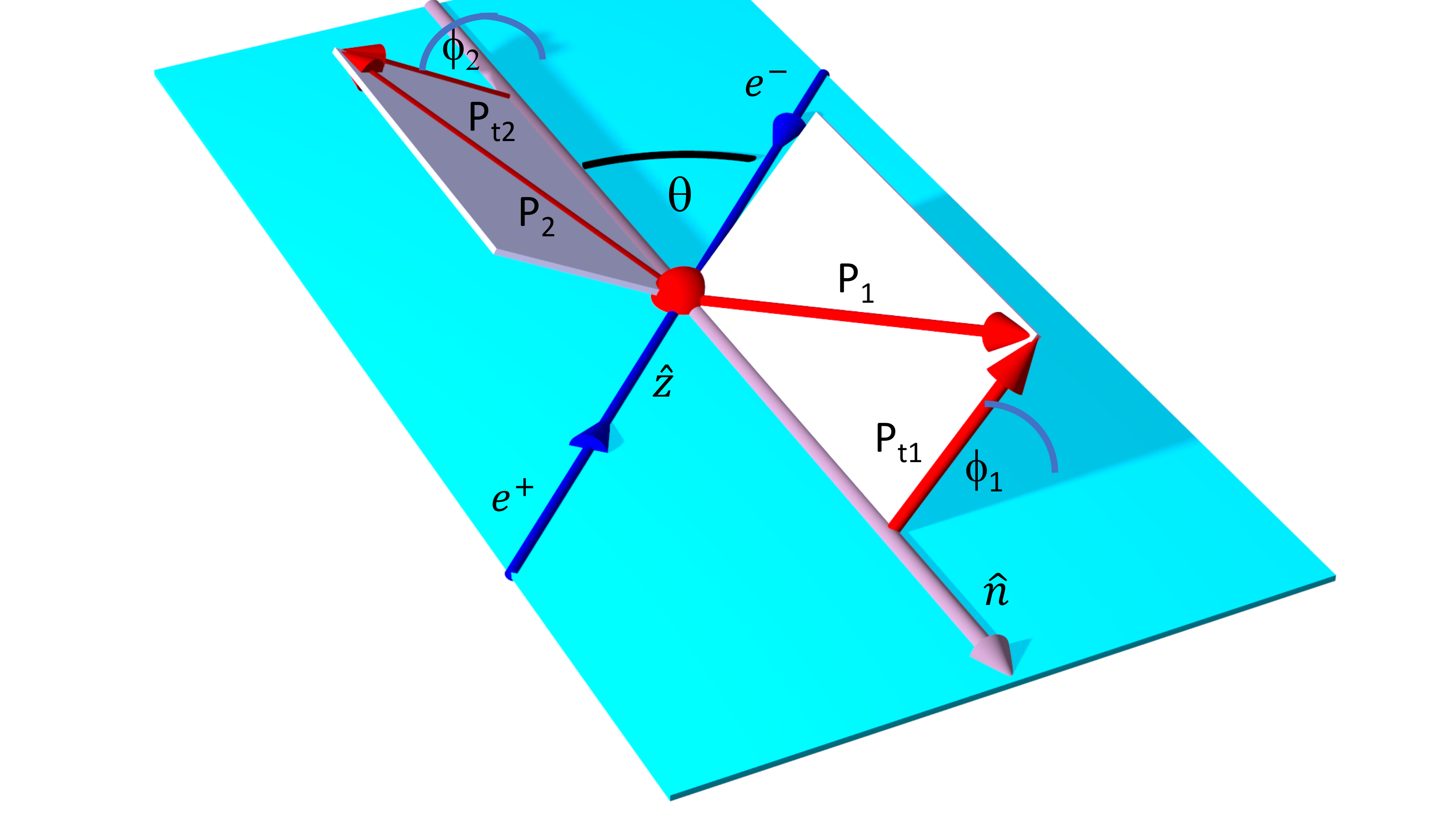}
\caption{Coordinate system used for this measurement\label{fig:coo}. The thrust axis is denoted $\boldsymbol{\hat{n}}$ and forms the angle $\theta$ in the c.m.~system with the beam axis (blue, color online). The thrust axis and beam axis span the event plane. The back-to-back hadrons with momenta $\boldsymbol{P}_{i}\: (i=1,2)$ form the azimuthal angles $\phi_{i}$ with the event plane. The transverse momenta of the hadrons with respect to the thrust axis are denoted $\boldsymbol{P}_{ti}$.}
\end{figure*}

The corresponding cross section for inclusive back-to-back production of two hadrons can be expressed as
\begin{widetext}
\begin{align}
\lefteqn{\frac{d\sigma(e^+e^-\rightarrow h_1 h_2 \mid_{\text{back-to-back}} +X)}{dy dz_1dz_2 d\boldsymbol{P}^2_{t1} d\boldsymbol{P}^2_{t2} d\phi_1d\phi_2}\propto} \nonumber \\
& \sum_{q,\bar{q}} \frac{3\alpha^2}{Q^2}\frac{e_q^2}{4}z^2_1z^2_2 \left\lbrace \left(\frac{1}{2}-y+y^2 \right) \, D_1^{q/h_1}(z_1,\boldsymbol{P}^2_{1\perp})\otimes D_1^{\bar{q}/h_2}(z_2,\boldsymbol{P}^2_{2\perp})\right. \nonumber \\  
& \left. + y(1-y) \, \cos(\phi_1+\phi_2) \, H_1^{\bot q/h_1}(z_1,\boldsymbol{P}^2_{1\perp})\otimes H_1^{\bot \bar{q}/h_2}(z_2,\boldsymbol{P}^2_{2\perp})\right\rbrace,
\label{eqn:cross_subsection_ee}
\end{align}
\end{widetext}
with $\otimes$ signifying convolutions over transverse momenta.
The invariant $y=(P_1 \cdot l)/(P_1 \cdot (l+l'))$ can be calculated from the 4-momenta of $h_1$, the electron, and the positron, \(P_1\), $l$, and $l'$, respectively.
The dependence on the quark polarization appearing in Eq.~\eqref{eqn:FF1} is now contained in the dependence on the azimuthal angles $\phi_1$ and $\phi_2$, which are measured between the hadron planes and the event plane as shown in Fig.~\ref{fig:coo}. 
The observable transverse momenta of the hadrons with respect to the thrust axis, which is defined below in Eq.~\eqref{eq:thrust}, are denoted $\boldsymbol{P}_{ti}$ and serve as a proxy for the parton level $\boldsymbol{P}_{i \perp}$.
 
Equation~\eqref{eqn:cross_subsection_ee} can be written more compactly as
\begin{equation}
d\sigma \sim A(y) D_1^{q/h_1} {D}_1^{\bar{q}/h_2} \, + \, B(y) \cos(\phi_1+\phi_2)H^{\bot q/h_1}_{1}{H}^{\bot \bar{q}/h_2 }_{1} .
\label{eqn:FF2}
\end{equation}
In the $e^+e^-$ center-of-mass (c.m.)~system, used in the following for all calculations, the kinematic factors $A$ and $B$ can be expressed as $A=\frac{1}{4}(1+\cos^2\theta)$ and $B=\frac{1}{4}(\sin^2\theta)$. The angle $\theta$ is the angle between the $q\bar{q}$ axis and the beam axis~\cite{Boer:2008fr}.
Since the transverse projection of the polarization can be calculated in QED as $(\sin^2\theta)/(1+\cos^2 \theta)$, the appearance of these factors is a reflection of the transverse-polarization dependence of \CFF.
In a leading-order partonic picture, the angles $\phi_i$ would be measured around the $q\bar{q}$ axis. As this quantity is not accessible, it is approximated by using the thrust axis. 
The thrust axis is defined as the unit vector $\hat{\boldsymbol{n}}$ that maximizes the thrust $T$:
\begin{equation}
\label{eq:thrust}
T=\sum_p\frac{|\boldsymbol{P_p\cdot\hat{n}}|}{|\boldsymbol{P_p}|}.
\end{equation}
The sum runs over all charged tracks and photons in the event. 

Using the thrust axis, it can be determined whether or not the hadrons $h_1$ and $h_2$ in a given pair are in different hemispheres (``back-to-back'') by requiring for their respective three-momenta $\boldsymbol{P}_{i}$:
\begin{equation}
(\boldsymbol{P_{1} \cdot \hat{n}})(\boldsymbol{{P}_{2}\cdot \hat{n}}) < 0 .
\end{equation}
The azimuthal angles $\phi_i$ are calculated as 
\begin{align}
\lefteqn{\phi_i=\sgn\left\lbrace\hat{\boldsymbol{n}}\cdot \left[ (\boldsymbol{\hat{z}}\times\hat{\boldsymbol{n}})\times(\hat{\boldsymbol{n}}\times{\boldsymbol{P}_{i}})\right] \right\rbrace\times}\nonumber\\ &\arccos\left(\frac{\hat{\boldsymbol{z}}\times\hat{\boldsymbol{n}}}{|\hat{\boldsymbol{z}}\times\hat{\boldsymbol{n}}|}\times\frac{\hat{\boldsymbol{n}}\times{\boldsymbol{P}_{i}}}{|\hat{\boldsymbol{n}}\times{\boldsymbol{P}_{i}}|}\right).
\label{eqn:collinsangledefine2}
\end{align}
Here, $\boldsymbol{\hat{z}}$ is the unit vector along the $e^+$ beam direction.

%


In the following, the Collins angle of a hadron pair is defined as $\phi_{12} \equiv \phi_1+\phi_2$. In terms of $\phi_{12}$, the hadron pair yield over all events for a given kinematic bin is given by $N_{12}\equiv N_{12}(\phi_{12})$. The normalized yield is computed from $N_{12}$ by dividing by the average yield: $R_{12}(\phi_{12})=(N_{12}(\phi_{12}))/(\langle  N_{12}\rangle)$. Considering only a $\cos(\phi_{12})$ modulation, $R_{12}$ can be parameterized as $R_{12}=1+a_{12}(\theta,z_1,z_2, \boldsymbol{P}^2_{t1},\boldsymbol{P}^2_{t2})\cos(\phi_{12})$, with the azimuthal asymmetry~\footnote{The parameters of the functional forms of the single ratios are denoted with small letters while capital letters are used for the parametrization of the later-introduced double ratios.}
\begin{align}
\lefteqn{a_{12}(\theta,z_1,z_2, \boldsymbol{P}^2_{t1},\boldsymbol{P}^2_{t2})=}\nonumber\\
&\frac{\sin^2\theta}{1+\cos^2\theta}
\frac{\sum\limits_{q,\bar{q}}e^2_qH^{\bot q/h_1}_1(z_1,\boldsymbol{P}^2_{1\perp})\otimes H^{\bot \bar{q}/h_2}_1(z_2,\boldsymbol{P}^2_{2\perp})}{\sum\limits_{q,\bar{q}}e^2_qD^{q/h_1}_1(z_1,\boldsymbol{P}^2_{1\perp})\otimes D^{\bar{q}/h_2}_1(z_2,\boldsymbol{P}^2_{2\perp})}.
\end{align} 
Note that in the expression for $a_{12}$ above,  the full dependence of the asymmetry $a_{12}$ on $\theta$, $z_i$, and $\boldsymbol{P}^2_{ti}$ is kept. In the measurements presented in this work, at most two variables are kept differential, the other ones are integrated over their accepted ranges.

Measured azimuthal distributions can be strongly distorted due to acceptance and  radiation effects. To remedy those effects the double ratio (DR) method can be used. A DR is the ratio of normalized distributions from different kinds of hadron pairs. Under the assumption that the effects are quark-/hadron-flavor independent, they largely cancel in double ratios~\cite{Seidl:2008xc,AsymmetryInEE,CollinsInSIDISandEE}. In the previous charged-pion analysis~\cite{Abe:2005zx,Seidl:2008xc,BabarCharged}, one double ratio was defined as the ratio of the normalized yield of unlike-sign  ($\pi^+\pi^-$) to that of like-sign pairs ($\pi^+\pi^+$ and $\pi^-\pi^-$).
In the current analysis this is extended to include neutral mesons:
\begin{equation}
\label{eqn:FF6}
\begin{aligned}
\mathcal{R}_{12}^{\pi^0}=\frac{R^{0\pm}_{12}}{R^L_{12}}&=\frac{\pi^0\pi^++\pi^0\pi^-}{\pi^+\pi^++\pi^-\pi^-},\\
\mathcal{R}_{12}^{\eta}=\frac{R^{\eta\pm}_{12}}{R^L_{12}}&=\frac{\eta\pi^++\eta\pi^-}{\pi^+\pi^++\pi^-\pi^-}.
\end{aligned}
\end{equation}
Here, $R^{0\pm}_{12}$ $(R^{\eta\pm}_{12},R^L_{12})$ denote the normalized yields of $\pi^0\pi^++\pi^0\pi^-$ $(\eta\pi^++\eta\pi^-,\pi^+\pi^++\pi^-\pi^-)$ pairs and the '$+$' sign between different combinations means that both pair combinations are considered for the yields.
For charged pions, asymmetries of like-sign pairs (L), unlike-sign pairs (U), or pairs that are summed over both charges (C) can be considered.
From these combinations the following two double ratios have traditionally been constructed:
\begin{equation}
\label{eqn:FF7}
\begin{aligned}
\mathcal{R}_{12}^{UL}=\frac{R^{U}_{12}}{R^L_{12}}&=\frac{\pi^+\pi^-+\pi^-\pi^+}{\pi^+\pi^++\pi^-\pi^-},\\
\mathcal{R}_{12}^{UC}=\frac{R^{U}_{12}}{R^C_{12}}&=\frac{\pi^+\pi^-+\pi^-\pi^+}{\pi^+\pi^++\pi^-\pi^-+\pi^+\pi^-+\pi^-\pi^+}.\\
\end{aligned}
\end{equation}
Analogue to the definition of $R^L_{12}$ for like-sign pairs, $R^U_{12}$ and $R^C_{12}$ denote the normalized yields of the unlike-sign and charge-summed pairs.
From $R^C_{12}$ and $R^L_{12}$ the double ratio 
\begin{equation}
\begin{aligned}
\label{eqn:RCL}
    \mathcal{R}_{12}^{CL}=\frac{R^{C}_{12}}{R^L_{12}}&=\frac{\pi^+\pi^++\pi^-\pi^-+\pi^+\pi^-+\pi^-\pi^+}{\pi^+\pi^++\pi^-\pi^-}
    \end{aligned}
\end{equation}
is constructed, which is interesting in the context of neutral pions as being equal to the $\pi^0$ double ratio $\mathcal{R}_{12}^{\pi^0}$ due to isospin symmetry~\cite{Efremov:2006qm}.

The double ratios~\eqref{eqn:FF6}-\eqref{eqn:RCL} contain the fragmentation functions of interest in various combinations. To simplify expressions, fragmentation functions are often categorized into favored and disfavored, depending on whether or not the fragmenting-quark flavor is part of the valence structure of the hadron formed. For pions, employing charge and isospin symmetry, the non-strange FFs are~\cite{FoundationsofpQCD,Efremov:2006qm}
\begin{equation}
\begin{aligned}
\label{eqn:FF4}
D^{fav} &\equiv D^{u/{\pi^+}}=D^{d/{\pi^-}}=D^{\bar{u}/{\pi^-}}=D^{\bar{d}/{\pi^+}},\\
D^{dis} &\equiv D^{u/{\pi^-}}=D^{d/{\pi^+}}=D^{\bar{u}/{\pi^+}}=D^{\bar{d}/{\pi^-}}, \\
D^{u/{\pi^0}} &= D^{\bar{u}/{\pi^0}}=D^{d/{\pi^0}}=D^{\bar{d}/{\pi^0}}=\frac{1}{2}(D^{dis}+D^{fav}).
\end{aligned}
\end{equation}
%
Besides up and down quarks, the contribution of strange quarks is considered here.~\footnote{Charm is qualitatively different due to its mass and the dominance of weak decay channels in pion production. In particular the Collins effect for charm quarks is expected to be small and found so in charm enhanced data samples at Belle and BaBar~\cite{Abe:2005zx,Seidl:2008xc,BabarCharged}.}
Employing the same symmetry arguments as before, the probability for strange-quark fragmentation is the same for all pion states, thus
\begin{equation}
\begin{aligned}
D^{dis}_{s\rightarrow\pi}&=D^{s/{\pi^-}}=D^{s/{\pi^+}}=D^{s/{\pi^0}}\\
&=D^{\bar{s}/{\pi^-}}=D^{\bar{s}/{\pi^+}} =D^{\bar{s}/{\pi^0}}.
\label{eq:pistrange}
\end{aligned}
\end{equation}
In a similar way the number of FFs for $\eta$ production can be reduced to 
\begin{equation}
\begin{aligned}
D^{u/{\eta}}&=D^{d/{\eta}}=D^{\bar{u}/{\eta}}=D^{\bar{d}/{\eta}}=\frac{1}{2}\left(D^{fav_\eta}+D^{dis_\eta}\right), \\
D_{s\rightarrow\eta}&=D^{s/{\eta}}=D^{\bar{s}/{\eta}}.
\label{eqn:FFetaquark}
\end{aligned}
\end{equation}
Since strange quarks are part of the \(\eta\) valence structure, the respective fragmentation function is not disfavored as is the case of the $\pi^0$ fragmentation functions.
 
The various double ratios
can then be expressed in terms of these FFs~\cite{Efremov:2006qm}. Using only the first term of a Taylor expansion in $\cos(\phi_{12})$ one obtains
\begin{widetext}
\begin{equation}
\begin{aligned}
\mathcal{R}^{UL}_{12}&\approx1+\cos(\phi_{12})\frac{\sin^2(\theta)}{1+\cos^2(\theta)}\\
&\times\bigg\{
  \frac{5(H^{\bot,fav}_1 \otimes H^{\bot,fav}_1 + H^{\bot,dis}_1 \otimes H^{\bot,dis}_1) 
            + 2H^{\bot,dis}_{1,s\rightarrow\pi} \otimes H^{\bot,dis}_{1,s\rightarrow\pi}}
       {5(D^{fav}_1 \otimes D^{fav}_1 + D^{dis}_1 \otimes D^{dis}_1) + 2D^{dis}_{1,s\rightarrow\pi} \otimes D^{dis}_{1,s\rightarrow\pi}}\\
&\quad -\frac{10 H^{\bot,fav}_1 \otimes H^{\bot,dis}_1
            + 2 H^{\bot,dis}_{1,s\rightarrow\pi} \otimes H^{\bot,dis}_{1,s\rightarrow\pi}}
       {10 D^{fav}_1 \otimes D^{dis}_1 + 2 D^{dis}_{1,s\rightarrow\pi} \otimes D^{dis}_{1,s\rightarrow\pi}} \bigg\},
\label{eqn:allratiosexpress2}
\end{aligned}
\end{equation}
\begin{equation}
\begin{aligned}
\mathcal{R}^{UC}_{12}&\approx1+\cos(\phi_{12})\frac{\sin^2(\theta)}{1+\cos^2(\theta)}\\
&\times\bigg\{
  \frac{5(H^{\bot,fav}_1 \otimes H^{\bot,fav}_1 + H^{\bot,dis}_1 \otimes H^{\bot,dis}_1) 
            + 2H^{\bot,dis}_{1,s\rightarrow\pi} \otimes H^{\bot,dis}_{1,s\rightarrow\pi}} 
       {5(D^{fav}_1 \otimes D^{fav}_1 + D^{dis}_1 \otimes D^{dis}_1) + 2D^{dis}_{1,s\rightarrow\pi} \otimes D^{dis}_{1,s\rightarrow\pi}}\\
&\quad -\frac{5(H^{\bot,fav}_1+H^{\bot,dis}_1) \otimes (H^{\bot,fav}_1+H^{\bot,dis}_1) 
            + 4H^{\bot,dis}_{1,s\rightarrow\pi} \otimes H^{\bot,dis}_{1,s\rightarrow\pi}}
       {5(D^{fav}_1+D^{dis}_1) \otimes (D^{fav}_1+D^{dis}_1) + 4D^{dis}_{1,s\rightarrow\pi} \otimes D^{dis}_{1,s\rightarrow\pi}} \bigg\},
\label{eqn:allratiosexpress3}
\end{aligned}
\end{equation}
and in particular 
\begin{equation}
\begin{aligned}
\mathcal{R}_{12}^{\pi^0}&=\frac{R^{0\pm}_{12}}{R^L_{12}}\approx1+\cos(\phi_{12})\frac{\sin^2(\theta)}{1+\cos^2(\theta)} \\
&\times\bigg\{
  \frac{5(H^{\bot,fav}_1+H^{\bot,dis}_1) \otimes (H^{\bot,fav}_1+H^{\bot,dis}_1) 
            + 4H^{\bot,dis}_{1,s\rightarrow\pi} \otimes H^{\bot,dis}_{1,s\rightarrow\pi}}
       {5(D^{fav}_1+D^{dis}_1) \otimes (D^{fav}_1+D^{dis}_1) + 4D^{dis}_{1,s\rightarrow\pi} \otimes D^{dis}_{1,s\rightarrow\pi})}\\
&\quad -\frac{10 H^{\bot,fav}_1 \otimes H^{\bot,dis}_1 
            + 2H^{\bot,dis}_{1,s\rightarrow\pi}H^{\bot,dis}_{1,s\rightarrow\pi}}
       {10 D^{fav}_1 \otimes D^{dis}_1  + 2D^{dis}_{1,s\rightarrow\pi} \otimes D^{dis}_{1,s\rightarrow\pi}} \bigg\}.
\label{eqn:FF5}
\end{aligned}
\end{equation}


Using Eq.~\eqref{eqn:FFetaquark} results in the following expression for the \(\eta\) double ratio:
\begin{equation}
\begin{aligned}
\mathcal{R}_{12}^{\eta}&=\frac{R^{\eta\pm}_{12}}{R^L_{12}}\approx1+\cos(\phi_{12})\frac{\sin^2(\theta)}{1+\cos^2(\theta)} \\
&\times\bigg\{
  \frac{5(H^{\bot,fav_\eta}_1+H^{\bot,dis_\eta}_1) \otimes (H^{\bot,dis}_1+H^{\bot,fav}_1) 
            +4H^{\bot}_{1,s\rightarrow\eta} \otimes H^{\bot,dis}_{1,s\rightarrow\pi}}
       {5(D^{\bot,fav_\eta}_1+D^{\bot,dis_\eta}_1) \otimes (D^{\bot,dis}_1+D^{\bot,fav}_1) 
        + 4D_{1,s\rightarrow\eta} \otimes D^{dis}_{1,s\rightarrow\pi})}\\
&\quad -\frac{10 H^{\bot,fav}_1 \otimes H^{\bot,dis}_1  
            + 2H^{\bot,dis}_{1,s\rightarrow\pi} \otimes H^{\bot,dis}_{1,s\rightarrow\pi}}
       {10 D^{\bot,fav}_1 \otimes D^{\bot,dis}_1 + 2D^{dis}_{1,s\rightarrow\pi} \otimes D^{dis}_{1,s\rightarrow\pi}} \bigg\}.
\label{eqn:FF5eta}
\end{aligned}
\end{equation}
\end{widetext}
%

In the measurement presented here, a parametrization of the form $1+A_{12} \cos(\phi_{12})$ is fitted to the double ratios.
The amplitude $A_{12}$ of the $\cos(\phi_{12})$ modulation is the azimuthal asymmetry that is presented for various meson combinations and binnings in $z$ and $P_t$.

\section{Experiment}
\label{sec:experiment}
The Belle experiment~\cite{BelleDetector} at the KEKB storage ring~\cite{KEKB} recorded about 1~ab$^{-1}$ of $e^+e^-$ annihilation data. The data were taken mainly at the $\Upsilon(4S)$ resonance at $\sqrt{s}=10.58$~GeV,     but also at other $\Upsilon(1S)$ to $\Upsilon(5S)$  resonances and at
    a continuum setting of $\sqrt{s}=10.52$~GeV. This analysis used data
    from all these sources for a total integrated luminosity of $980.4$~$\textrm{fb}^{-1}$. 
The Belle instrumentation used in this analysis includes a central drift chamber (CDC) and a silicon vertex detector, which provide precision tracking for tracks in $0.30$~rad$ < \theta_\textrm{Lab}< 2.62$~rad, 
and electromagnetic calorimeters (ECL)~\cite{ECL} covering the same region. The complete ECL consists of 8736 CsI(Tl) counters, which are subdivided into the barrel region ($0.56$~rad $< \theta_\textrm{Lab} < 2.25$~rad) and the endcaps. This analysis uses the barrel ECL for the reconstruction of $\pi^0$ and $\eta$ mesons. 
Particle identification is performed using information on dE/dx in the CDC, a time-of-flight system in the barrel, aerogel Cherenkov counters in the barrel and the forward endcap, as well as a muon and $K_L$ identification system embedded in the flux return steel outside the
    superconducting solenoid coils. The magnet provides a 1.5~T magnetic field.
Using these systems, the selection of charged pions in the barrel, which is used in this analysis, achieves a purity of 97\% over all kinematic bins.

\section{Analysis}
\label{sec:analysis}

\begingroup
 \begin{table*}[tp]
  \begin{ruledtabular}  
\caption{Constraints applied in the analysis. The ones that are different in this analysis compared to previous Belle Collins analyses~\cite{Abe:2005zx,Seidl:2008xc} are set in bold. (See text for description.)\label{tab:cuts}}
    \begin{tabular}{ll} 
%
Description & Constraint\\
\hline
Minimum visible energy $E_\textrm{vis}$  &$E_\textrm{vis}>7$~GeV\\
Thrust $T$ & $T>0.8$\\
Opening angle $\alpha_O$ of reconstructed meson \em{w.r.t.} $\boldsymbol{\hat{n}}$\qquad \phantom{f} &{\boldmath $\alpha_O< 0.3$~{\bf rad}}\\
Thrust axis polar angle $\theta$& {\boldmath $1.34$~{\bf rad~}$< \theta<2.03$~{\bf rad}}\\
Minimum photon energy $E_{\gamma,\pi^0}$  for $\pi^0$ &{\bf \boldmath $E_{\gamma,\pi^0}>50$~MeV} \\
Minimum photon energy  $E_{\gamma,\eta}$  for $\eta$ &{\bf \boldmath $E_{\gamma,\eta}>150$~MeV} \\
Opening angle $\alpha_{O,\gamma}$ for photons \em{w.r.t.} $\boldsymbol{\hat{n}}$&{\boldmath $\alpha_{O,\gamma}<0.5$~{\bf rad}} \\
\end{tabular}
  \end{ruledtabular}
 \end{table*}
\endgroup

As in previous similar Belle extractions of azimuthal asymmetries of hadrons and di-hadron pairs~\cite{Abe:2005zx,Seidl:2008xc,Vossen:2011fk}, hadronic events are selected by requiring a minimum visible energy of 7~GeV and a thrust $T > 0.8$. 
These constraints reduce the contribution of $\tau$ leptons and $B$ mesons to below 1\% and allow    the inclusion of all on- and off-resonance data in the analysis.
A number of fiducial constraints are applied in the c.m.~system with the goal to minimize effects from variations of the acceptance of the detector on the extracted asymmetries.
For this reason only mesons reconstructed from tracks and photons in the barrel region of the detector are considered. Table~\ref{tab:cuts} lists the fiducial as well as the other constraints applied. 
This work expands the previous charged-pion analysis~\cite{Abe:2005zx,Seidl:2008xc} to $\pi^0$ and $\eta$ mesons, which requires adaptation of several differing or additional selection requirements. They are highlighted in Table~\ref{tab:cuts}.
No correction of the asymmetries for these kinematic restrictions are applied, i.e., the asymmetries extracted are averages in the so-defined phase space.

To minimize the impact of the fiducial constraints on the extracted asymmetry, a hierarchical set of opening-angle constraints on photons, hadron momenta, and the thrust axis is applied. This ensures that the detector acceptance of all mesons is radially symmetric around the thrust axis and the acceptance in $z$ and $P_t$ of charged and neutral mesons is approximately equal. All photons used for the reconstruction of $\pi^0$ and $\eta$ mesons have a maximal opening angle of 0.5~rad from the thrust axis. All charged and reconstructed neutral mesons used in the asymmetry computation are required to have a maximal opening angle of 0.3~rad from the thrust axis in the c.m.~system. Finally, dictated by the geometric acceptance of the ECL, the thrust-axis polar angle is restricted to  \( 1.34 \text{~rad} < \theta < 2.03 \text{~rad} \)  to ensure the radial symmetry of the acceptance for photons inside the barrel around the thrust axis. 
To reconstruct $\pi^0$ and $\eta$ mesons, pairs of photons are used for which a minimum energy of 50~MeV and 150~MeV, respectively, is required to reduce background due to combinatorics.

\begin{figure*} 
  \centering     
  \subfigure[{ }$\pi^0$ invariant-mass fit, $0.2<z<0.3$]{\label{fig:pi0crystalfit_1}\includegraphics[clip, trim=0.4cm 0.2cm 2.8cm 1.1cm,width=.48\textwidth]{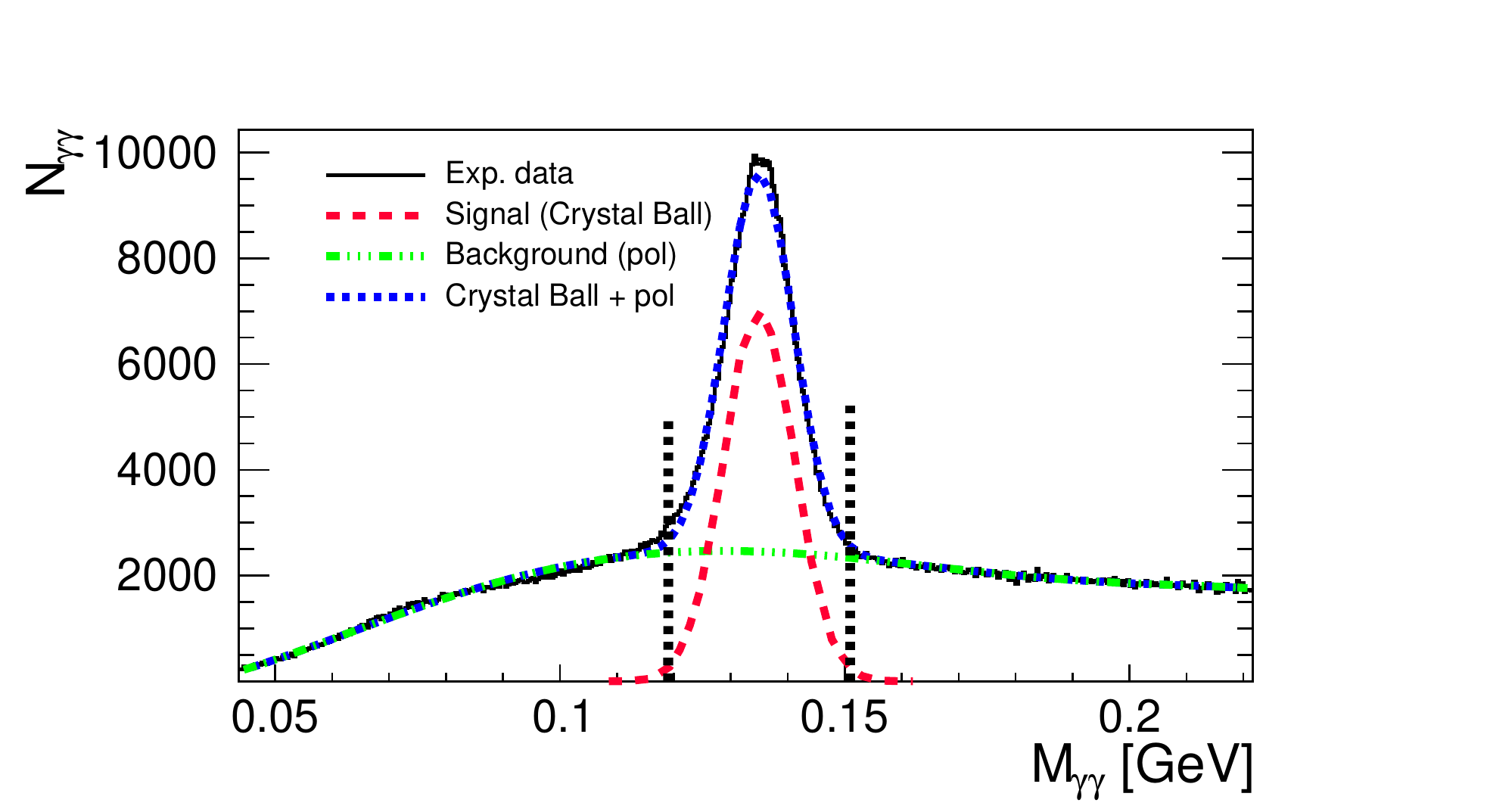}}
  \subfigure[{ }$\pi^0$ invariant-mass fit, $0.6<z<0.7$]{\label{fig:pi0crystalfit_2}\includegraphics[clip, trim=0.4cm 0.2cm 2.8cm 1.1cm,width=.48\textwidth]{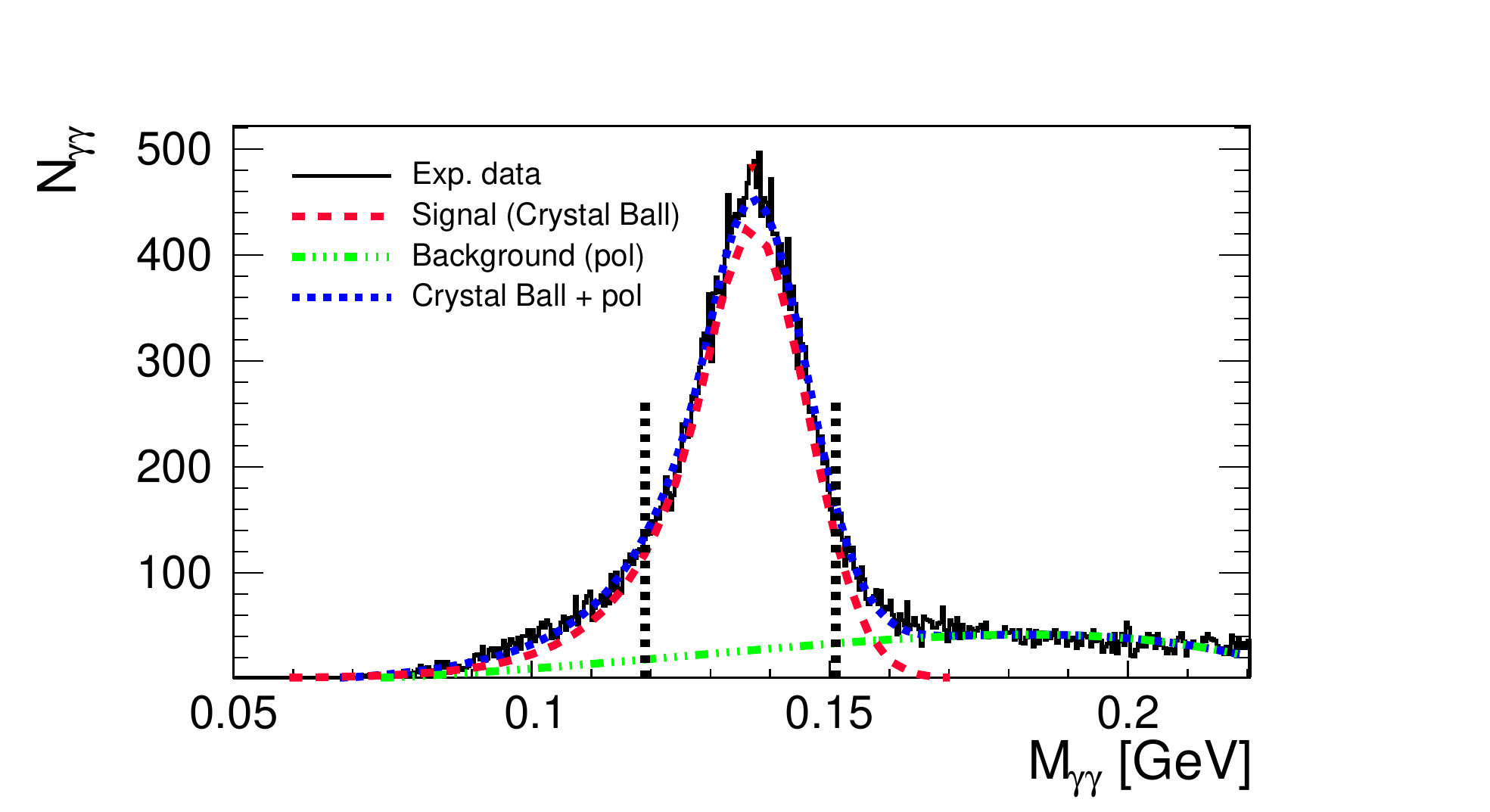}}
    \subfigure[{ }$\eta$ invariant-mass fit, $0.3<z<0.4$]{\label{fig:etacrystalfit_1}\includegraphics[clip, trim=0.4cm 0.2cm 2.8cm 1.1cm,width=.48\textwidth]{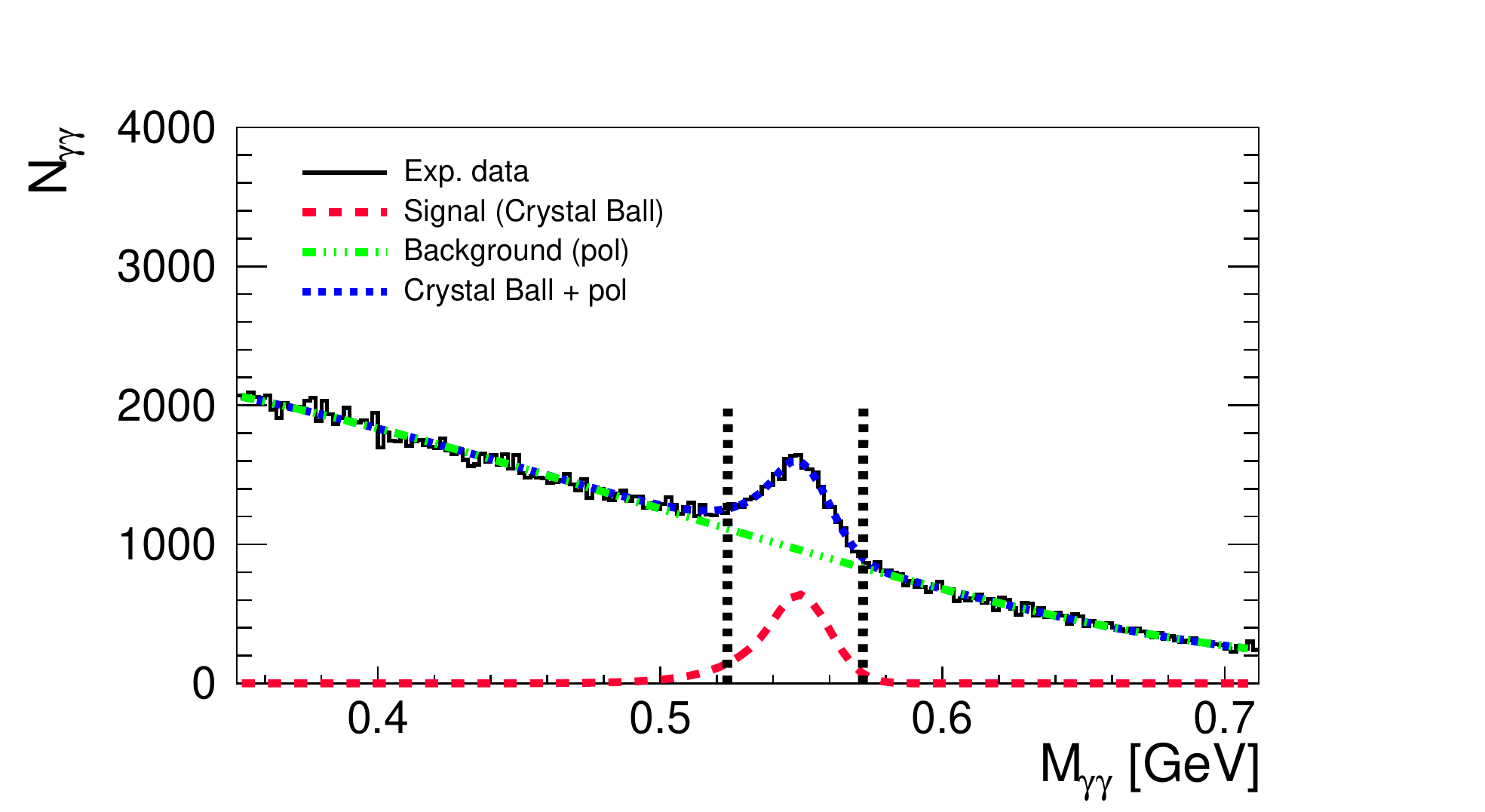}}
  \subfigure[{ }$\eta$ invariant-mass fit, $0.6<z<0.7$]{\label{fig:etacrystalfit_2}\includegraphics[clip, trim=0.4cm 0.2cm 2.8cm 1.0cm,width=.48\textwidth]{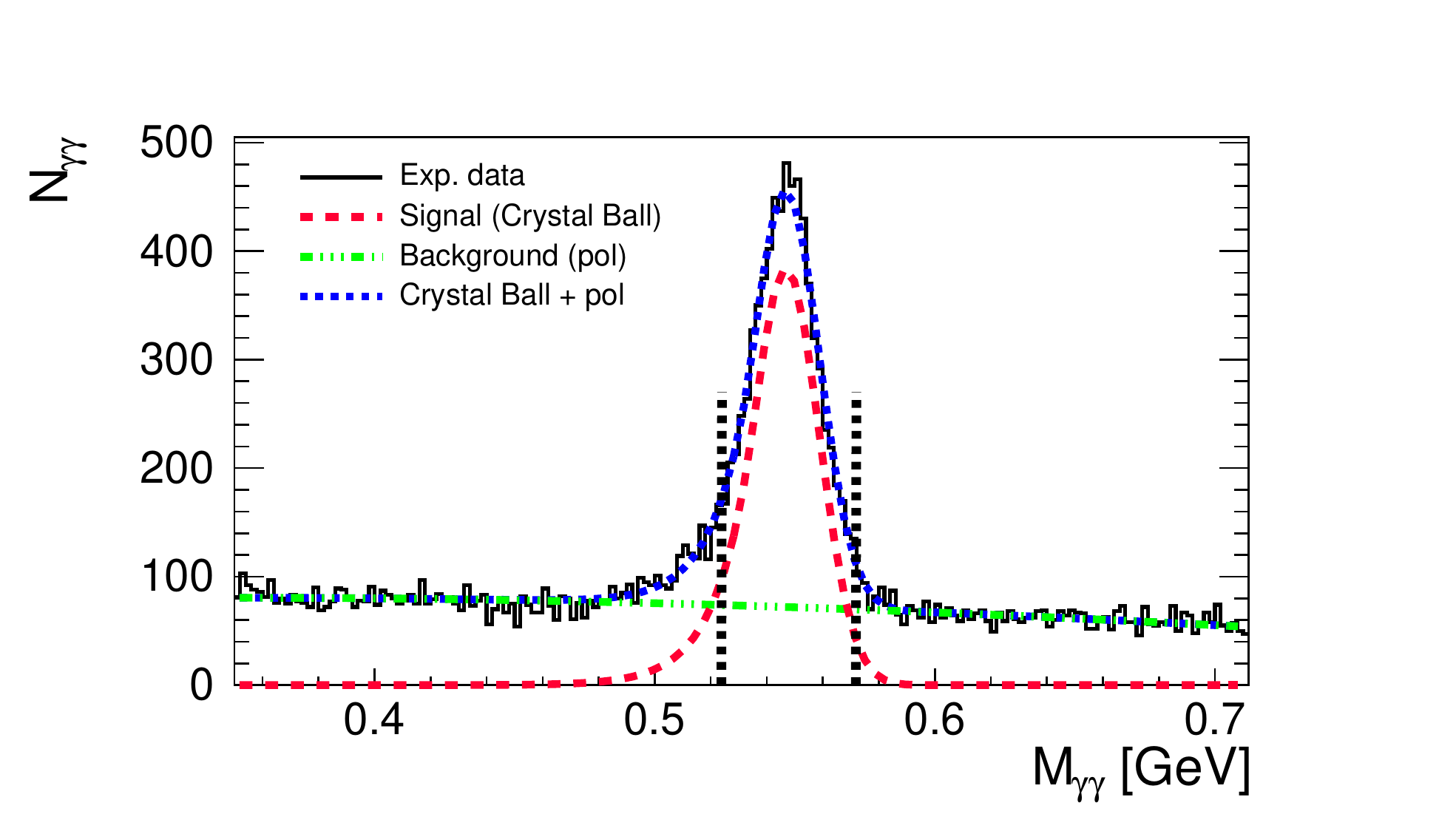}}
  \caption{Typical two-photon invariant-mass distributions, fit using a Crystal-Ball function for the signal and a polynomial background function, for $\pi^0$ (top plots) and $\eta$ (bottom plots) mesons. In each plot, the green  dash-dotted line represents the fitted background using a polynomial of fifth order, the  red dashed line the fitted signal, and the blue dotted line is the combined background and signal fit. The combined fit agrees well with the experimental data in black. The vertical dashed lines indicate the boundaries used in the analysis for signal events.}
  \label{fig:crystalfit}
\end{figure*}

The yields of $\pi^0$ and $\eta$ mesons in each kinematic bin are extracted from a fit to the two-photon invariant-mass distribution, with a Crystal-Ball~\cite{CrystalBallFunc} function for the signal and a fifth-order polynomial for the background. The signal to background ratio determined in this way is then used to correct the measured raw asymmetry for the background contribution in the respective kinematic bin in the way described below.
Some exemplary fits for $\pi^0$ and $\eta$ mesons are shown in Fig.~\ref{fig:crystalfit}. The measured invariant-mass distributions from experimental data were compared with those from simulations.
The simulations used in this analysis employ Pythia~\cite{Sjostrand:2006za} and EvtGen~\cite{Lange:2001uf} for various physics processes not including the polarization-dependent Collins effect, and GEANT3~\cite{Brun:1987ma} for the detector effects.  
For low-$z$ bins some disagreement between the shape of the invariant-mass distributions of reconstructed $\pi^0$s in experimental data and simulation was observed. Therefore an almost non-parametric method, which does not rely on the fit of the signal, was evaluated as well. 
The method is based on the observation that the background, defined as any pair of electromagnetic clusters in the ECL that do not come from the same $\pi^0$, is well described by the simulation in the sideband region both in magnitude and shape. 
Hence, instead of fitting the entire invariant-mass spectrum with a background and a signal component, a background description using a quadratic function fitted to 20 points in the upper and lower sidebands obtained from MC, respectively, is used. Once determined in this way, the background is subtracted from the measured invariant-mass spectrum leaving the remaining yield as the signal.
The difference between the two extraction methods for the final asymmetry is small, typically less than one per mille in absolute asymmetry value, and is added to the systematic uncertainties.


Using the reconstructed $\pi^0$ and $\eta$ mesons, as well as charged pions that are reconstructed using the Belle tracking and particle identification subsystems described in \hyperref[sec:experiment]{Sec.~\ref{sec:experiment}}, pairs of ``back-to-back'' hadrons are constructed. This is done by assigning a hemisphere to each meson in the event based on the projection on the thrust axis $\boldsymbol{\hat{n}}$ and then considering all combinations of hadrons in the first hemisphere with those in the second.
Utilizing the thrust axis, the azimuthal angles $\phi_1$ and $\phi_2$ for these ``back-to-back'' pairs of mesons are computed using Eq.~(\ref{eqn:collinsangledefine2}).

Double ratios of $\phi_{12}$-dependent yields 
are constructed for the various meson pairs. A cosine function is fitted to the data in order to extract raw asymmetries binned in various combinations of  $z_i$ and $P_{ti}$. Here, $i=1$ always refers to the neutral meson in the pair when applicable. For pairs of charged pions, the assignment of the first and second pion in a pair is random. 
Since smearing effects are largest and the Collins effect is smallest at low $z$, a constraint of $z_1>0.2$ is used, with the exception of the results that are binned in both $z_1$ and $z_2$, where $z_i>0.1$ is used. 
The bin boundaries for the $P_t$ binning are $0,0.15,0.3,0.5,$  and $3$~GeV.
For the binning in $z_i$, bin boundaries differ between results only binned in $z_1$ and those binned in both $z_1$ and $z_2$.
In the former case, bins of $[0.2-0.3], [0.3-0.4],[0.4-0.5],[0.5-0.6],[0.6-0.7],[0.7-1.0]$ and in the latter case, bins of $[0.1-0.2], [0.2-0.3],[0.3-0.5],[0.5-0.7],[0.7-1.0]$ are used. 
For the $\eta$, due to its higher mass, an additional constraint of $z>0.3$ is added for all mesons in the respective pairs.

To arrive at the final asymmetries, several corrections are applied to the raw asymmetries as explained below.

First, the raw asymmetries for $\pi^0$ and $\eta$ mesons are corrected for the contribution from the combinatorial background. The background contribution is determined by calculating asymmetries using $\gamma$ pairs with a reconstructed mass in the sideband region of the $\pi^0$ ($\eta$) invariant-mass distribution. Given the limited statistics in this region, four values of the asymmetry are calculated, two in the lower sideband and two in the upper sideband. The observed background asymmetries on both sides of the $\pi^0$ ($\eta$) signal are consistent with each other and we use a linear fit to extract the contribution of the background to the asymmetry in the signal region using the signal-to-background ratio extracted from the fits to the invariant-mass spectra described earlier.

Second, false asymmetries, determined from simulations, are subtracted. Since the simulation does not contain the Collins effect, any residual asymmetry is a systematic error. 
These residual asymmetries are consistent with zero within their statistical uncertainties, which are added to our final systematic uncertainties. The relative contribution of these uncertainties ranges from the sub-percent level at low $z$ to a few percent at high $z$.

Finally, the  asymmetries are corrected for thrust-smearing and bin-migration effects. The smearing of the reconstructed $z$ values is negligible due to the excellent momentum reconstruction of the Belle apparatus. In contrast, bin migration is significant for the reconstructed $P_t$.
The reason for this is that $P_t$ is defined with respect to the thrust axis, the latter suffering from sizable misreconstruction due to particles missed outside of the detector acceptance.

To estimate and correct for the effect of the smearing in $P_t$, a reweighted simulation sample was used. Reweighting the existing simulation is necessary, as the original simulation does not contain the Collins effect. 
The procedure used weights for each reconstructed hadron pair by assigning a weight $w_i=1+A\cos(\phi^i_{12})$, where $A$ is the amplitude of the injected Collins effect and $\phi^i_{12}$ the Collins angle of the $i^\text{th}$ pair.

The goal of the reweighting of the simulation is the reproduction of the shape of the double-ratio asymmetries observed in the data. The $P_t$ dependence of the extracted asymmetries, discussed in more detail in \hyperref[sec:results]{Sec.~\ref{sec:results}}, is well described by a linear function in each $z$ bin. Therefore, a $(P_ {t1},P_{t2})$-dependent amplitude of the form $A(P_{t1},P_{t2})=1+a_{N,D}P_{t1}P_{t2}$ was chosen for the reweighting in each $z$ bin.  The observed double ratios determine the amplitudes of modulation in the numerator ($a_N$) and denominator ($a_D$) only up to a common scaling factor. 
 The dependence of the smearing factor on this scaling factor and on reasonable variations of the ratio $a_N/a_D$ was observed to be negligible. 
 Using this reweighted simulation, a correction factor $f_S$ for each bin is calculated as the ratio of the input double-ratio asymmetries and the reconstructed double-ratio asymmetries. For the former, the generated kinematics of the detected hadrons are used and the thrust axis is computed taking all generated particles in the event into account, including those that are outside of the acceptance of the spectrometer.

The statistical uncertainties in $f_S$ contribute to the final systematic uncertainty. Values for $f_S$ are between $f_S=1.2$ and $f_S=1.3$, with the exception of the kinematic boundaries in the lowest $P_t$ bin or when both particles in the pair are in the highest $z$ bin. Here, the hadrons are close to the thrust axis, enhancing smearing effects, and the correction factor takes values between $f_S=1.4$ and $f_S=1.5$, depending on the particle species. The relative uncertainty on $f_S$ is again driven by the Monte Carlo statistics and is below 2\% in the single-$z$ binning, while for the binning in the $z$ values of both hadrons it is below 3\% for most bins, but reaches 10\% for the highest $(z_1$,$z_2)$ bin. 


The applied corrections for smearing effects, background contributions, and false asymmetries can be summarized by
\begin{equation}
A_{12}=(A_\textrm{raw,bg-corrected}-A_\textrm{MC}) \, f_S\, .
\label{eq:masterCorrection}
\end{equation}
Here, $A_\textrm{raw,bg-corrected}$ is the raw asymmetry after background correction.
$A_\textrm{MC}$ is the false asymmetry measured in simulation.
Finally, the asymmetry is corrected for smearing using the smearing correction $f_S$.
Similarly, systematic uncertainties that arise from the statistical uncertainties on the smearing effects, the background contribution, and the false asymmetries can be summarized by
\begin{equation}
\sqrt{ (A_{12})^2 \left(\frac{\delta{f_S}}{f_S} \right)^2 + (f_S \, \delta F)^2 + (f_S \, \delta {A_\textrm{MC}})^2} \, .
\label{eq:masterSys}
 \end{equation}
 Here, $\delta F$ is the systematic uncertainty stemming from the differences in extracted raw asymmetries using the two different fit procedures.

\section{Results and Discussion}
\label{sec:results}

Azimuthal asymmetries are measured for double ratios involving charged pions, neutral pions, and eta mesons. Their cosine amplitudes are extracted in various kinematic binnings including  $z$, \(P_{t}\), and a mixed \(z\)--\(P_{t}\) binning. Significantly non-zero cosine amplitudes are found for all double ratios examined, with magnitudes of mainly a few percent but reaching up to 20\% in certain kinematic corners, as pointed out further below.

\begin{figure*} [t]
\includegraphics[clip, trim=0.1cm 2.4cm 0.2cm 0.2cm,height=0.25\textheight,width=0.55\textwidth]{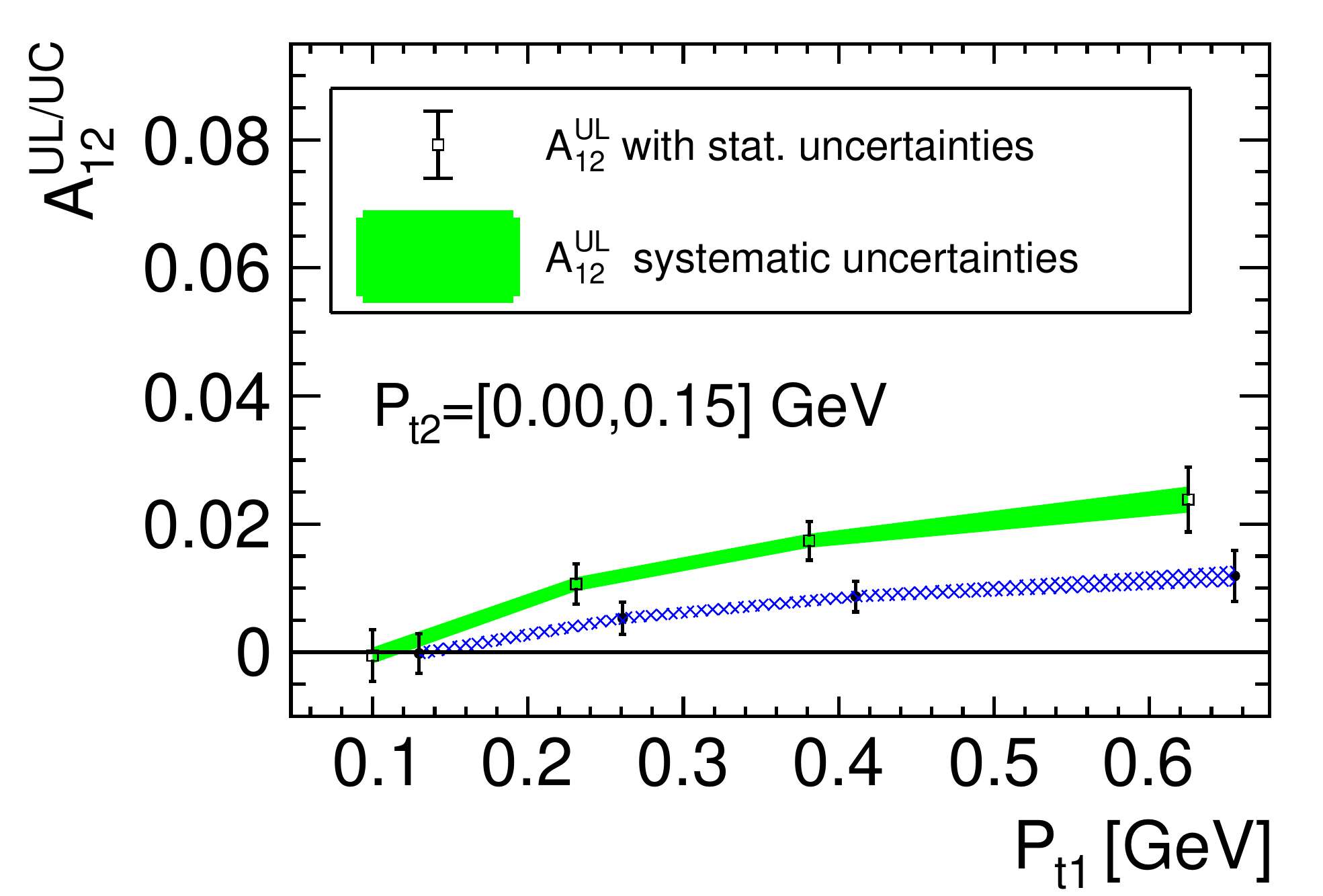}\includegraphics[clip, trim=4.4cm 2.4cm 0.2cm 0.2cm,height=0.25\textheight,width=0.44\textwidth]{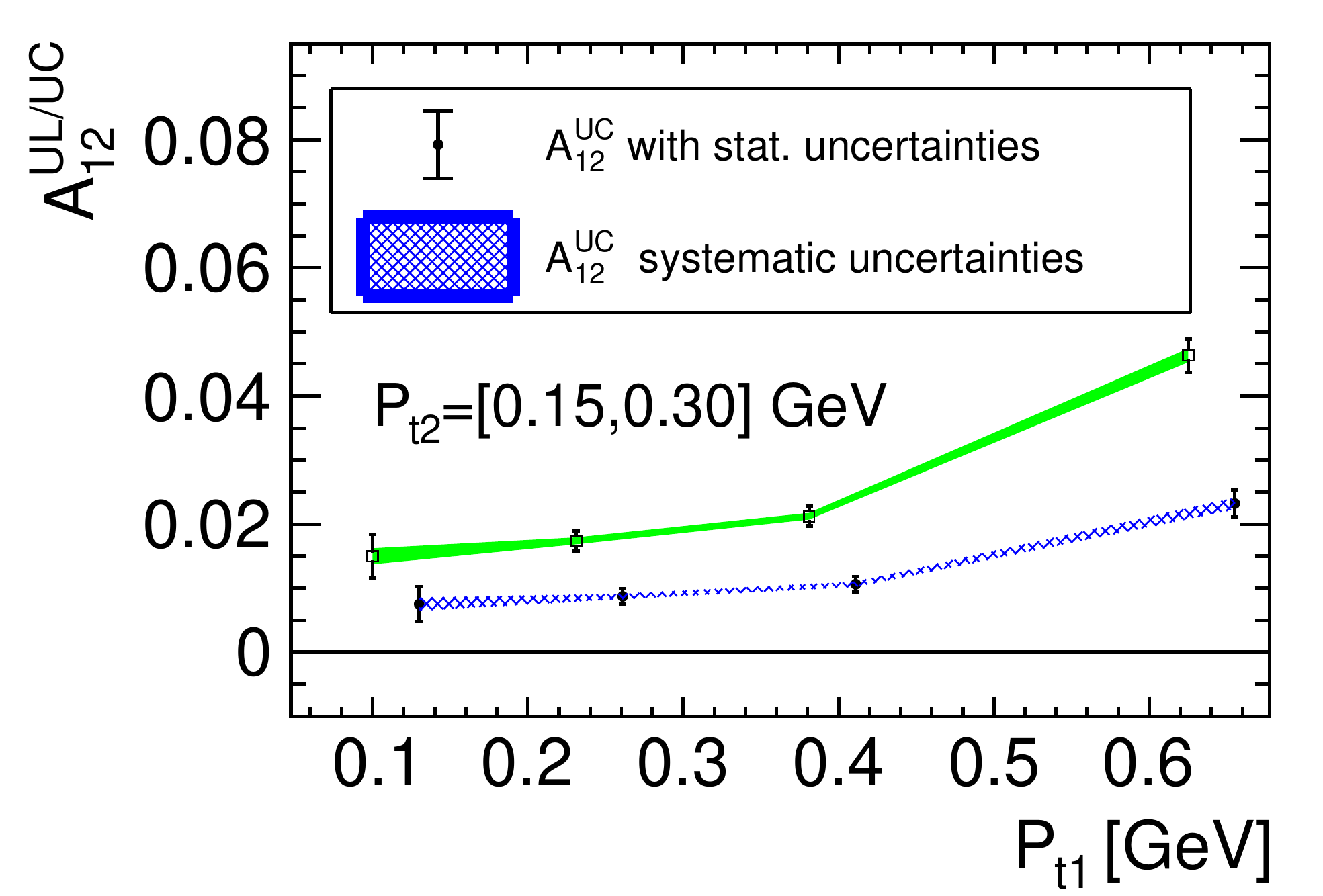}\\
\includegraphics[clip, trim=0.1cm 0.0cm 0.2cm 0.5cm,height=0.27\textheight,width=0.55\textwidth]{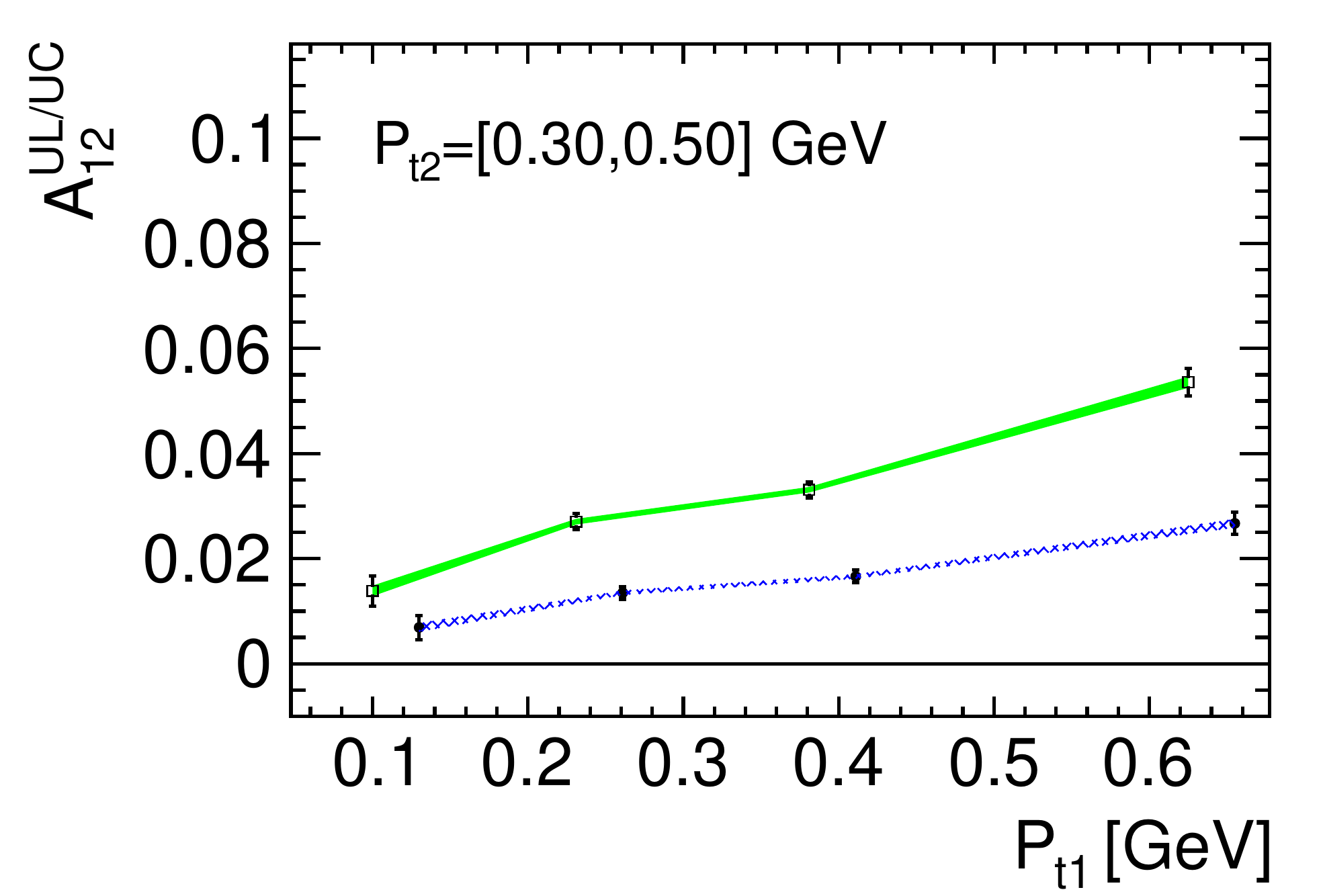}\includegraphics[clip, trim=4.4cm 0.0cm 0.2cm 0.5cm,height=0.27\textheight,width=0.44\textwidth]{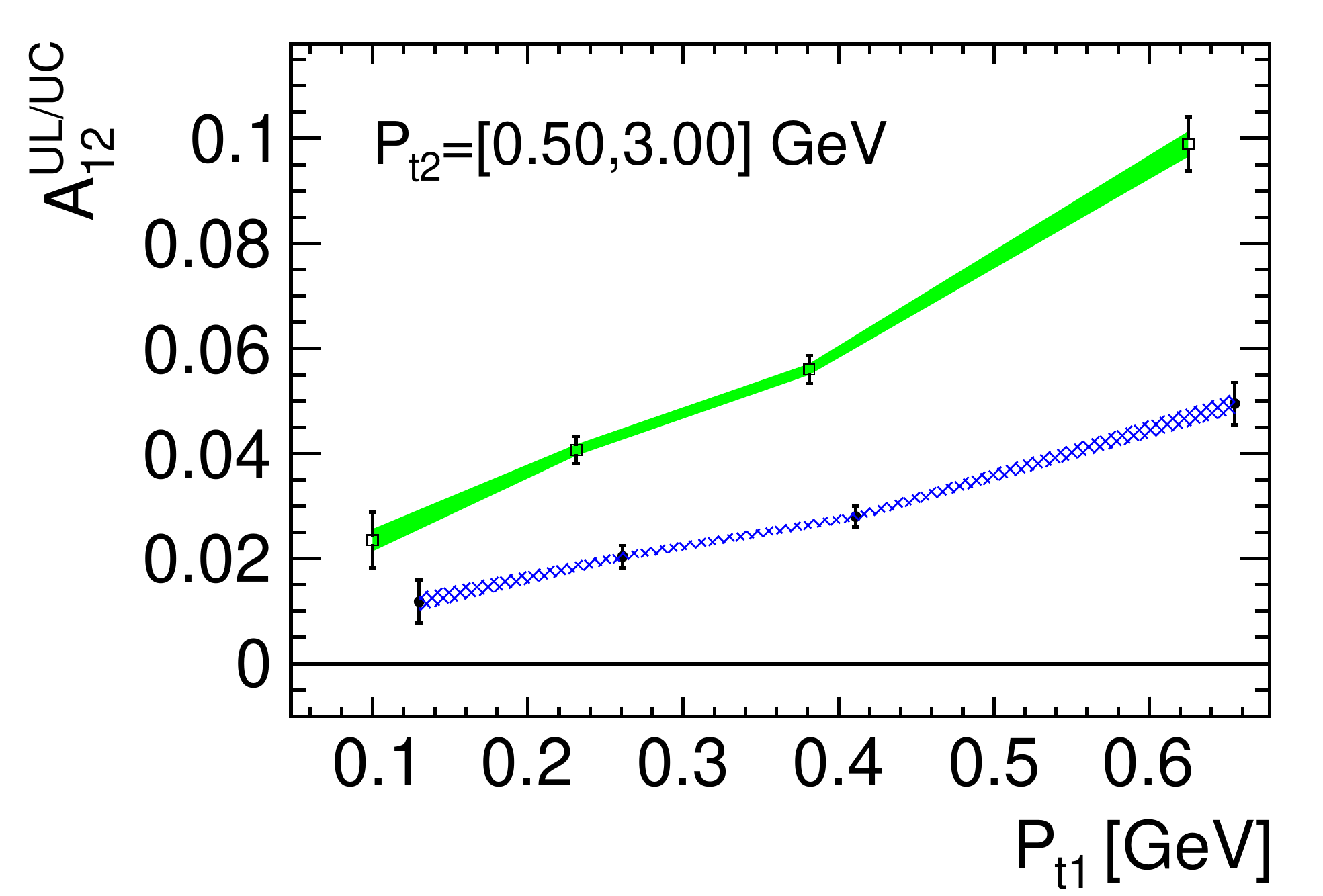}\\
\caption{$A^{UL}_{12}$ (squares) and $A^{UC}_{12}$ (circles) for charged-pion pairs versus~$P_{t1}$ for four bins in $P_{t2}$ (as labeled), integrating within the overall limits over \(z\). Error bars represent statistical uncertainties while the colored bands indicate systematic uncertainties.
\label{fig:resUlUcPt1Pt2}}	
\end{figure*} 

One novelty of the measurements presented here compared to previous Belle analyses~\cite{Abe:2005zx,Seidl:2008xc} is the inclusion of explicit transverse-momentum dependence of the asymmetries. This should help significantly to better constrain the transverse-momentum dependence of the Collins fragmentation function. 
Figure~\ref{fig:resUlUcPt1Pt2} shows the dependence of both  \(A_{12}^{UC}\) and \(A_{12}^{UL}\) on the transverse momentum of each of the  two pions, where the superscripts $UC$ and $UL$ denote the charge sign combination as defined in \eqref{eqn:FF7}.
In general, \(A_{12}^{UL}\) is found to be about double the size of \(A_{12}^{UC}\), consistent with previous analyses of these asymmetries~\cite{Abe:2005zx,Seidl:2008xc,BabarCharged}.
Both asymmetries exhibit a clear rise with increasing, \(P_{t1}\) and \(P_{t2}\) without showing any indication of leveling out at larger values of \(P_{t1}\) and \(P_{t2}\). In contrast, the largest asymmetry (in this projection) of around 10\% for \(A_{12}^{UL}\) is found in the last (\(P_{t1},P_{t2}\)) bin. This behavior is similar to what was found by BaBar~\cite{BabarCharged}, which can be explained perhaps by the limited reach in \(P_{t}\).
A direct quantitative comparison of these results with those by BaBar is hampered by the significantly different binning used here. Only in the case of the (\(z_{1},z_{2}\)) binning, a few bins at large \(z_{1}\) and \(z_{2}\) can be made out that have similar average \(z\) and \(P_{t}\). Still, the polar angular range of the thrust axis covered by the two measurements is quite different leading to a \(\sin^{2}\theta/(1+\cos^2\theta)\) scaling of the cosine modulations [cf.~Eqs.~\eqref{eqn:allratiosexpress2}-\eqref{eqn:FF5}] that are in variance with each other. However, those are simple scale factors that can be divided out, leaving asymmetries that can be directly compared. In the end, a discrepancy between Belle and BaBar is apparent that cannot be explained easily by charm contributions included here but corrected for at BaBar. Such discrepancy between Belle and BaBar is not new and was observed already before for the large-\(z\) region~\cite{Garzia:2016kqk}.
It is thought to be caused by differences in the applied constraints, e.g., differences in the methodology for removing \(\tau\) contributions.

\begin{figure*} [t]
\includegraphics[clip, trim=0.1cm 2.4cm 0.2cm 0.2cm,height=0.25\textheight,width=0.55\textwidth]{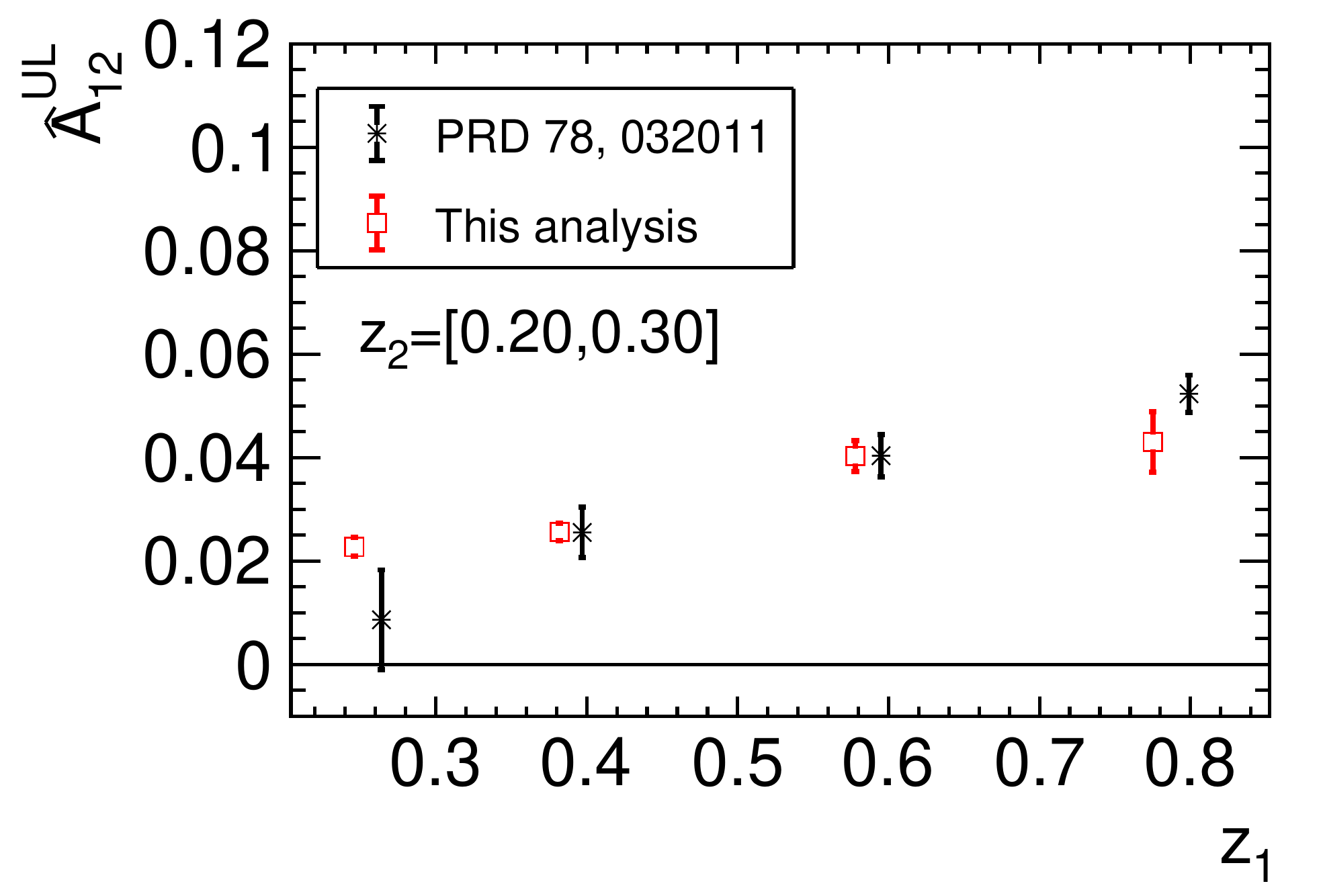}
\includegraphics[clip, trim=4.4cm 2.4cm 0.2cm 0.2cm,height=0.25\textheight,width=0.44\textwidth]{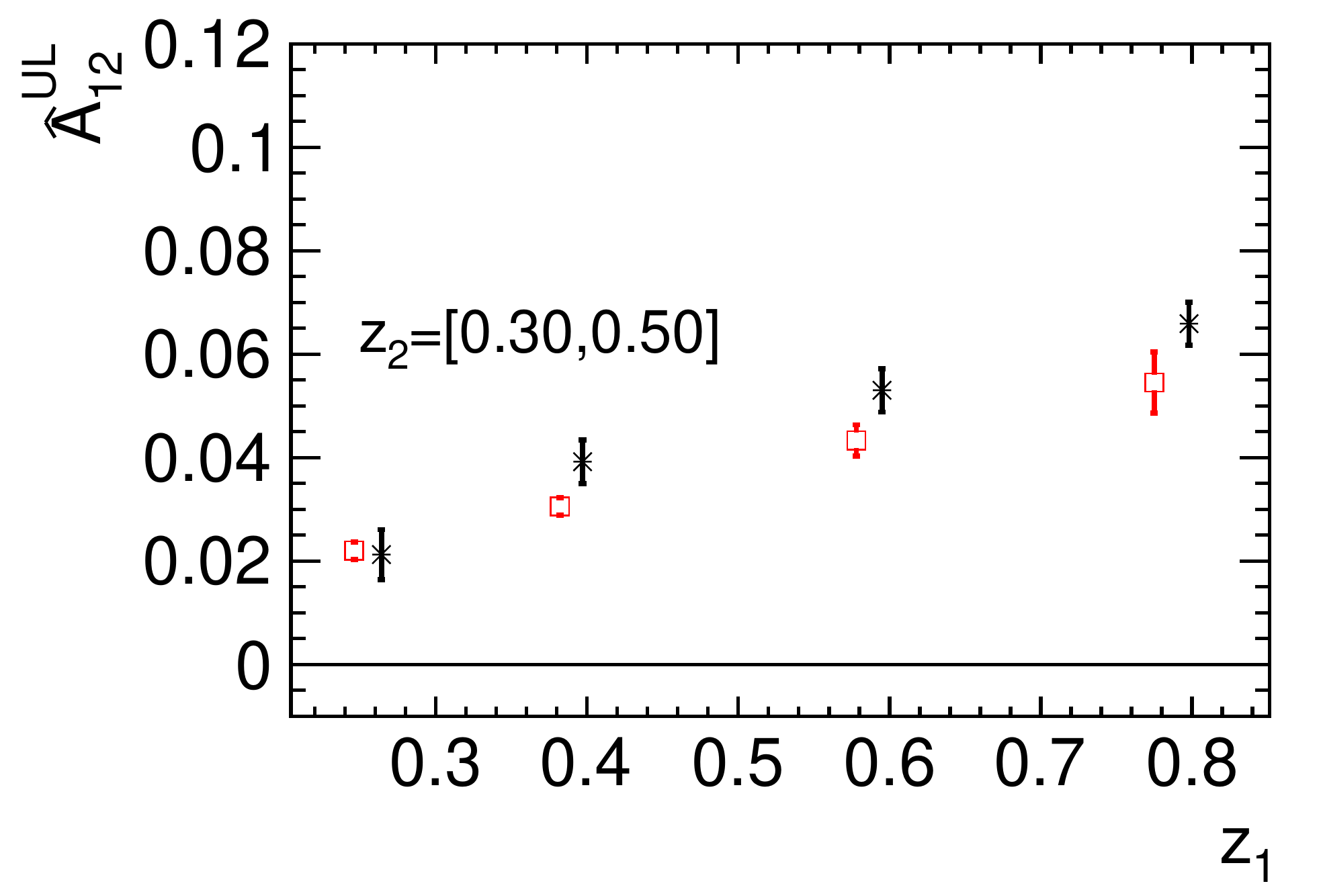}\\
\includegraphics[clip, trim=0.1cm 0.0cm 0.2cm 0.5cm,height=0.27\textheight,width=0.55\textwidth]{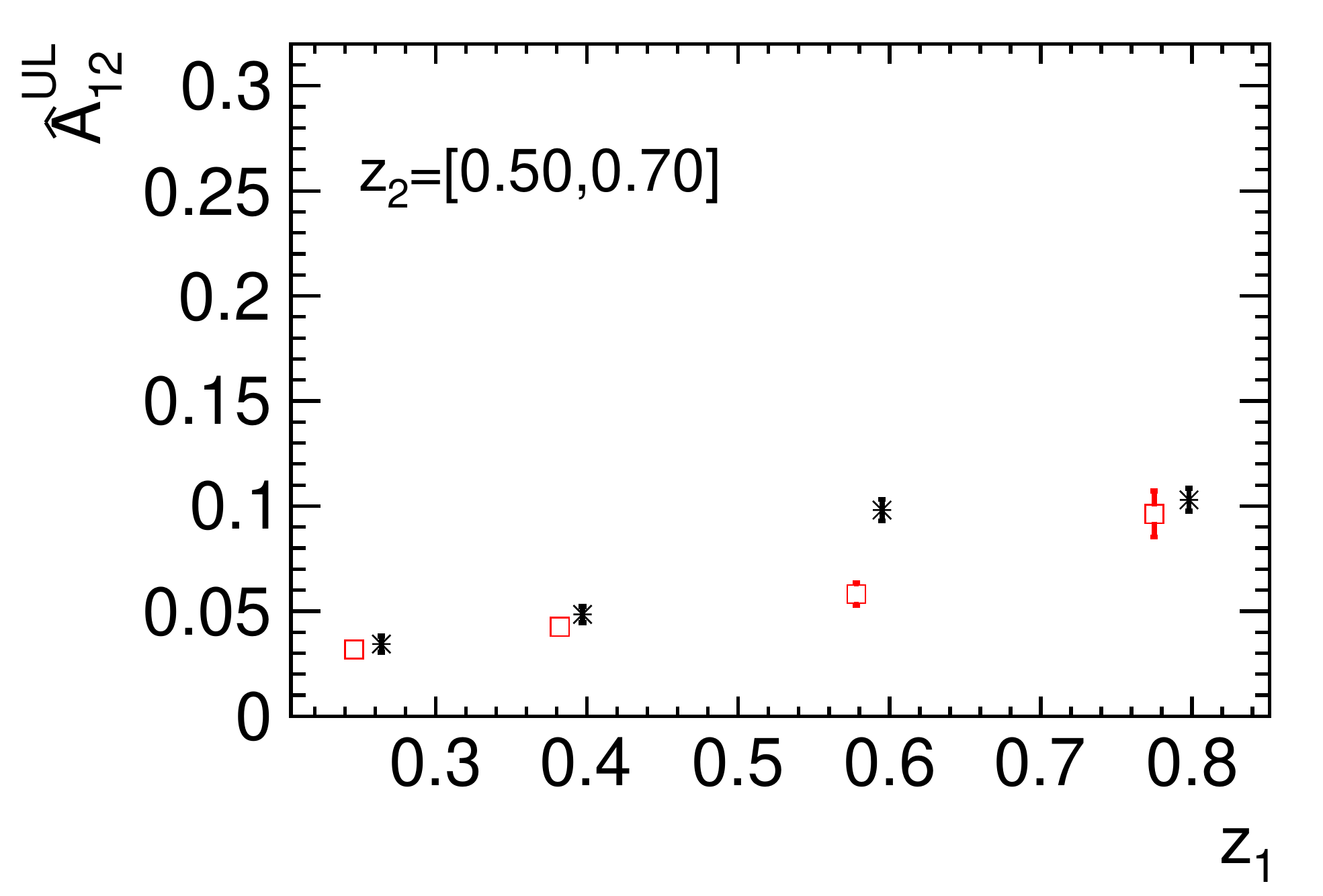}
\includegraphics[clip, trim=4.4cm 0.0cm 0.2cm 0.5cm,height=0.27\textheight,width=0.44\textwidth]{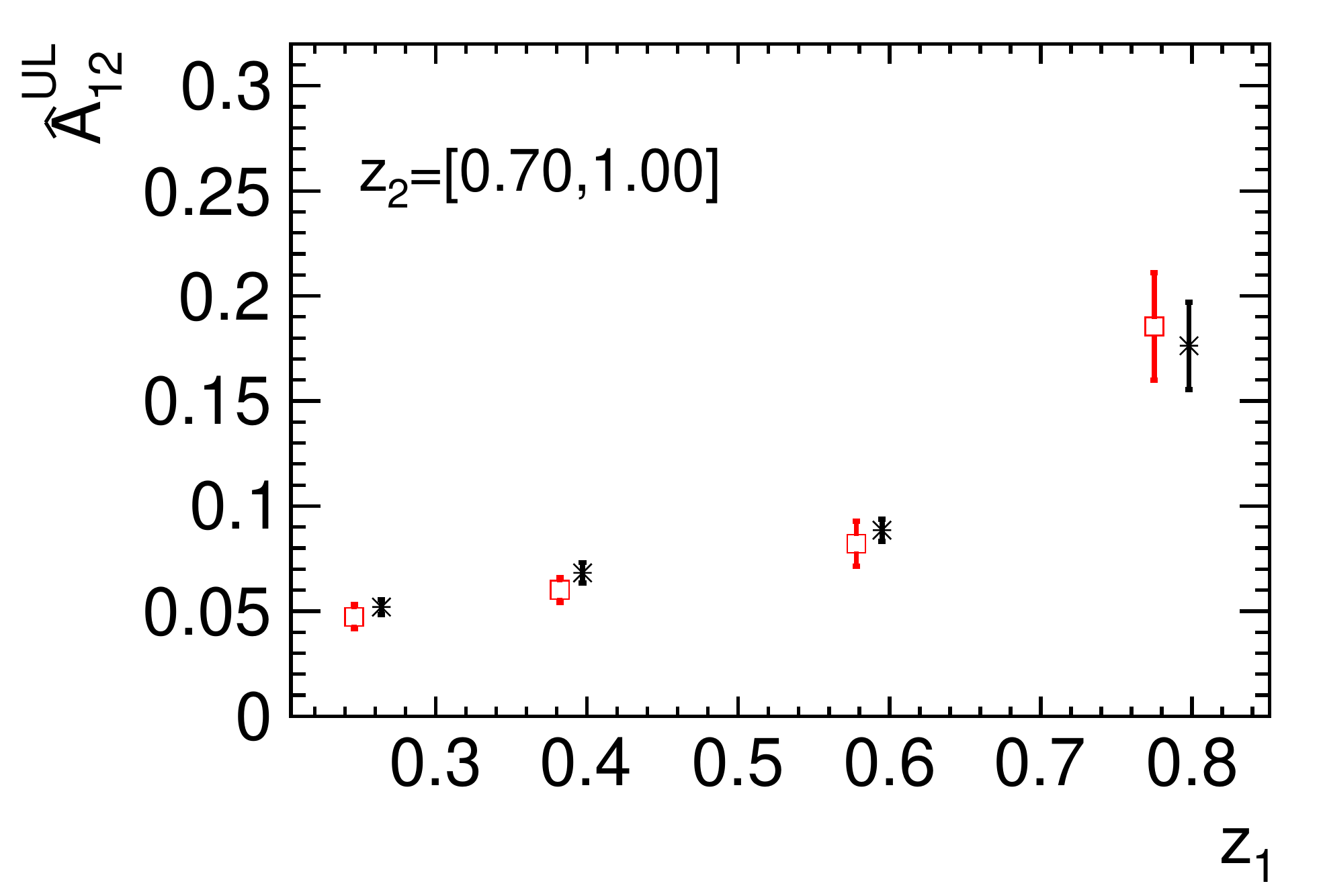}\\
\caption{\label{fig:belleComp} Comparison of the values for $A^{UL}_{12}$ extracted in this analysis and the previous Belle analysis in the ($z_1$,$z_2$) after undoing the different smearing corrections, and after correcting for the different average transverse polarization of the $q\bar{q}$ pairs in the two measurements due to differences in the \(\theta\) ranges probed. To make the comparison, the  contribution of charm quarks was corrected for by assuming a vanishing charm asymmetry. To avoid confusion with the corrected results, the symbol $\hat{A}^{UL}_{12}$ has been used to denote the asymmetry.
The lowest $z$ bin was omitted, since the previous analysis used a constraint of $z>0.2$. In the figure, data points of the previous analysis are offset horizontally by 0.02 for better visibility. \label{fig:resOldBelleComp}}
\end{figure*}

Since there are already published results from Belle for charged-pion pairs for the $(z_1,z_2)$ binning, which cover roughly the same kinematic region, a comparison between the results presented here and those from the previous publications~\cite{Abe:2005zx,Seidl:2008xc} is provided. 
The previous results use a smearing correction to correct back to the $q\bar{q}$ axis extracted from simulation. Since this is not an observable and can be defined cleanly only at leading order, this correction is replaced with a correction back to the thrust axis in the present analysis. Therefore the comparison is performed for asymmetries for which the smearing corrections are removed. This corresponds to a division by the mean smearing correction factor $1.66$ for the previous analysis whereas the available bin-by-bin correction is used for this analysis.
Further, the compared asymmetry values have been corrected for the kinematic factor $\sin^2(\theta)/(1+\cos^2(\theta))$ bin-by-bin, which differs between the two analyses as a result of the different fiducial constraints. The analysis in Ref.~\cite{Seidl:2008xc} uses a constraint on the $z$ projection of the thrust axis of $|T_z|<0.75$, which corresponds to $0.72\textrm{ rad}< \theta < 2.42\textrm{ rad}$.
Hence, for the previous analysis the mean kinematic factor is $0.77$ whereas it is $0.91$ for the presented analysis. 
The results after adjustments for both the smearing and kinematic factors for the asymmetry values and their uncertainties is the comparison shown in Fig.~\ref{fig:resOldBelleComp}. 

There are  two further noteworthy differences between the two analyses: 
(i) The previous analysis does not apply opening-angle constraints. 
One effect of this difference is that the sampled $P_t$ range is different, since high-$z$ hadrons tend to be closer to the thrust axis.

(ii) The previous Belle analysis corrects for the charm contribution using a $D^*$ sample. In this analysis, the charm contribution was not corrected for, since using the $D^*$ sample can introduce a bias in phase space and introduces larger uncertainties. Instead, the fractional contribution from charm to the event sample is given for each bin in~\hyperref[sec:App1]{\ref{sec:App1}}, so it can be used for a global extraction. 

For the comparison in Fig.~\ref{fig:belleComp}, it is assumed that the Collins signal coming from charm fragmentation vanishes. In that case, the charm contribution reduces to a simple dilution of the asymmetry of size \((1-f_{c})\), where \(f_{c}\) is the ratio of the number of events coming from \(c\bar{c}\) production compared to the sum from \(c\bar{c}\) and light quarks (\(uds\)), which in this analysis is extracted from Monte Carlo simulations (see~\hyperref[sec:App1]{\ref{sec:App1}} for more details). As such the dilution factor can be divided out. 

\begin{figure*} 
\begin{minipage}{0.55\textwidth}
\includegraphics[clip, trim=0cm 2.7cm 0.8cm 0cm,height=0.25\textheight,width=1.0\textwidth]{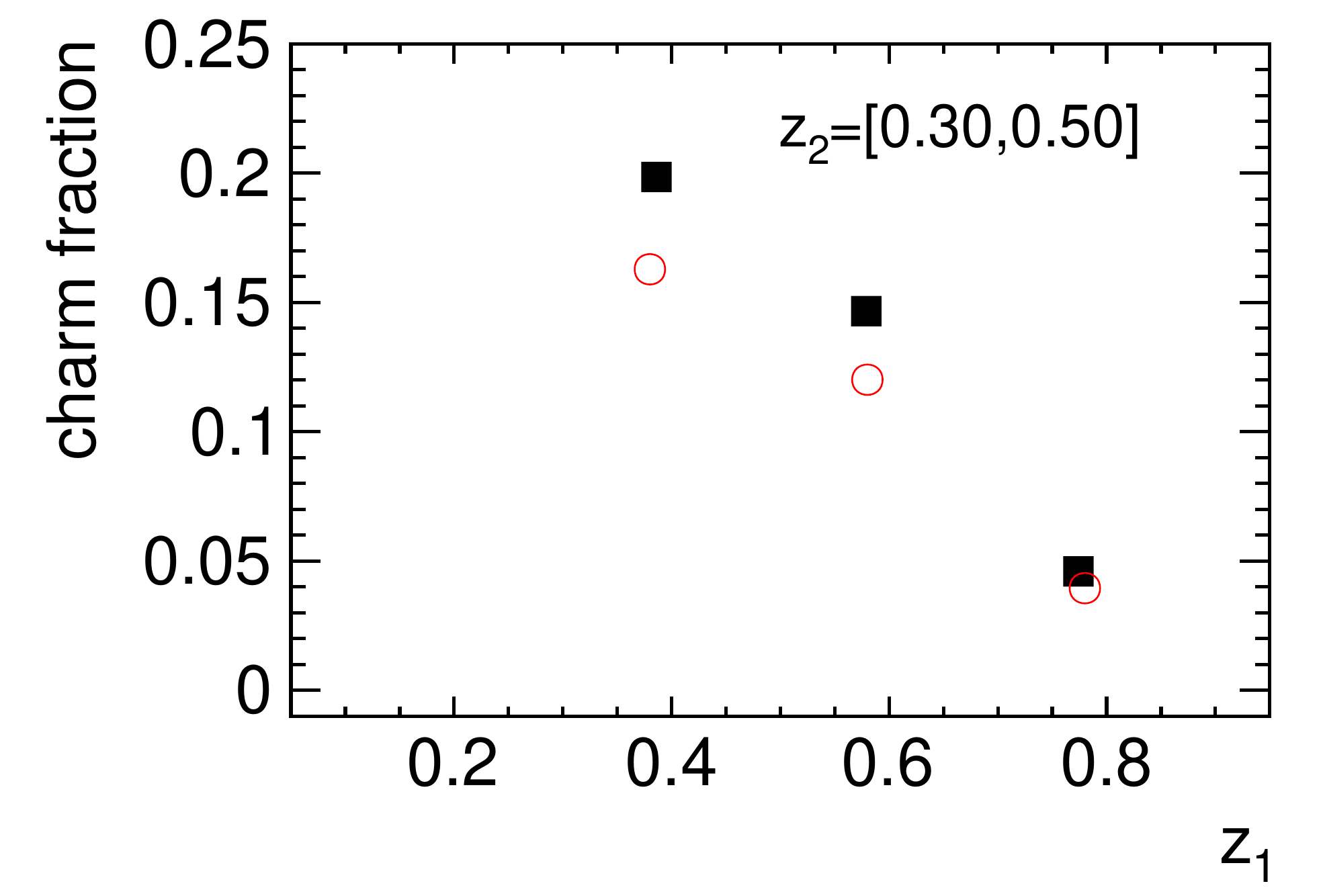}\\
\includegraphics[clip, trim=0cm 0cm 0.8cm 0cm,height=0.29\textheight,width=1.0\textwidth]{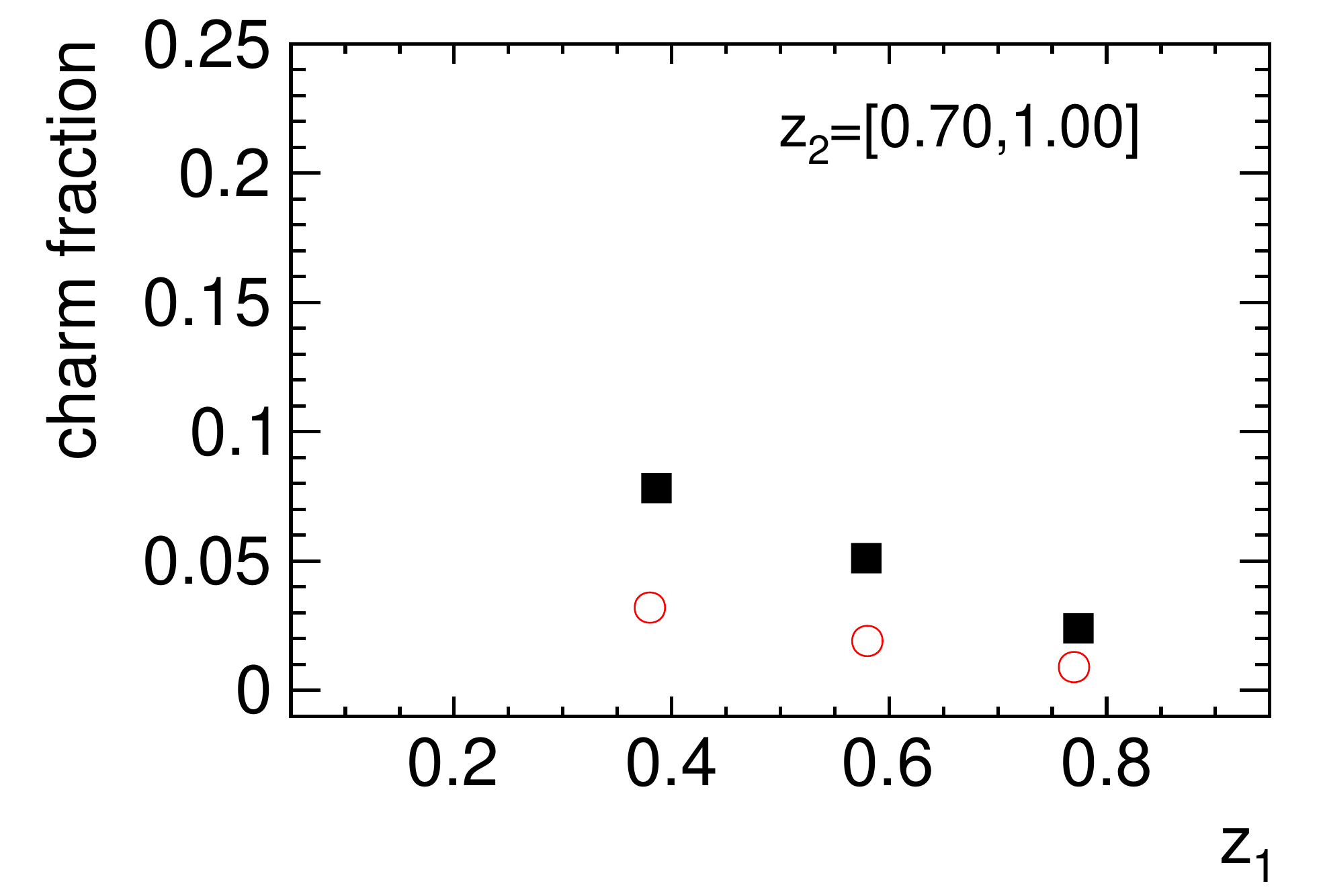}
\end{minipage}
\begin{minipage}{0.44\textwidth}
\includegraphics[clip, trim=4.4cm 0.0cm 0cm 0.305cm,height=0.305\textheight,width=1.0\textwidth]{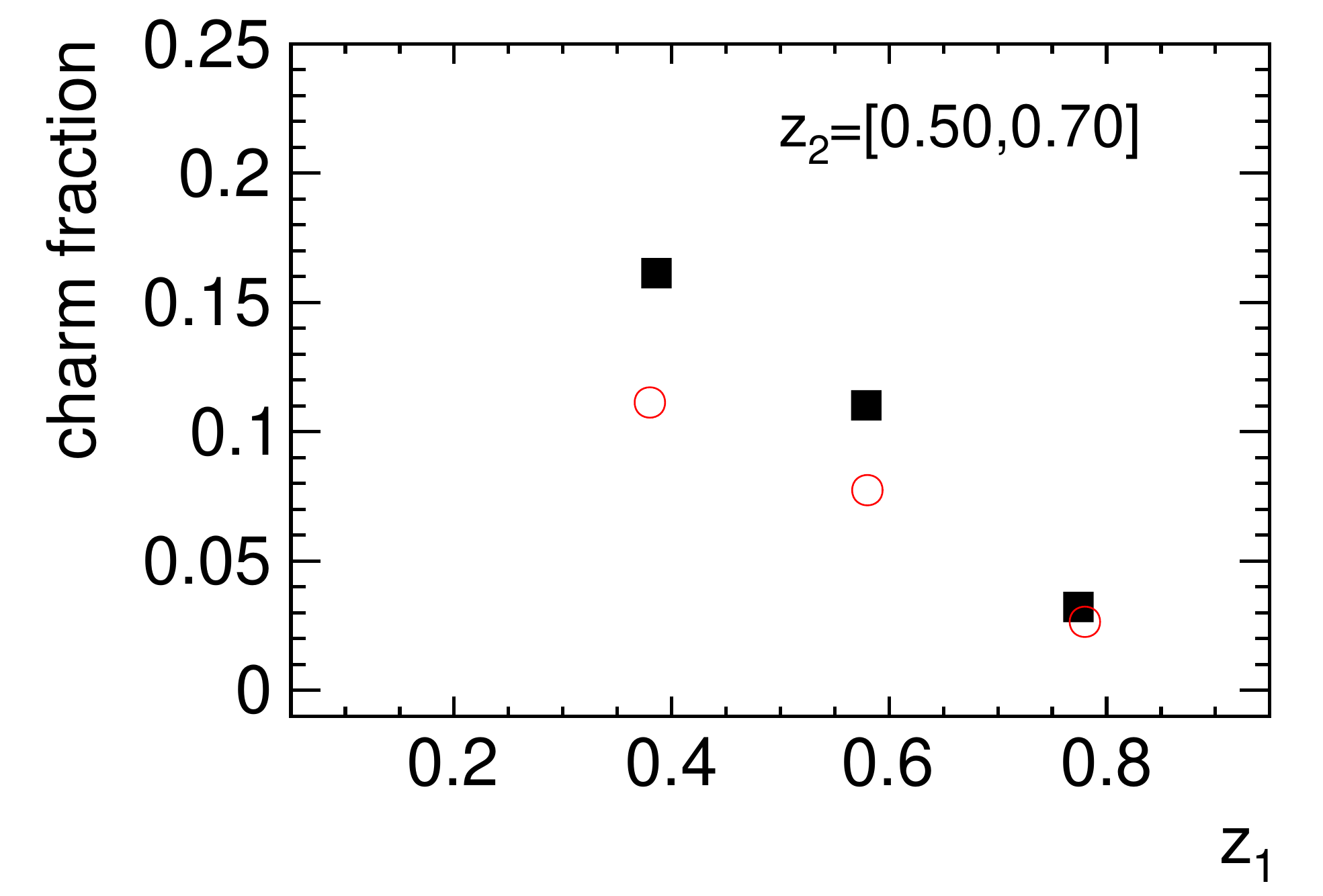}\\
\includegraphics[clip, trim=0cm 2cm 0cm 0cm,height=0.22\textheight,width=1.0\textwidth]{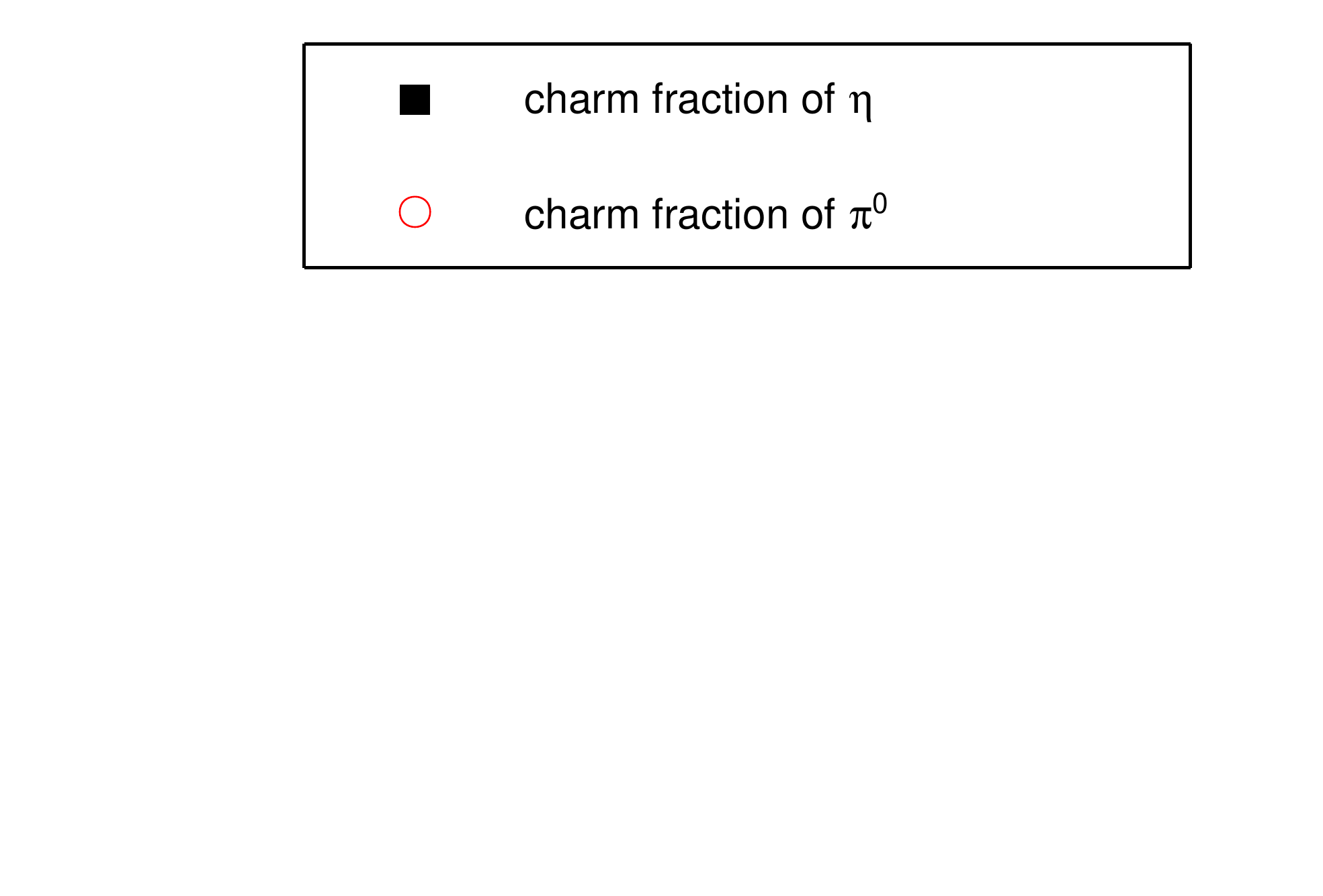}
\end{minipage}
\caption{Comparison of the charm fractions extracted from Monte Carlo for the events used to calculate the $A_{12}^{\pi^0}$ (red circles) and $A_{12}^{\eta}$ (black squares) asymmetries for the (\(z_1,z_2\)) binning.
\label{fig:charmFractComp}}
\end{figure*}

Before discussing the comparison with the previous Belle results, one word of caution on such a charm correction is in place here: The observable of interest in this analysis is the cosine moment of a double ratio, the latter being of the form \([1+a_{12}^{\text{hadron pair 1}}\cos(\phi_{12})]/[1+a_{12}^{\text{hadron pair 2}}\cos(\phi_{12})]\), which is Taylor-expanded to \(1+\cos(\phi_{12}) [a_{12}^{\text{hadron pair 1}} - a_{12}^{\text{hadron pair 2}}]\). 
Clearly, the charm correction sketched above works when both hadron pairs suffer the same amount of dilution. However, it does not work in general when the charm contribution is different for the two hadron pairs, as in that case the dilution factors do not factor out. While this is of a lesser problem for the \(\pi^{0}\) asymmetries presented here, as the charm fractions are similar for charged-pion pairs and those involving a \(\pi^{0}\) (cf.~Tables~\ref{tab:sinzcharmratio}-\ref{tab:zptcharmratio}), it is certainly more difficult to make this argument for the \(\eta\) asymmetries. It is also for that reason that both the \(\pi^{0}\) and \(\eta\) asymmetries discussed further below are not corrected for charm contributions.
Figure~\ref{fig:charmFractComp} shows an example comparison of the model-dependent charm fractions in the ($z_1$, $z_2$) used for the $A_{12}^{\pi^0}$ and $A_{12}^{\eta}$ asymmetries extracted from the Belle Monte Carlo. Here the superscripts refer to the charge combinations as defined in~\eqref{eqn:FF6}. The charm fractions become small and similar at large $z$, but deviate from each other for \(\pi^0\) and \(\eta\) at lower values of \(z\), where the charm fraction gets as large as 20\% in the case of \(\pi^\pm \eta\) pairs.

Coming  back to the comparison presented in Fig.~\ref{fig:belleComp}, in general a good agreement is visible
with the exception of one point in the third $z_1$ and $z_2$ bin, which seems to be an outlier.
However, a quantification of the agreement is difficult, since the uncertainties of the measurements are correlated. 
Disregarding this correlation and excluding the outlier, one arrives at a $\chi^2$ per degree of freedom of $1.2$.
The consistency between the results indicates that the assumption of a vanishing asymmetry for charm quarks is justified.

A second novelty of this measurement is the inclusion of double ratios involving neutral mesons, more specifically \(\pi^{0}\) and \(\eta\). 
The fragmentation functions for neutral pions are related to those of charged pions through isospin symmetry. Similarly, the \(\eta\) fragmentation functions can be related to those of pions through SU(3) flavor symmetry, which, however, is known to be violated due to the substantially larger mass of strange quarks.



\begin{figure*} 
\includegraphics[clip, trim=0cm 2.7cm 0.2cm 0cm,height=0.25\textheight,width=0.55\textwidth]{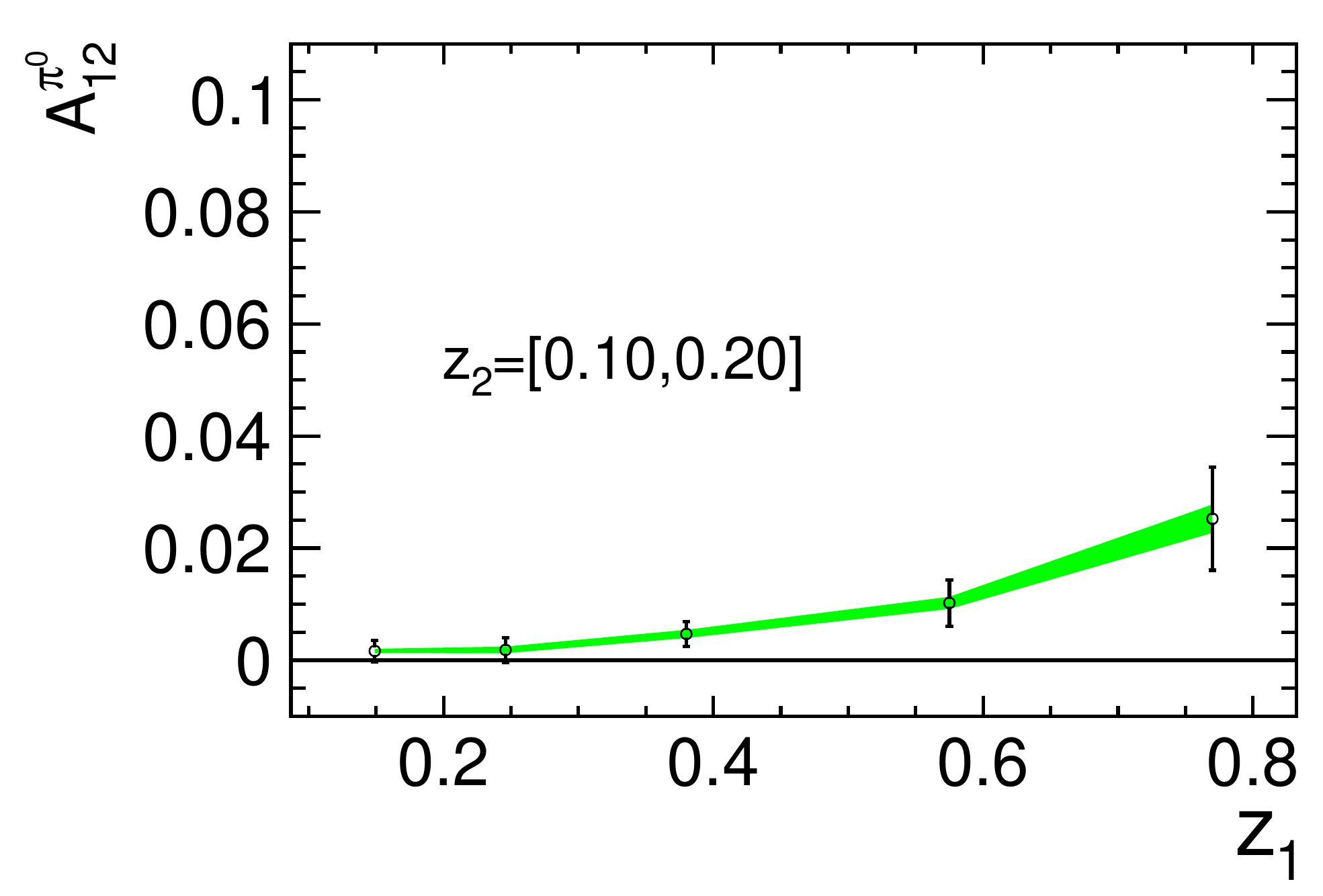}
\includegraphics[clip, trim=4.4cm 2.7cm 0cm 0cm,height=0.25\textheight,width=0.44\textwidth]{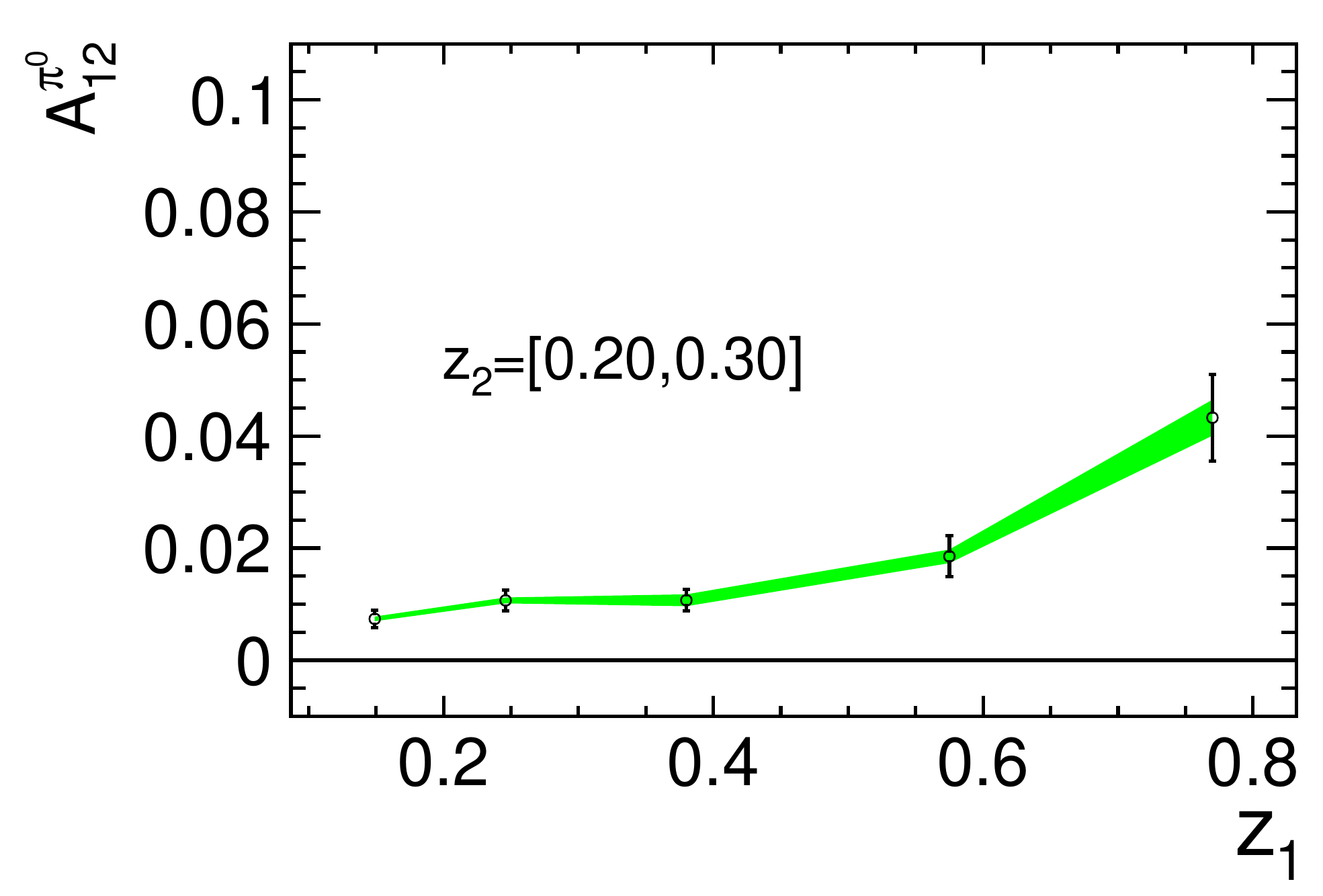}\\
\begin{minipage}{0.55\textwidth}
\includegraphics[clip, trim=0cm 2.7cm 0.2cm 0cm,height=0.25\textheight,width=1.0\textwidth]{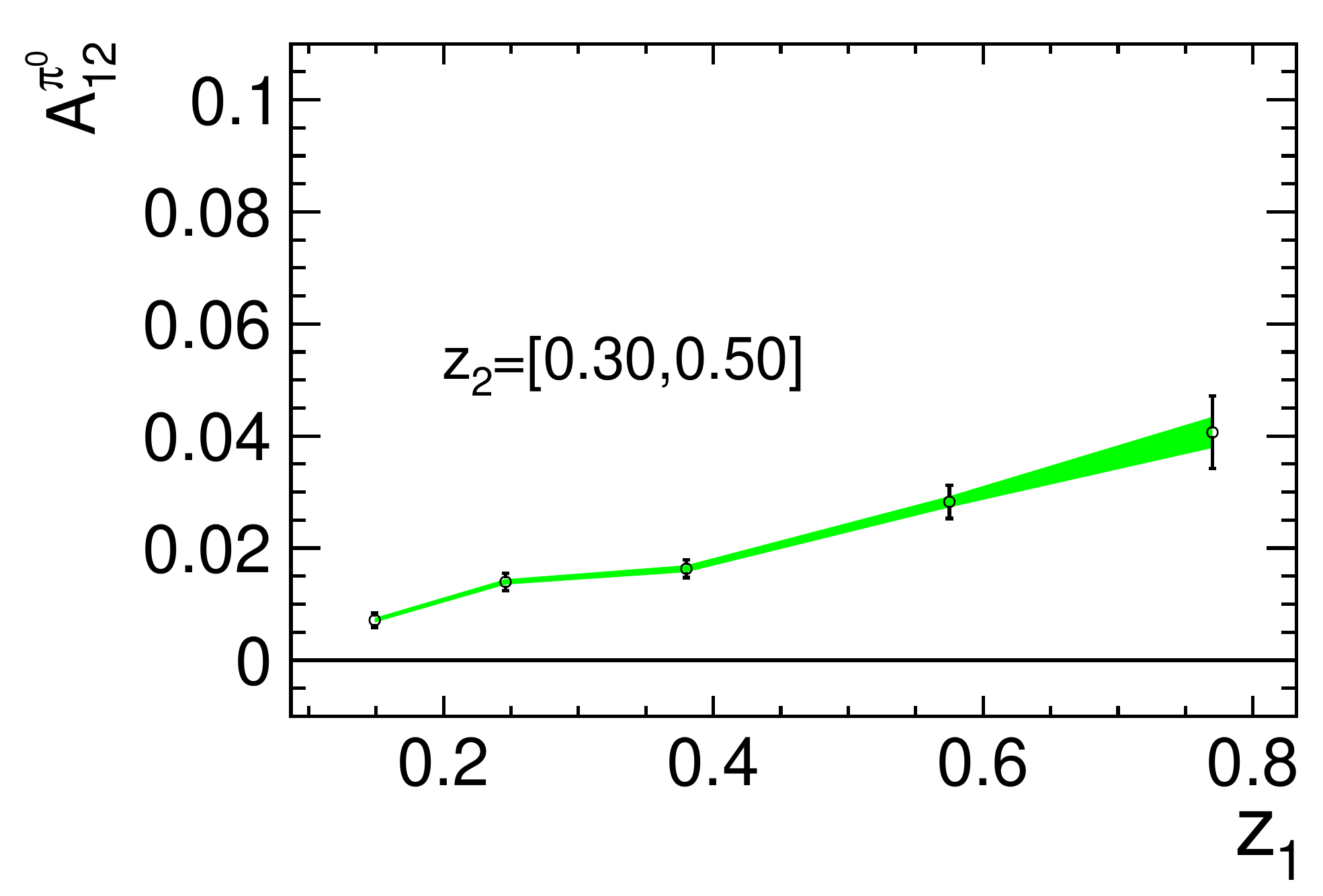}\\
\includegraphics[clip, trim=0cm 0cm 0.2cm 0cm,height=0.29\textheight,width=1.0\textwidth]{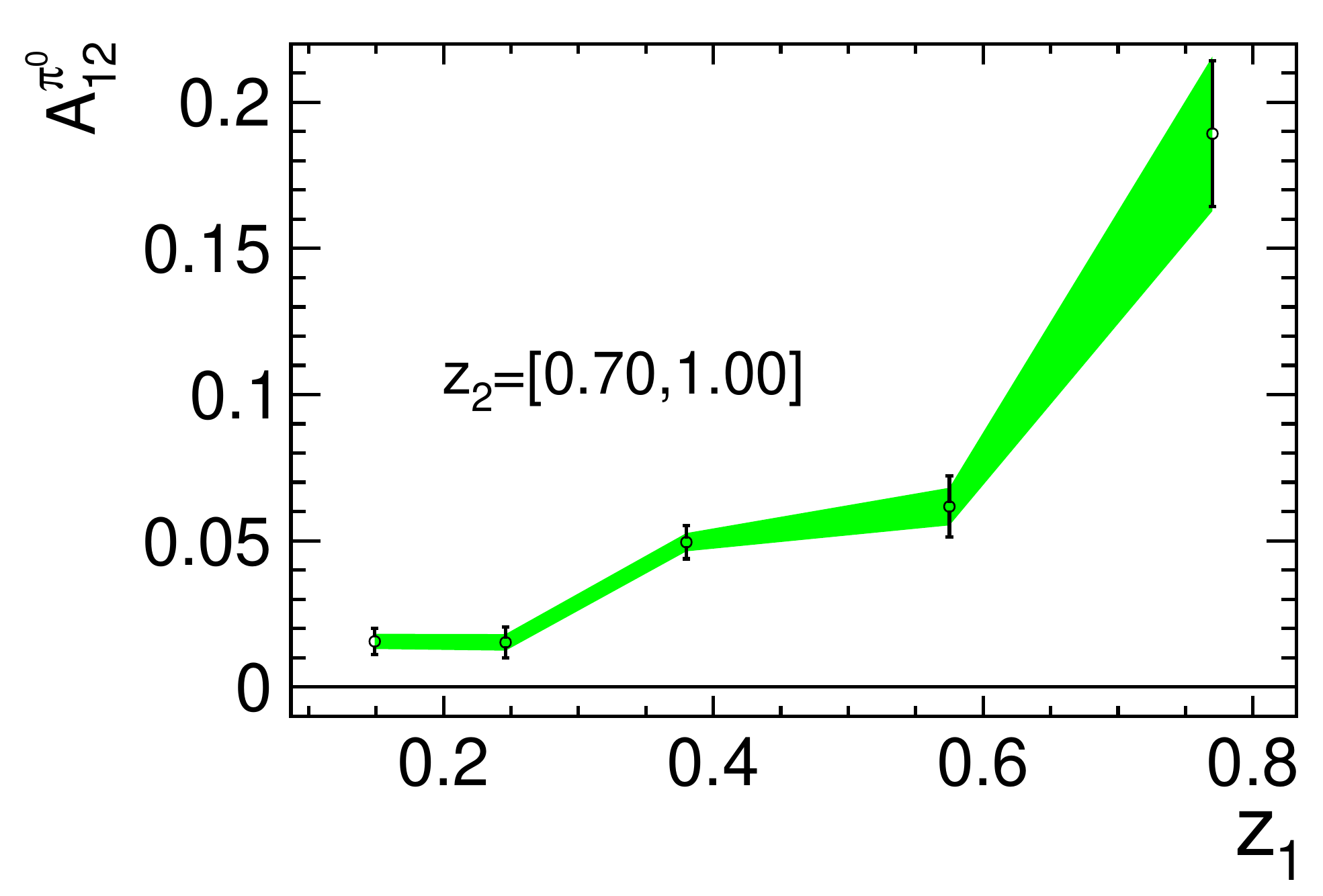}
\end{minipage}
\begin{minipage}{0.44\textwidth}
\includegraphics[clip, trim=4.4cm 0.2cm 0cm 0.6cm,height=0.293\textheight,width=1.0\textwidth]{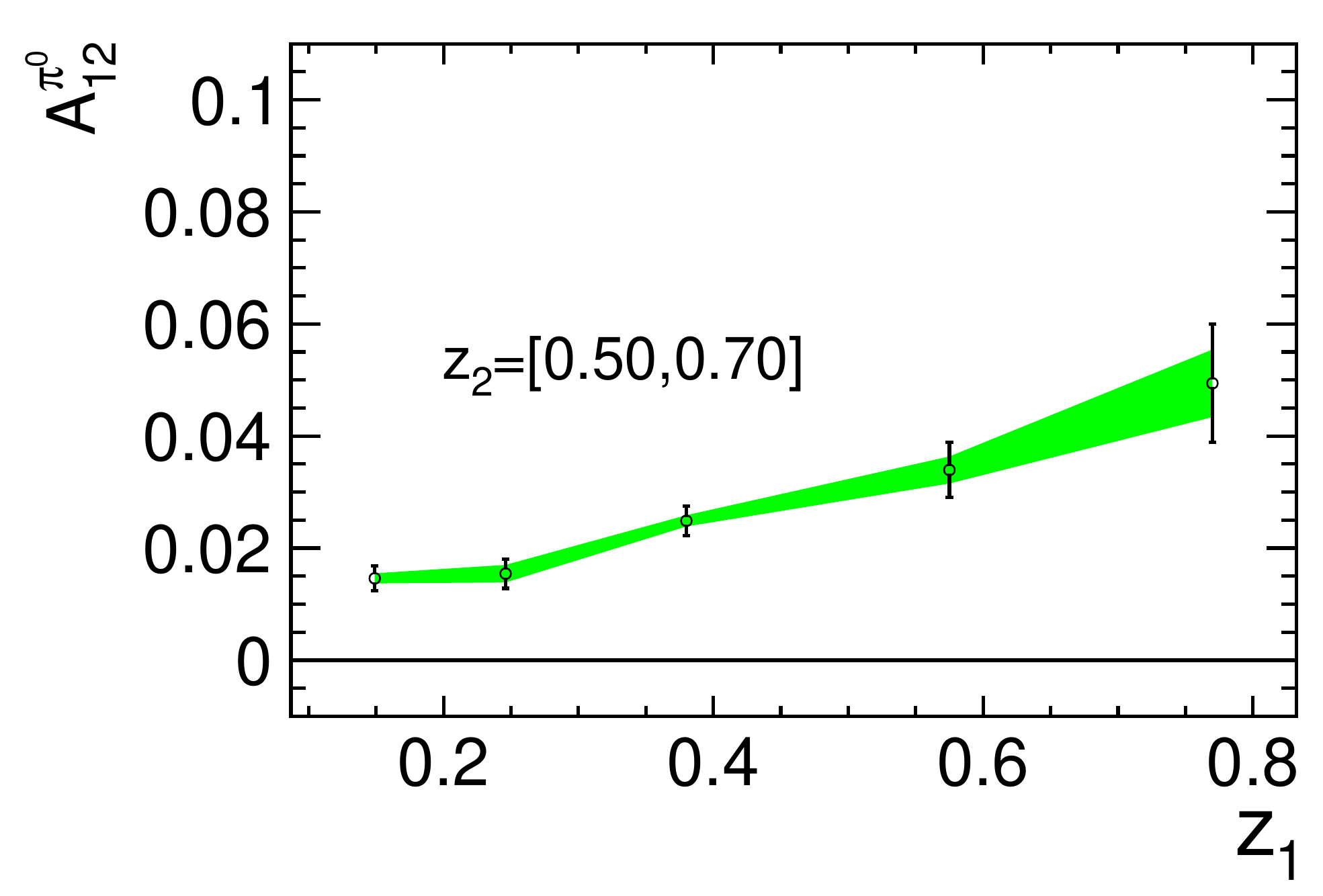}\\
\includegraphics[clip, trim=0cm 2cm 0cm 0cm,height=0.22\textheight,width=1.0\textwidth]{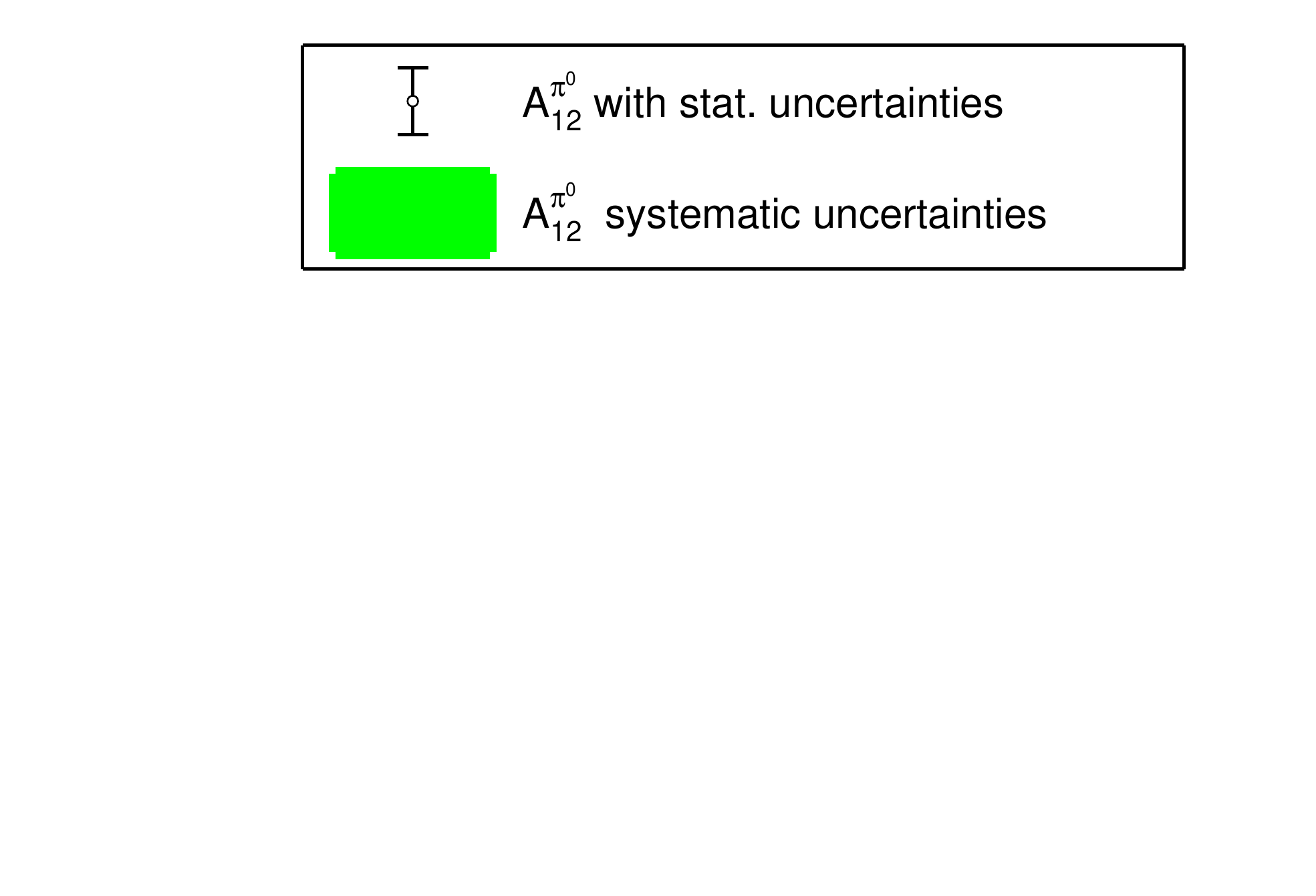}
\end{minipage}
\caption{Dependence of $A^{\pi^0}_{12}$ on $z_{1}$ for five bins in $z_{2}$ (as labeled), integrating within the overall limits over \(P_{t}\). Error bars represent statistical uncertainties while the colored bands indicate systematic uncertainties.\label{fig:resPi0VsZ1Z2}}
\end{figure*} 

Figure~\ref{fig:resPi0VsZ1Z2} displays the dependence of $A^{\pi^0}_{12}$ on $z_1$ and $z_2$. As expected from the charged-pion results, significant asymmetries that rise with  $z$ are observed. In the highest $(z_1,z_2)$ bin, for which one expects the largest correlation between the fragmenting quark, including its polarization and the final-state hadron, they are reaching $20$\%. In the lowest \(z\) bin, where a large amount of disfavored fragmentation contributes, the asymmetries are consistent with zero within statistical and systematic precision on the sub-percent level.


For the double ratios involving neutral mesons, the asymmetries do not have to be symmetric under interchange of the hadron subscript on \(z\) and \(P_{t}\) as the neutral meson in the numerator of the double ratios is identified as hadron 1 and the charged pion in the opposite hemisphere as hadron 2. As a result the \( z_{1} \) and \( P_{t1} \) dependences provide the most sensitivity to the \( \pi^{0} \) and \( \eta \) fragmentation functions.

\begin{figure*} 
\includegraphics[clip, trim=0.1cm 2.4cm 0.2cm 0.2cm,height=0.25\textheight,width=0.55\textwidth]{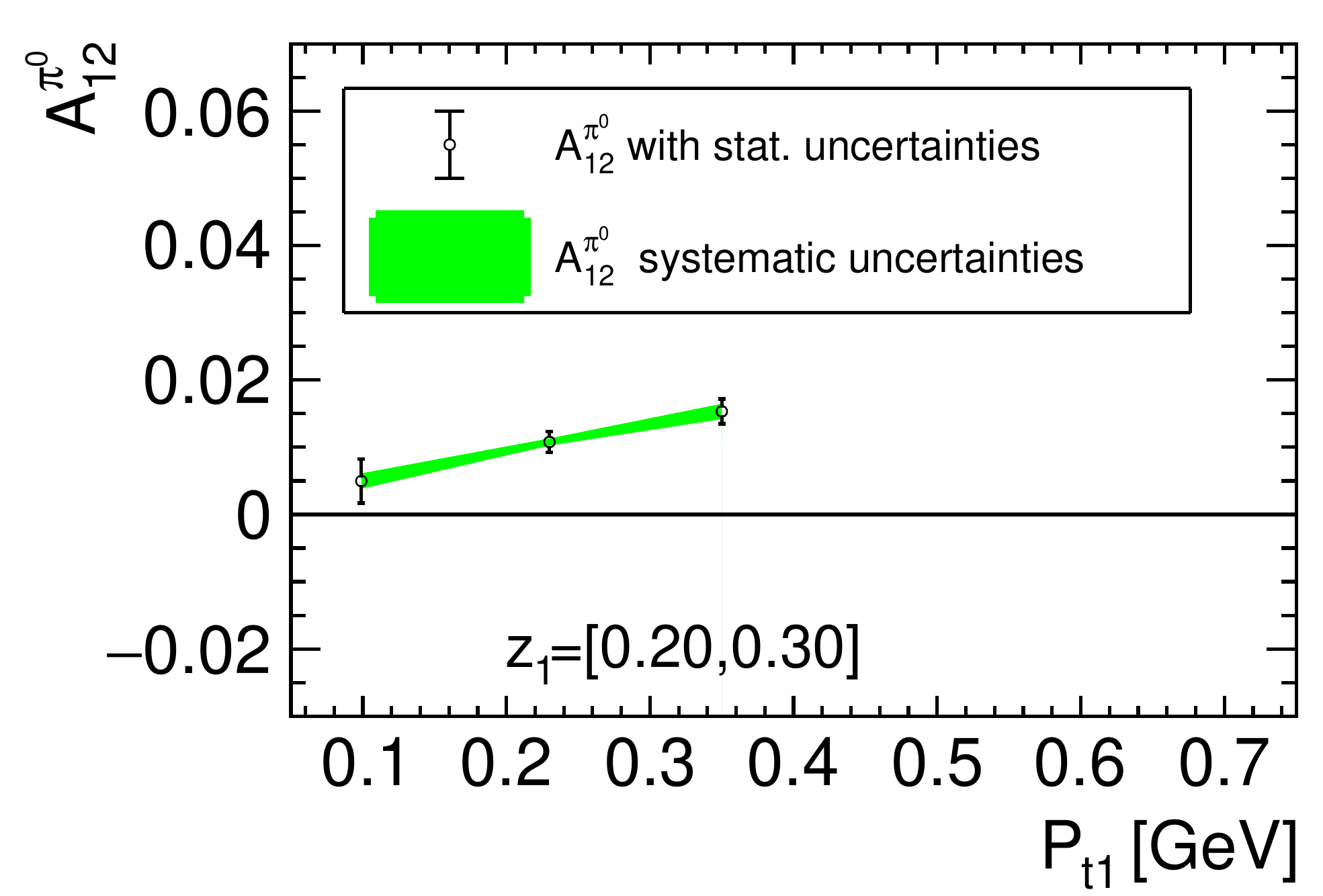}
\includegraphics[clip, trim=4.4cm 2.4cm 0.2cm 0.2cm,height=0.25\textheight,width=0.44\textwidth]{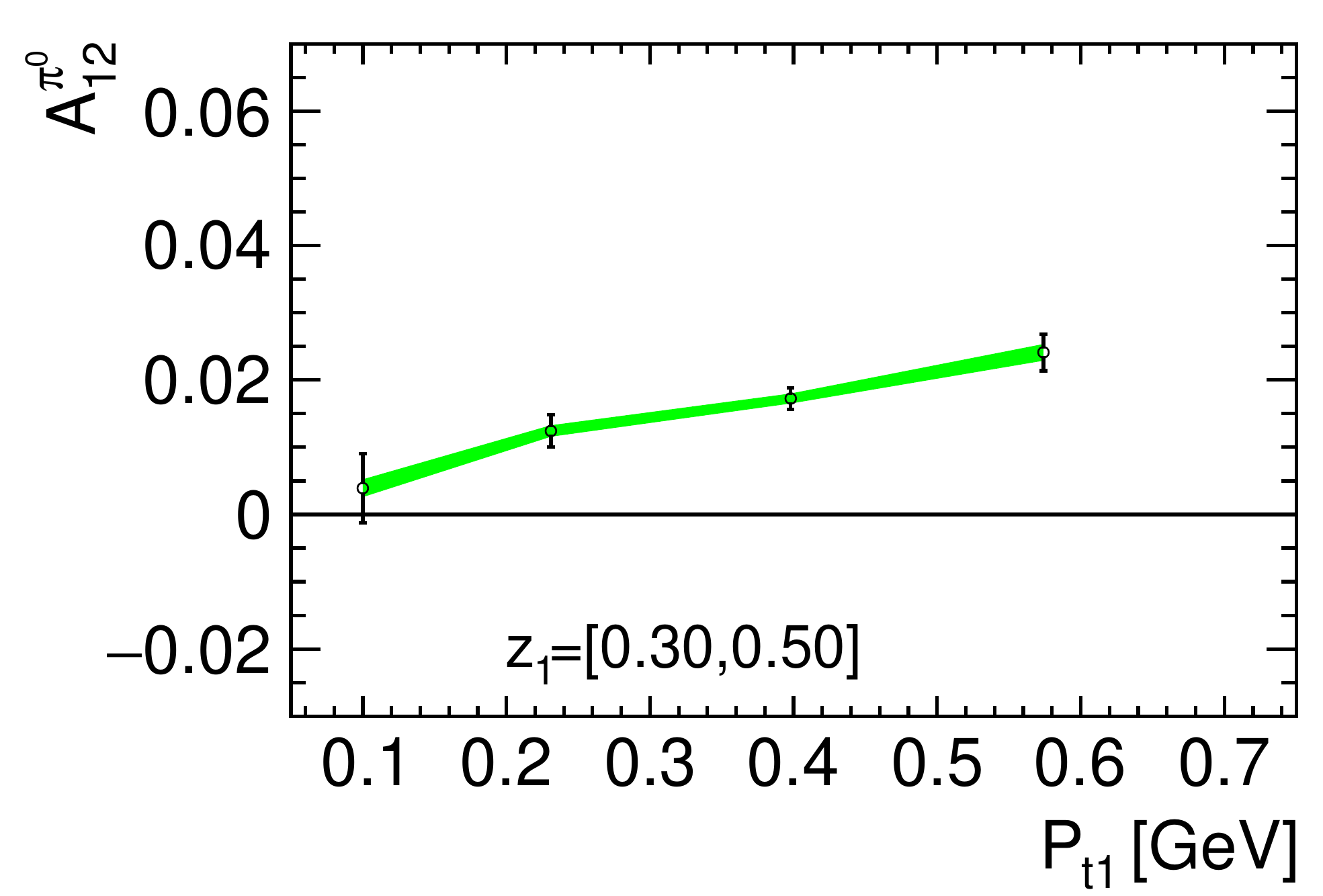}\\
\includegraphics[clip, trim=0.1cm 0.0cm 0.2cm 0.5cm,height=0.27\textheight,width=0.55\textwidth]{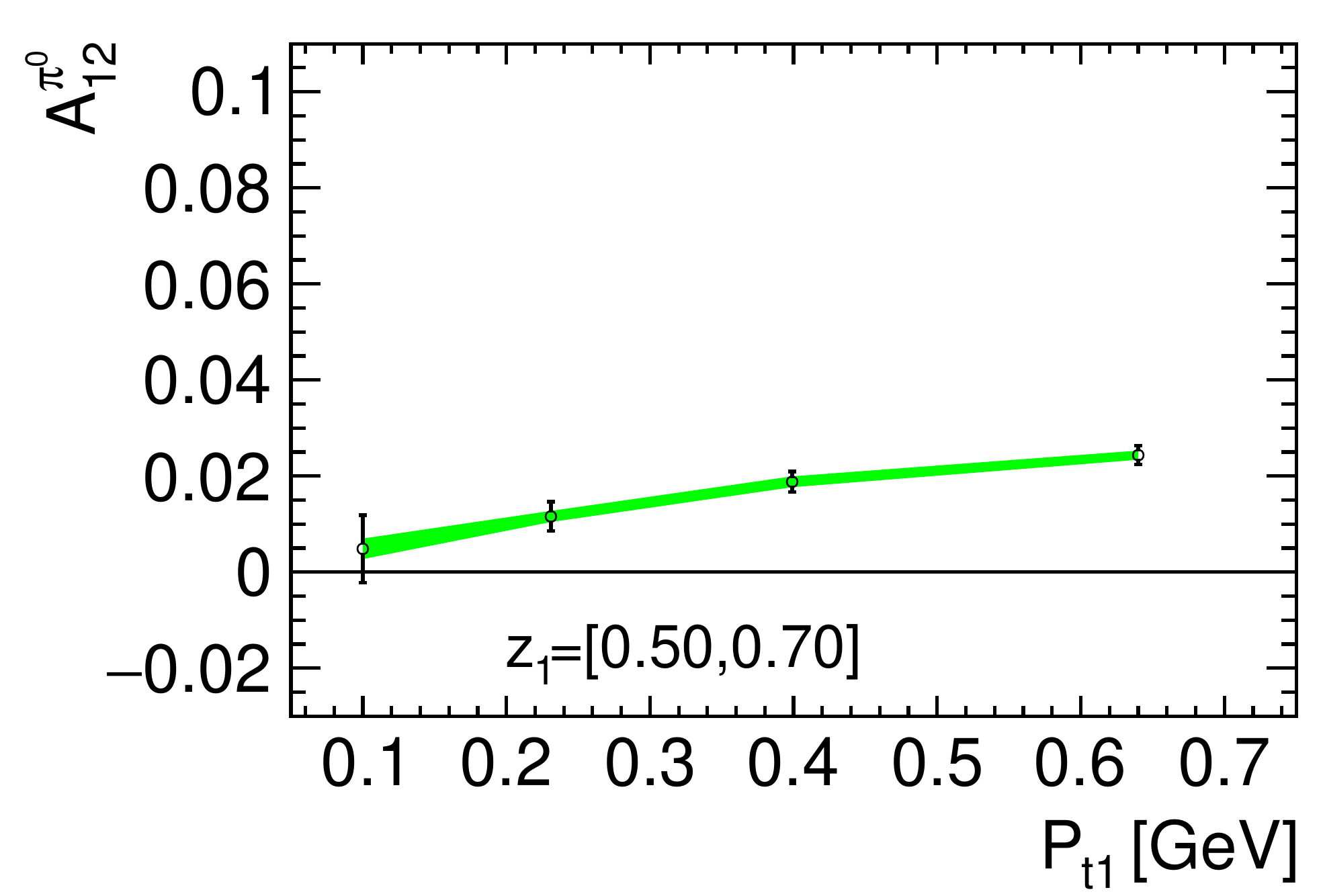}
\includegraphics[clip, trim=4.4cm 0.0cm 0.2cm 0.5cm,height=0.27\textheight,width=0.44\textwidth]{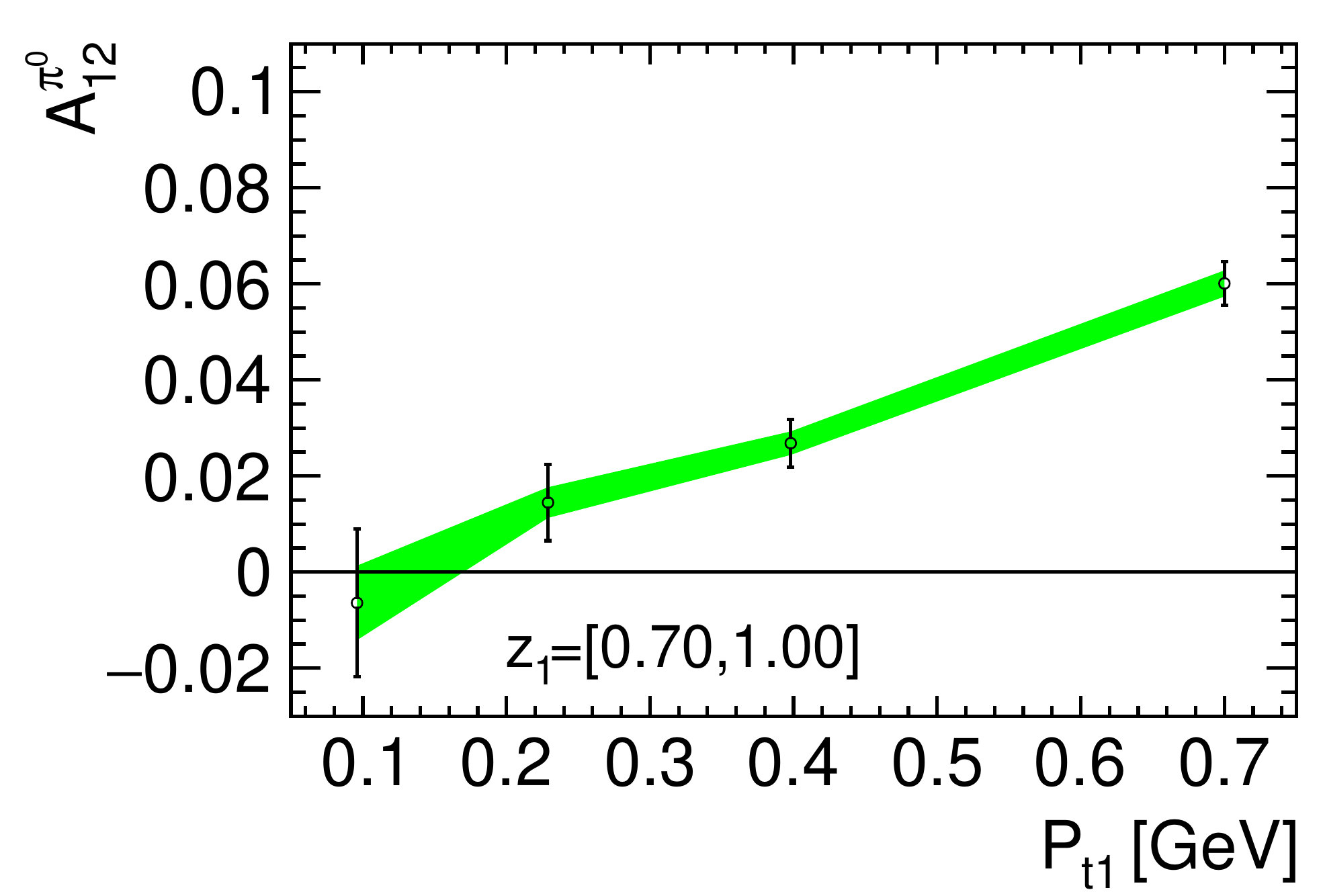}
\caption{Dependence of $A^{\pi^0}_{12}$ on $P_{t1}$ for four bins in $z_{1}$ (as labeled), integrating within the overall limits over the kinematics of the second hadron. Error bars represent statistical uncertainties while the colored bands indicate systematic uncertainties.\label{fig:resPi0VsZPt}}
\end{figure*}

\begin{figure*} 
\includegraphics[clip, trim=0.1cm 2.4cm 0.2cm 0.2cm,height=0.25\textheight,width=0.55\textwidth]{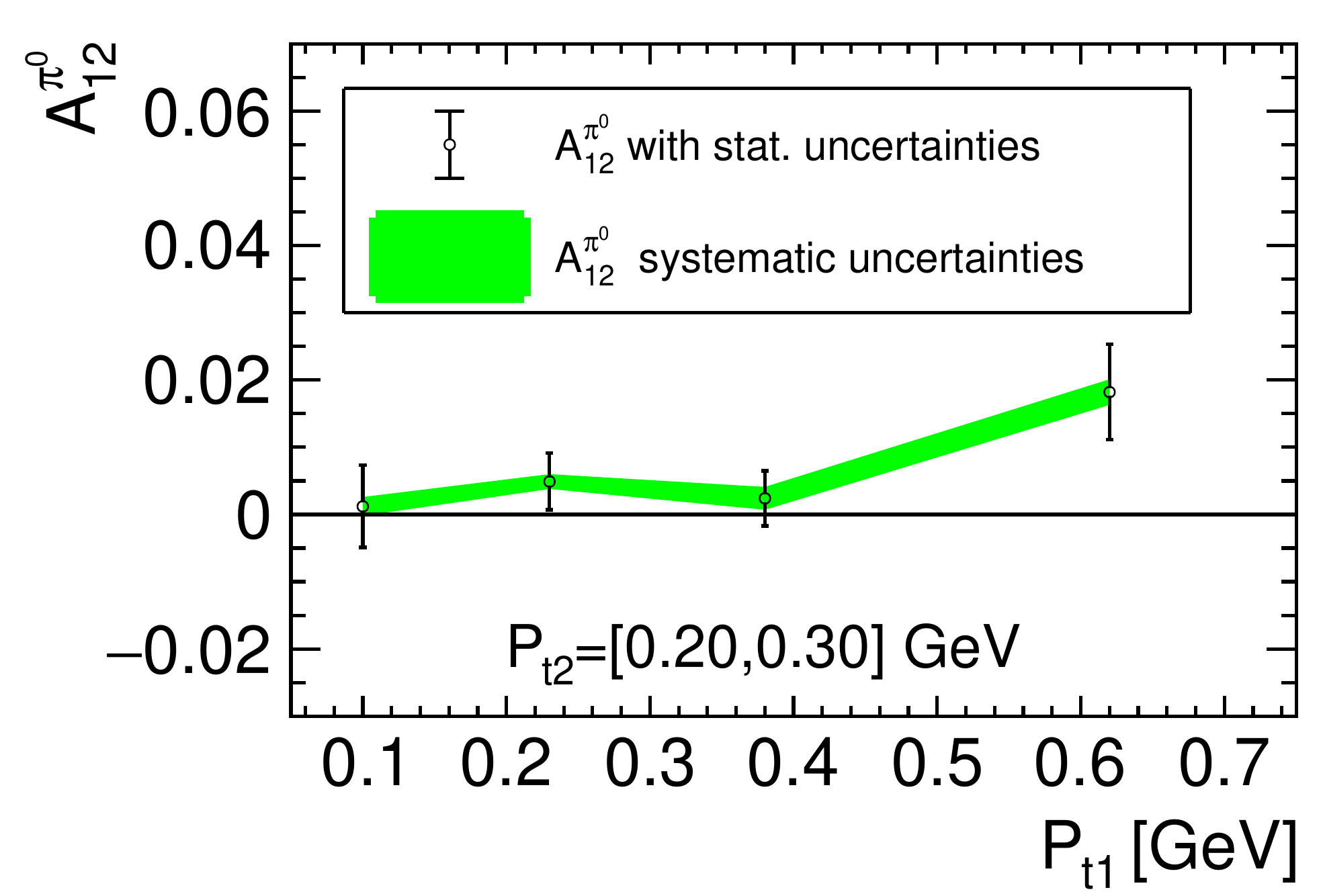}
\includegraphics[clip, trim=4.4cm 2.4cm 0.2cm 0.2cm,height=0.25\textheight,width=0.44\textwidth]{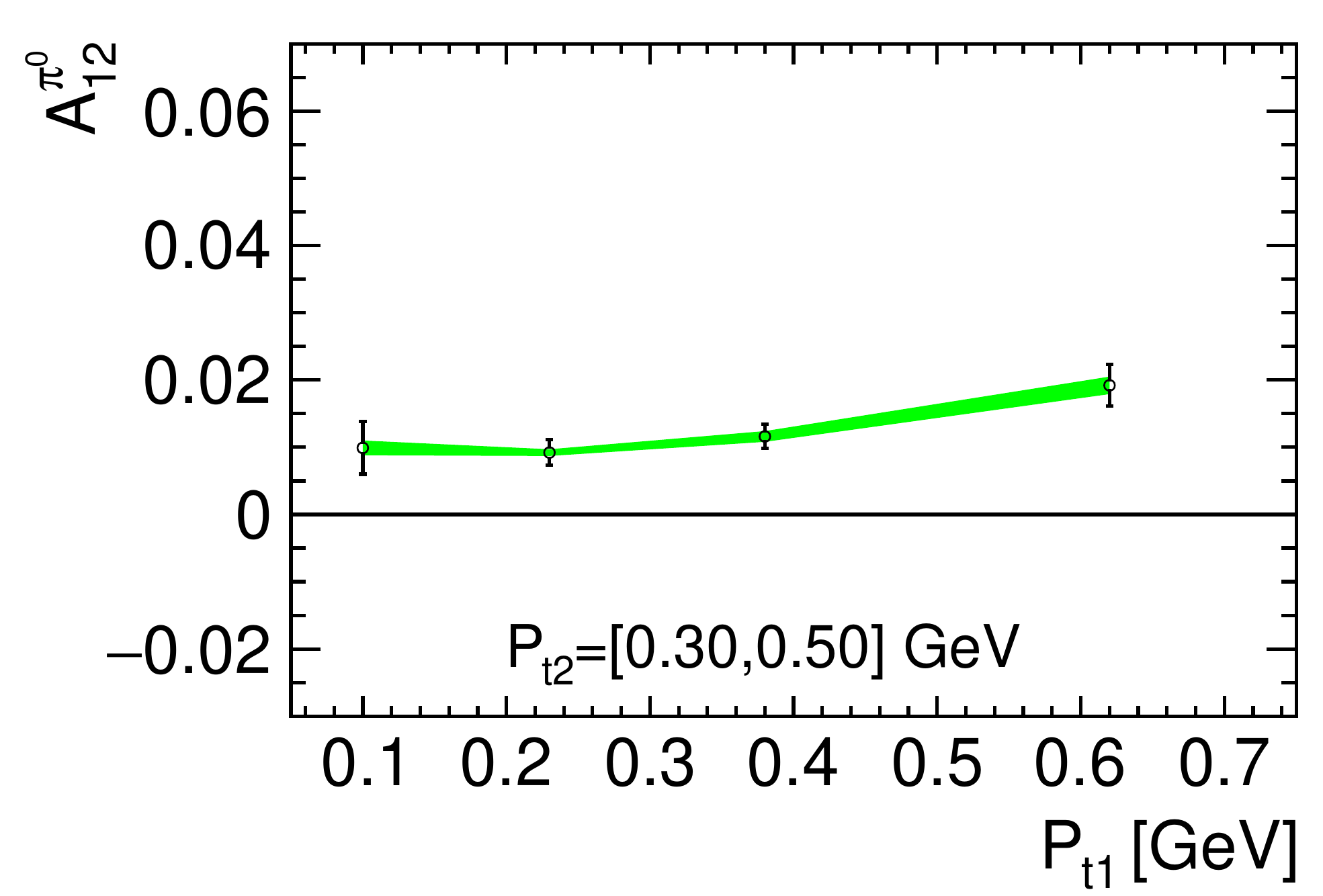}\\
\includegraphics[clip, trim=0.1cm 0.0cm 0.2cm 0.5cm,height=0.27\textheight,width=0.55\textwidth]{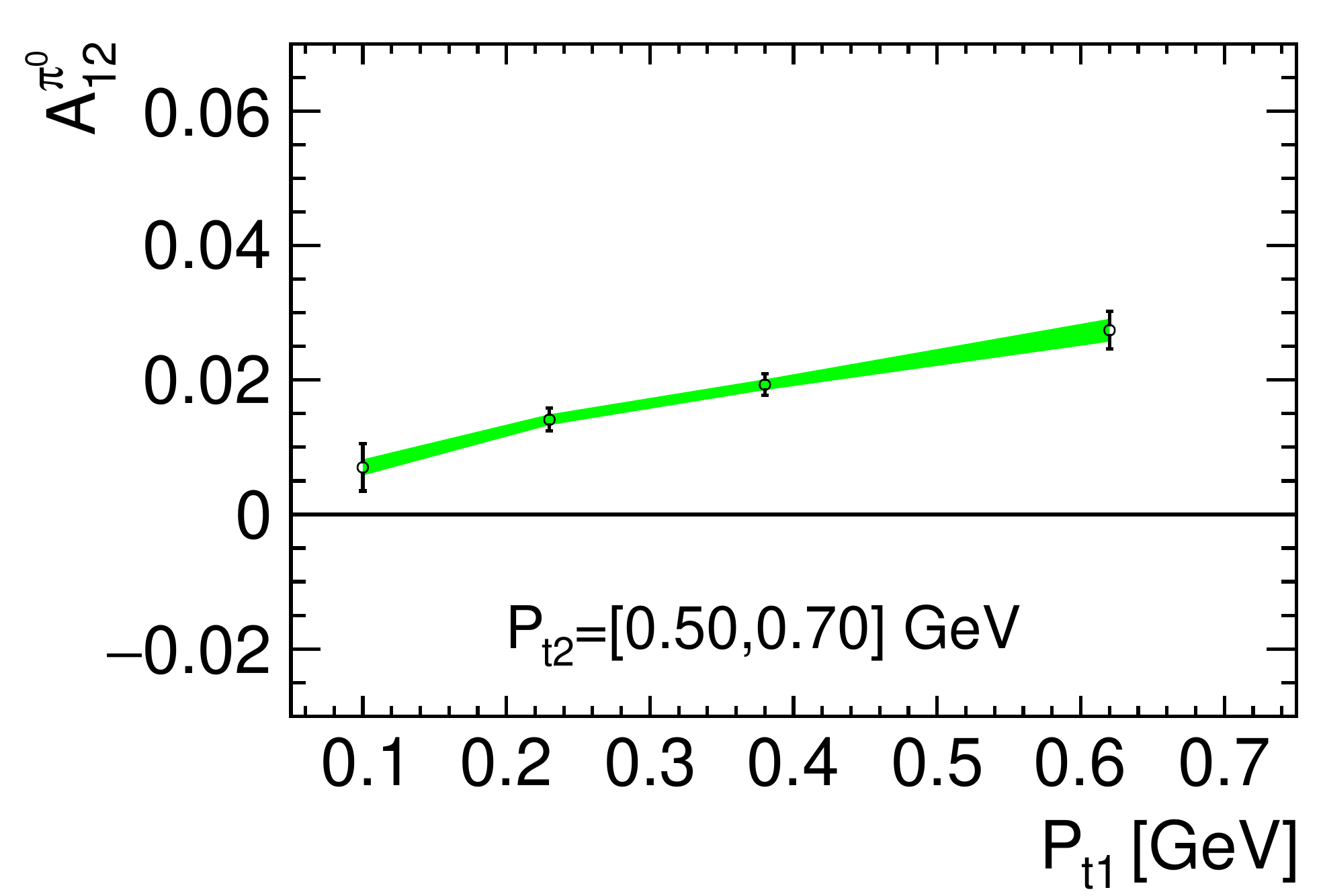}
\includegraphics[clip, trim=4.4cm 0.0cm 0.2cm 0.5cm,height=0.27\textheight,width=0.44\textwidth]{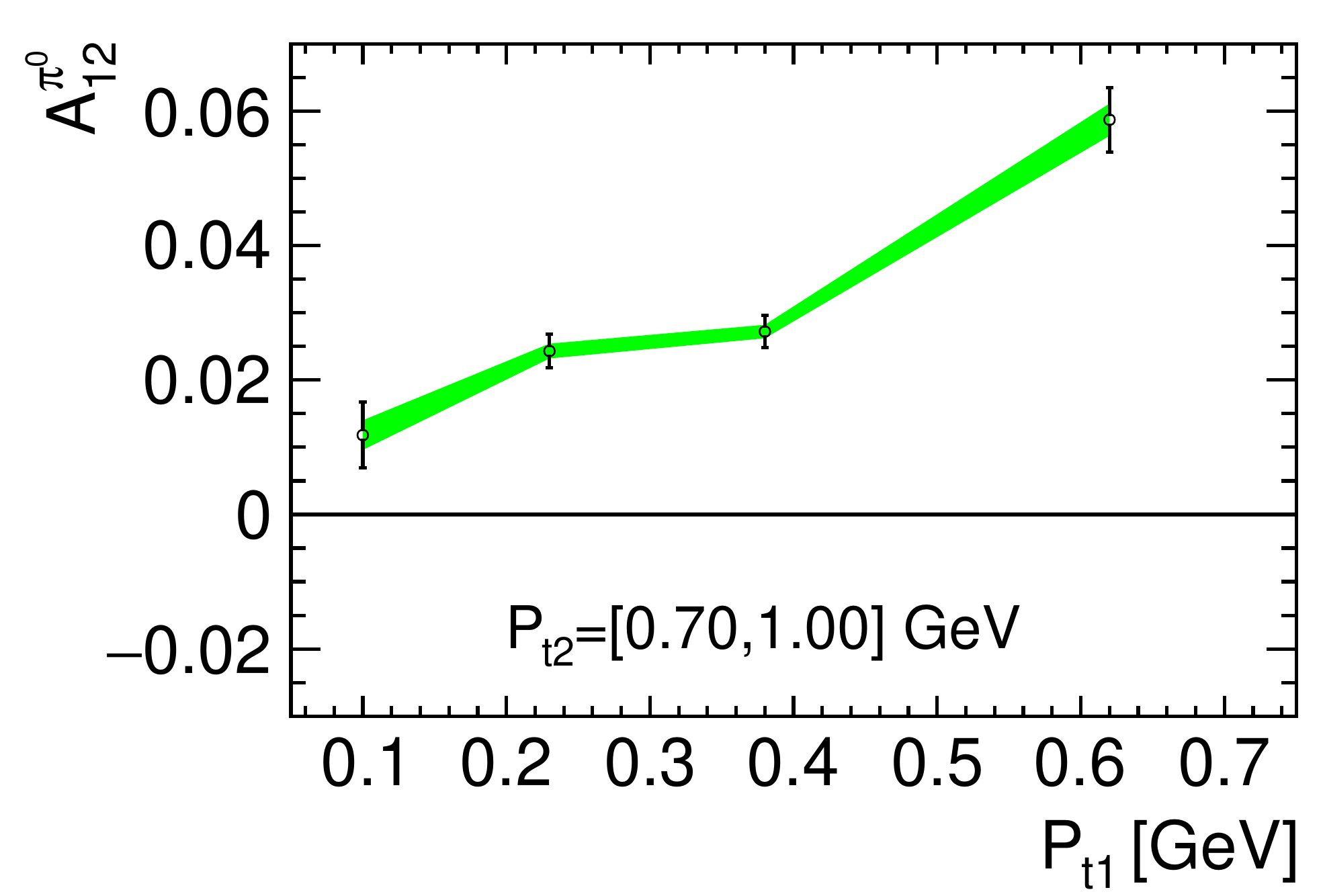}
\caption{Dependence of $A^{\pi^0}_{12}$ on $P_{t1}$ for four bins in $P_{t2}$ (as labeled), integrating within the overall limits over $z$. Error bars represent statistical uncertainties while the colored bands indicate systematic uncertainties.\label{fig:resPi0VsPt1Pt2}}
\end{figure*}

The transverse-momentum dependence is explored in both a mixed \(z_{1}\)--\(P_{t1}\) binning and a \(P_{t1}\)--\(P_{t2}\) binning. Figure~\ref{fig:resPi0VsZPt} shows the results for $A^{\pi^0}_{12}$ versus $z_1$ and $P_{t1}$, and  Fig.~\ref{fig:resPi0VsPt1Pt2} the results versus $P_{t1}$ and $P_{t2}$. For $P_{t1}$ approaching zero, the asymmetry vanishes. The continuous rise with $P_{t1}$ is consistent with a linear behavior. Higher values of $z_1$ are again associated with larger values of $A^{\pi^0}_{12}$, following the same behavior encountered for the charged-pion case. 

\begin{figure*} 

\begin{minipage}{0.55\textwidth}
\includegraphics[clip, trim=0cm 2.7cm 0.3cm 0cm,height=0.25\textheight,width=1.0\textwidth]{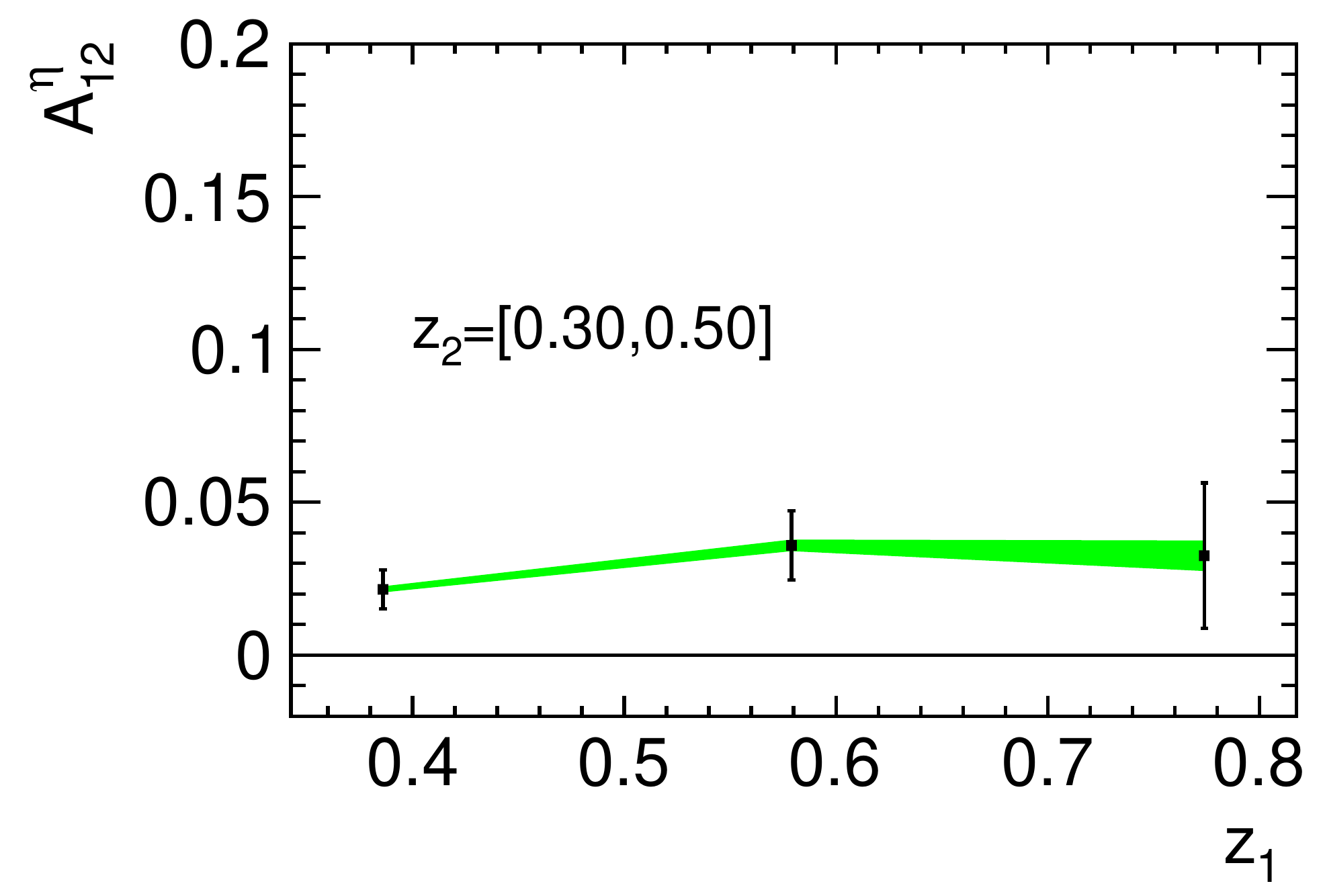}\\
\includegraphics[clip, trim=0cm 0cm 0.2cm 0cm,height=0.29\textheight,width=1.0\textwidth]{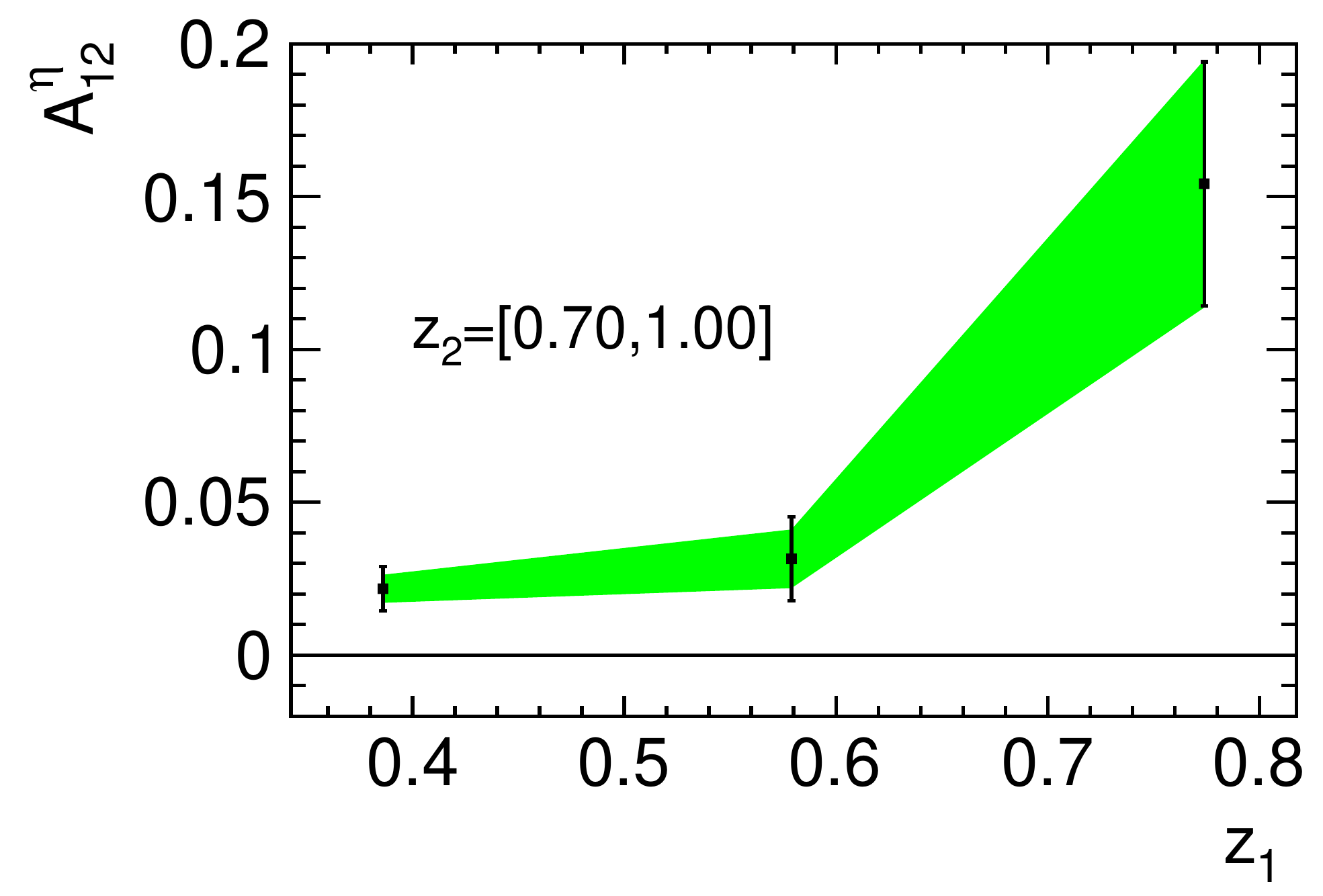}
\end{minipage}
\begin{minipage}{0.44\textwidth}
\includegraphics[clip, trim=4.3cm 0.2cm 0.3cm 0.6cm,height=0.293\textheight,width=1.0\textwidth]{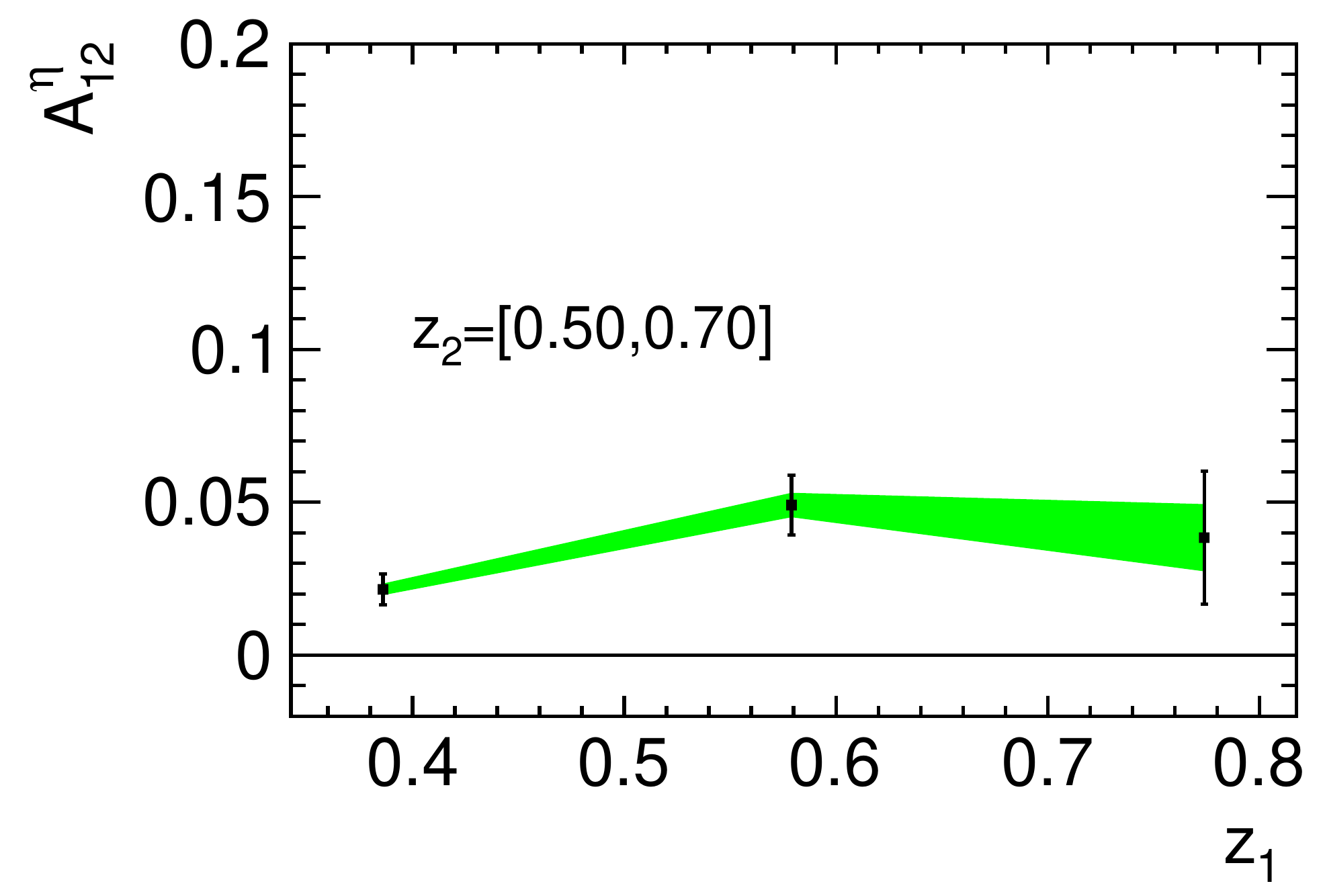}\\
\includegraphics[clip, trim=0cm 2cm 0cm 0cm,height=0.22\textheight,width=1.0\textwidth]{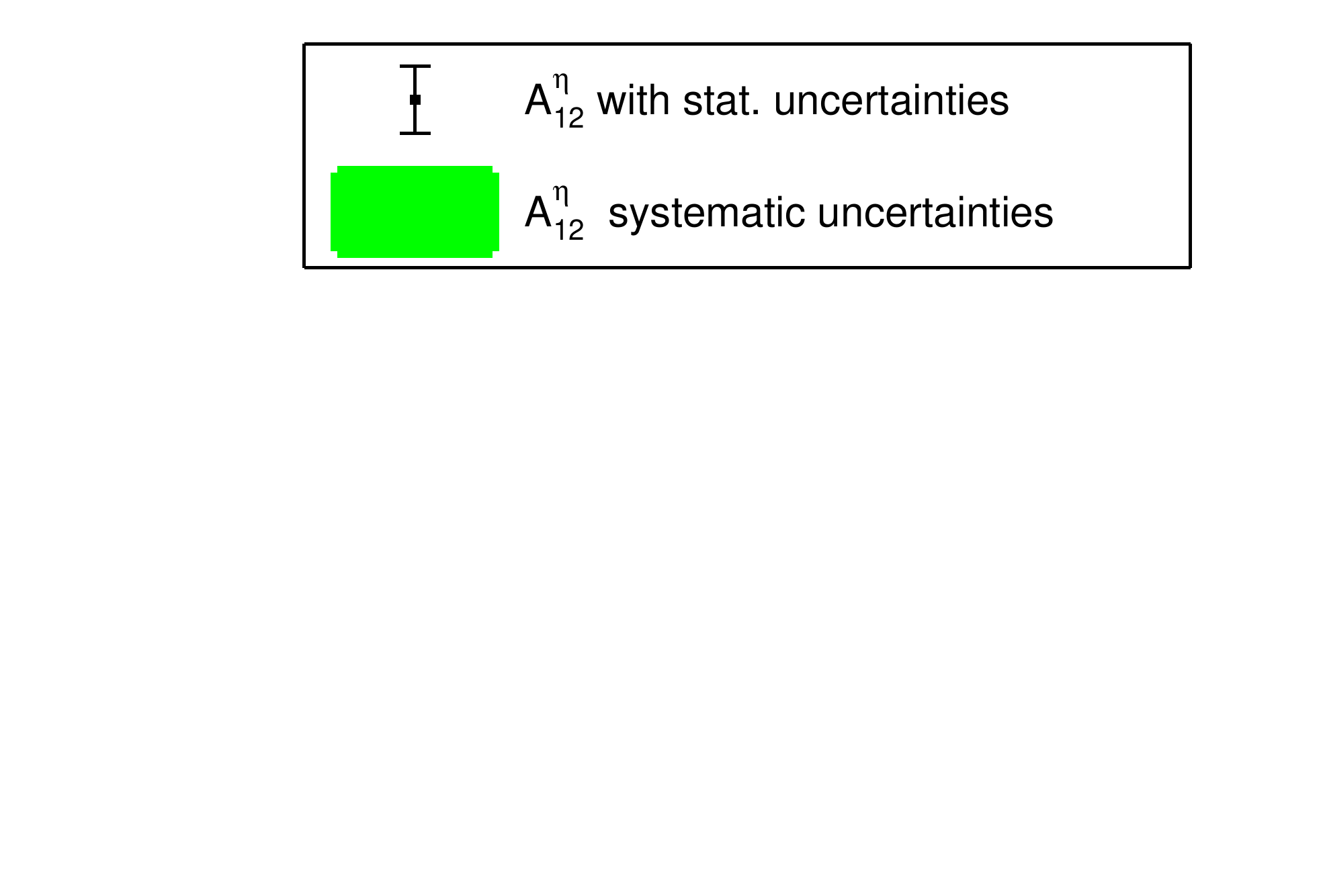}
\end{minipage}
\caption{Dependence of $A^{\eta}_{12}$ on $z_{1}$ for three bins in $z_{2}$ (as labeled), integrating within the overall limits over \(P_{t}\). Error bars represent statistical uncertainties while the colored bands indicate systematic uncertainties. 
\label{fig:resEtaZ1Z2}}
\end{figure*}

\begin{figure*} 
    \centering
    \includegraphics[width=0.55\textwidth]{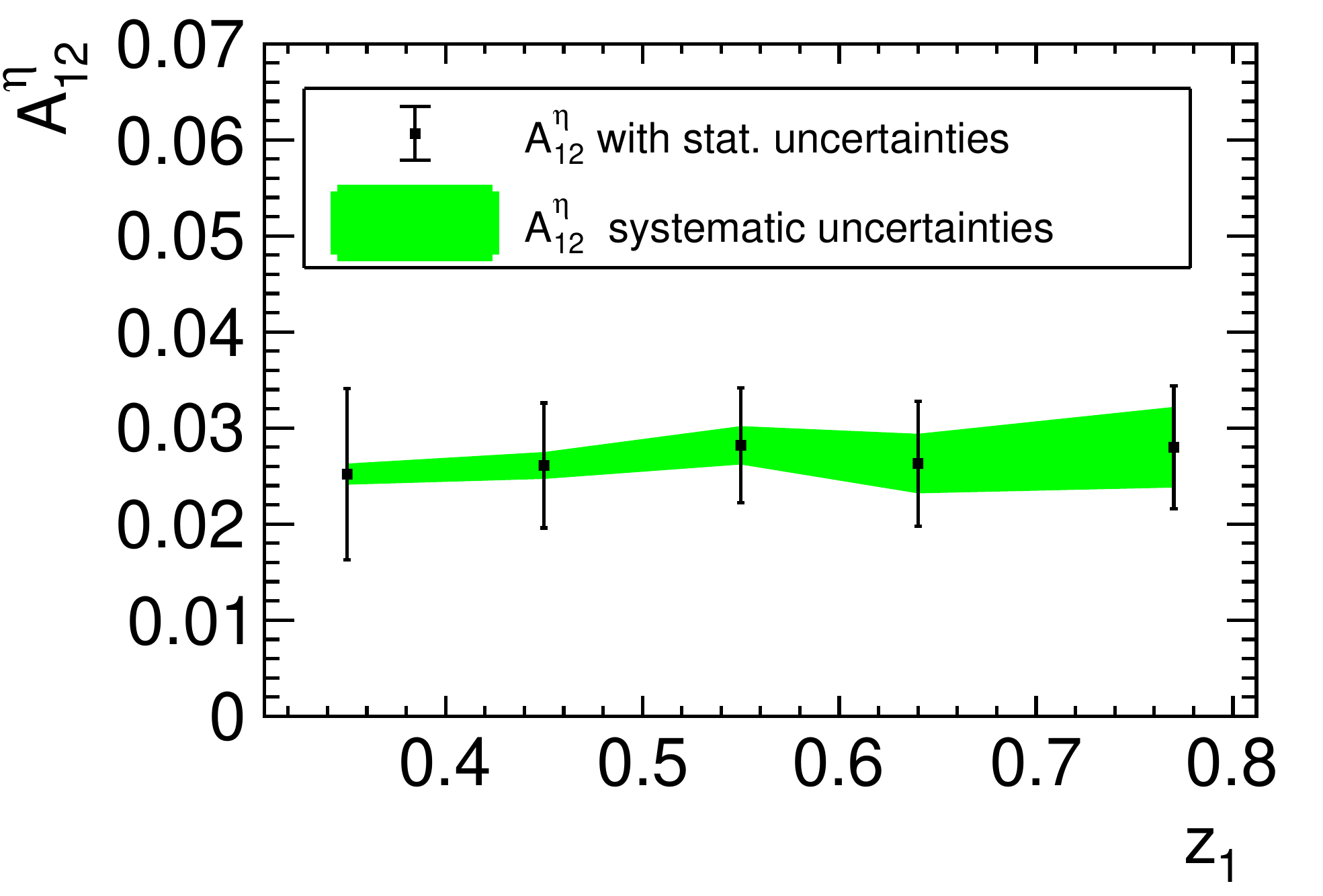}
    \caption{Dependence of $A^{\eta}_{12}$ on $z_{1}$, integrating within the overall limits over \(P_{t}\) and $z_{2}$.\label{fig:etaVsZ}}
\end{figure*}

The results for the $\eta$ asymmetries have significantly larger uncertainties than those from $\pi^0$. 
They are extracted from the Belle data imposing a minimum \(z\) of 0.3 for both the \(\eta\) and the charged pions involved in the construction of the double ratios.
Figure~\ref{fig:resEtaZ1Z2} shows the results of $A^{\eta}_{12}$ binned in $(z_1,z_2)$. 
The rise with \(z\) is much less pronounced than the one for charged and neutral pions. Indeed, for the sole $z_{1}$ dependence, integrating over \( P_{t1} \)  as well as the kinematics of the hadrons in the opposite hemisphere, the asymmetry appears almost constant as shown in Fig.~\ref{fig:etaVsZ}.

\begin{figure*} 
\begin{minipage}{0.55\textwidth}
\includegraphics[clip, trim=0cm 2.7cm 0.7cm 0cm,height=0.25\textheight,width=1.0\textwidth]{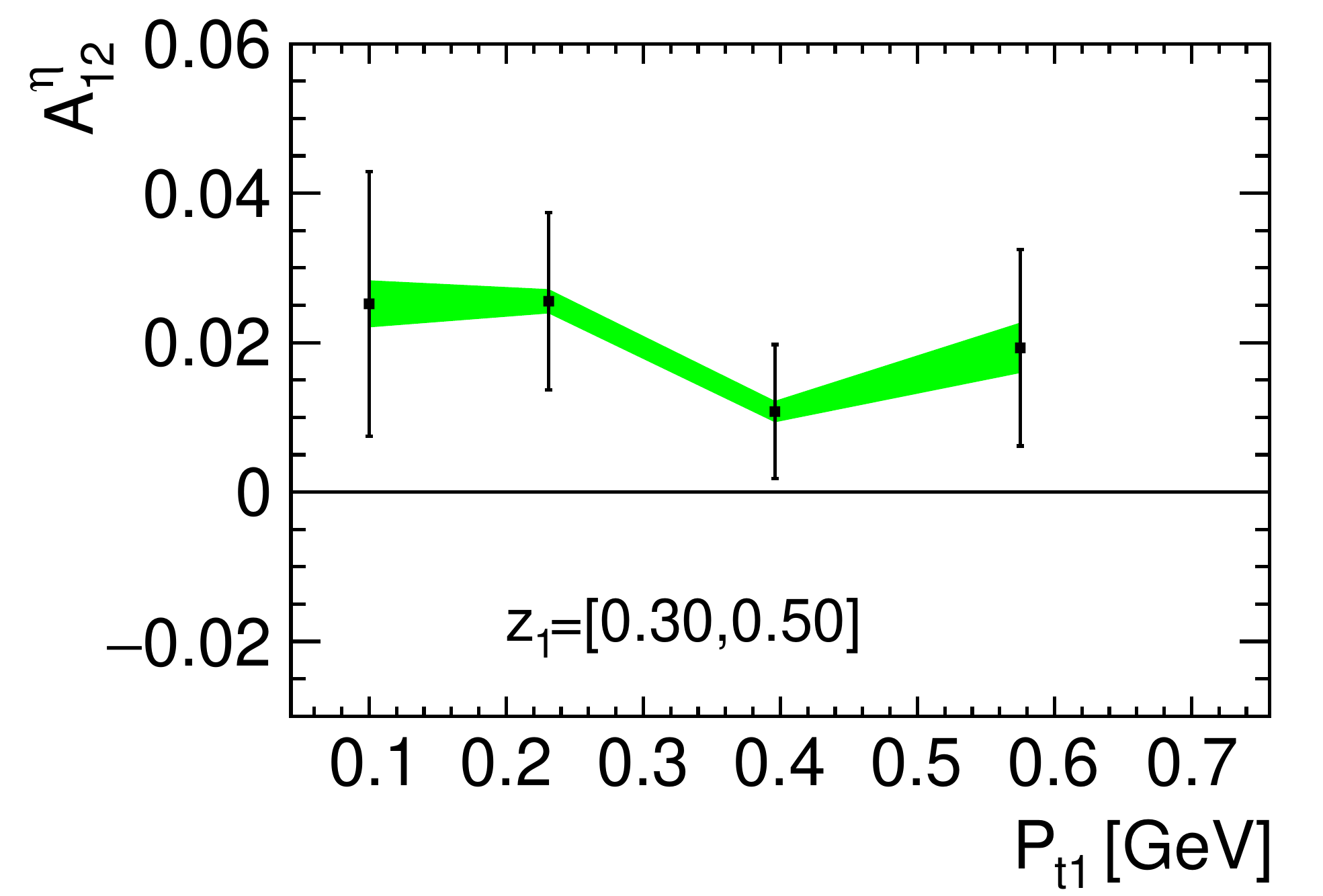}\\
\includegraphics[clip, trim=0cm 0cm 0.7cm 0.2cm,height=0.29\textheight,width=1.0\textwidth]{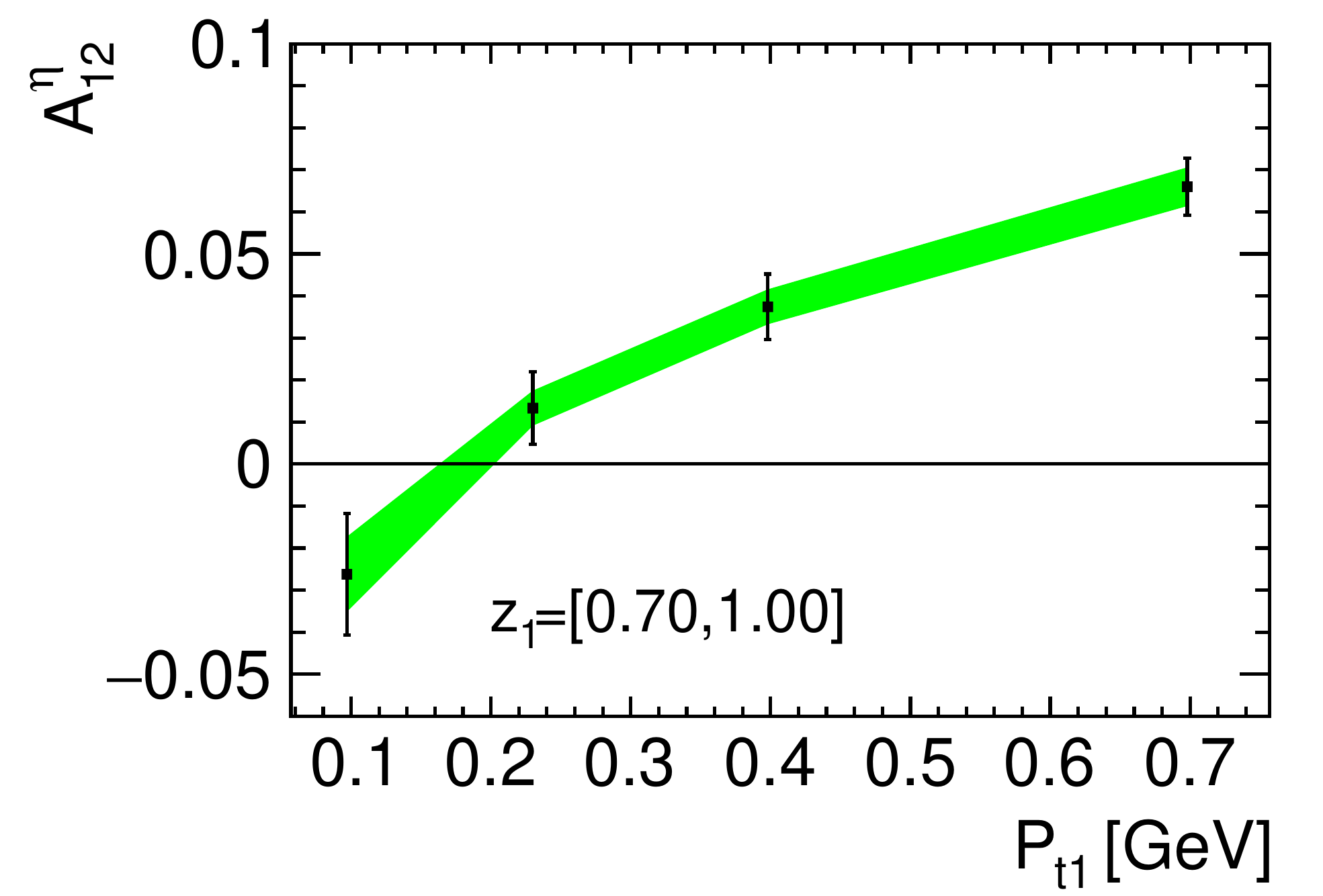}
\end{minipage}
\begin{minipage}{0.44\textwidth}
\includegraphics[clip, trim=4.3cm 0.0cm 0cm 0.305cm,height=0.305\textheight,width=1.0\textwidth]{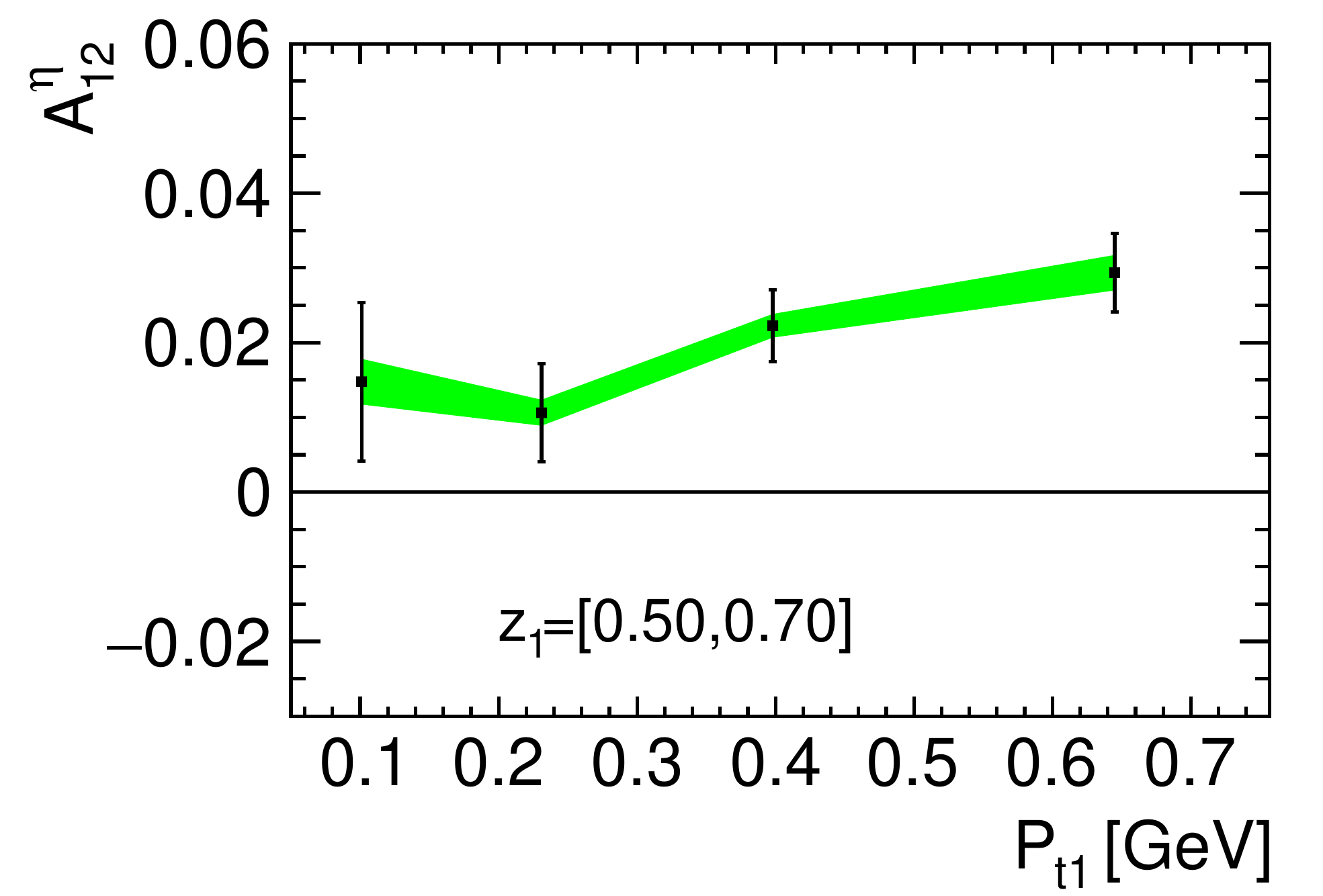}\\
\includegraphics[clip, trim=0cm 2cm 0cm 0cm,height=0.22\textheight,width=1.0\textwidth]{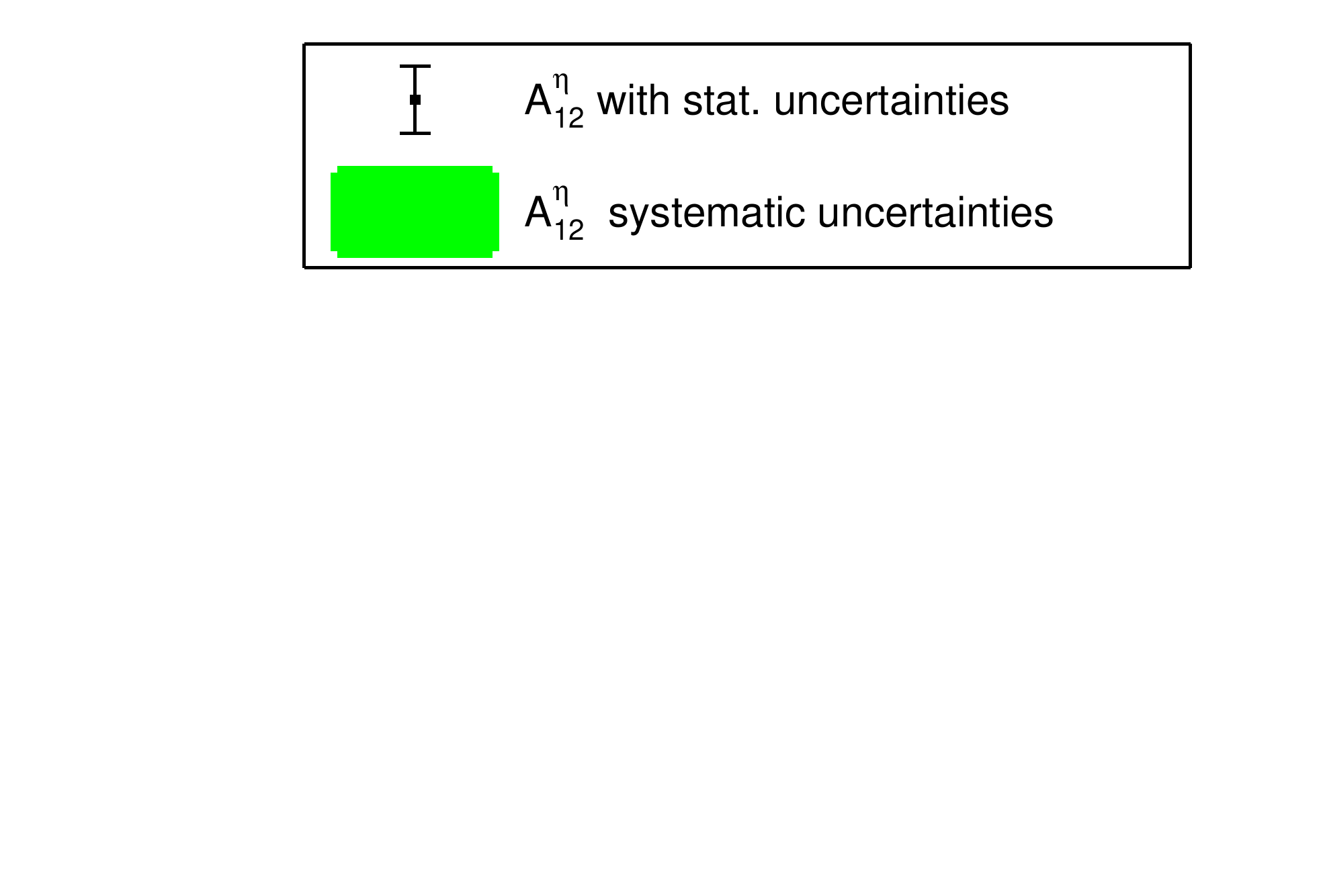}
\end{minipage}
\caption{Dependence of $A^{\eta}_{12}$ on $P_{t1}$ for three bins in $z_{1}$ (as labeled), integrating within the overall limits over the kinematics of the second hadron. Error bars represent statistical uncertainties while the colored bands indicate systematic uncertainties.
\label{fig:resEtaZPt}}
\end{figure*}

Figure~\ref{fig:resEtaZPt} shows the results of $A^{\eta}_{12}$ binned in $(P_{t1},P_{t2})$. A clear rise of the asymmetry with transverse momentum can be identified that reaches up to 0.05 for the largest values of \(P_{ti}\). 
Within large uncertainties, these results for $A^{\eta}_{12}$ are mostly consistent with those of  $A_{12}^{\pi^0}$.

\begin{figure*} 
\includegraphics[clip, trim=0.1cm 2.4cm 0.2cm 0.2cm,height=0.25\textheight,width=0.55\textwidth]{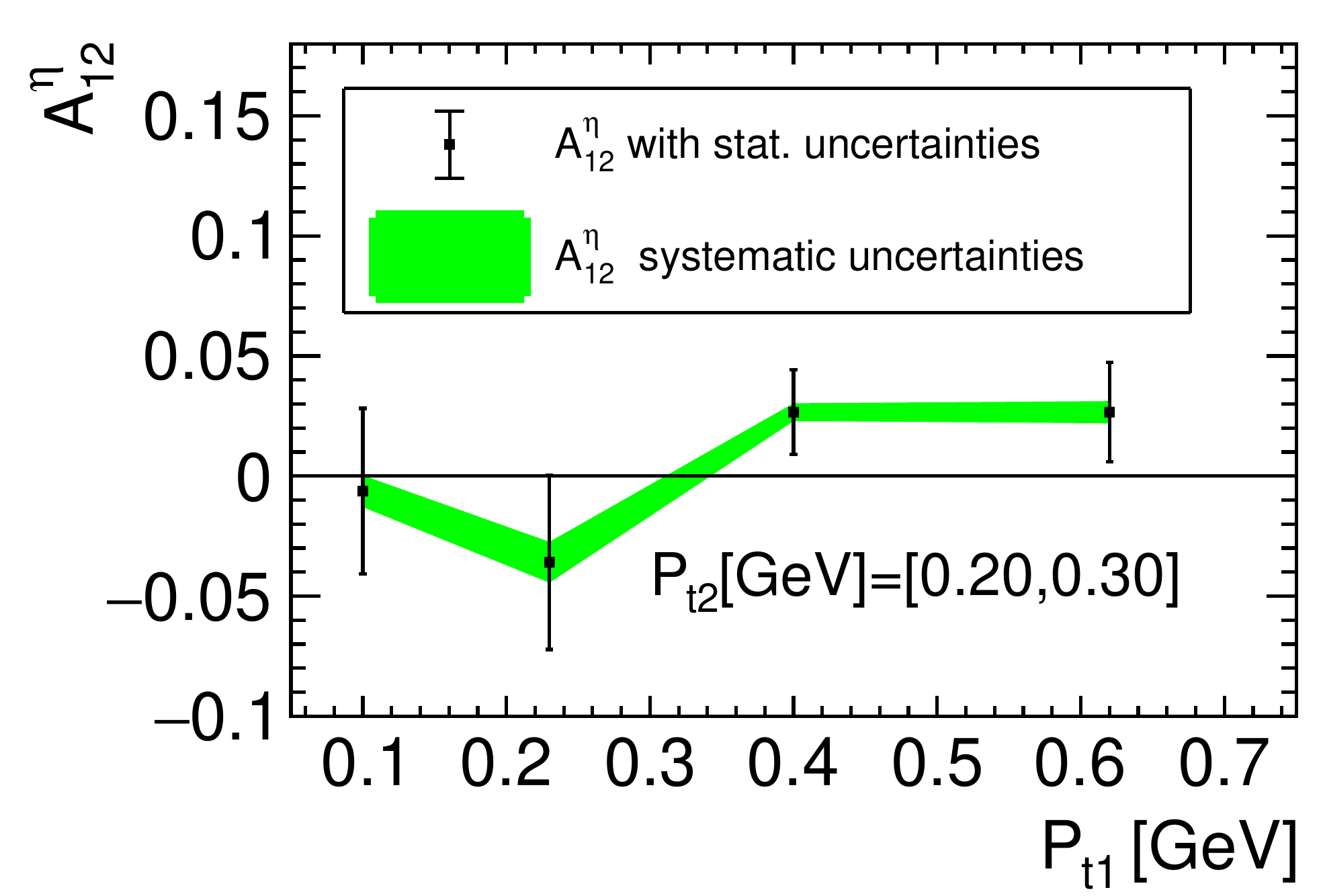}
\includegraphics[clip, trim=4.4cm 2.4cm 0.2cm 0.2cm,height=0.25\textheight,width=0.44\textwidth]{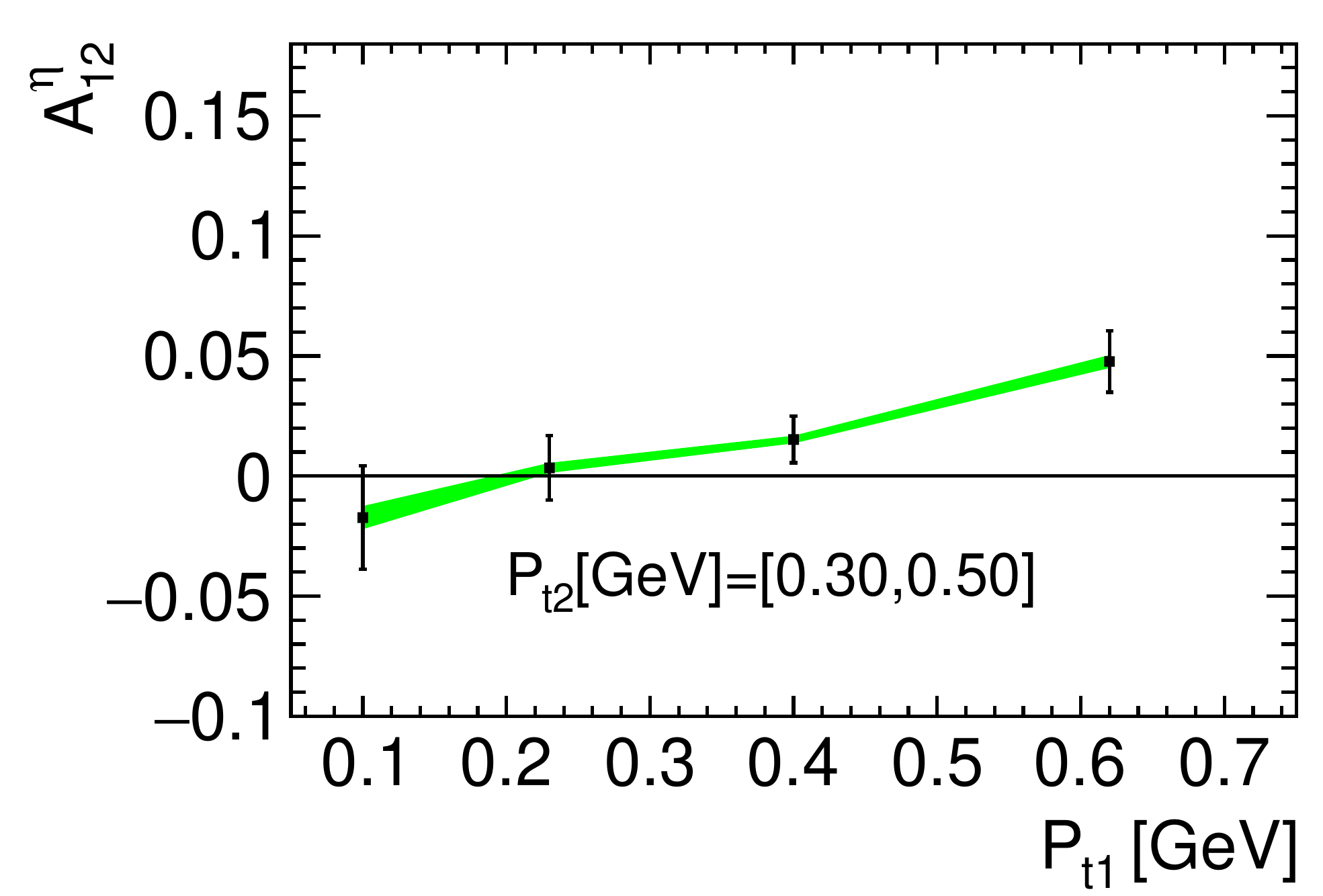}\\
\includegraphics[clip, trim=0.1cm 0.0cm 0.2cm 0.5cm,height=0.27\textheight,width=0.55\textwidth]{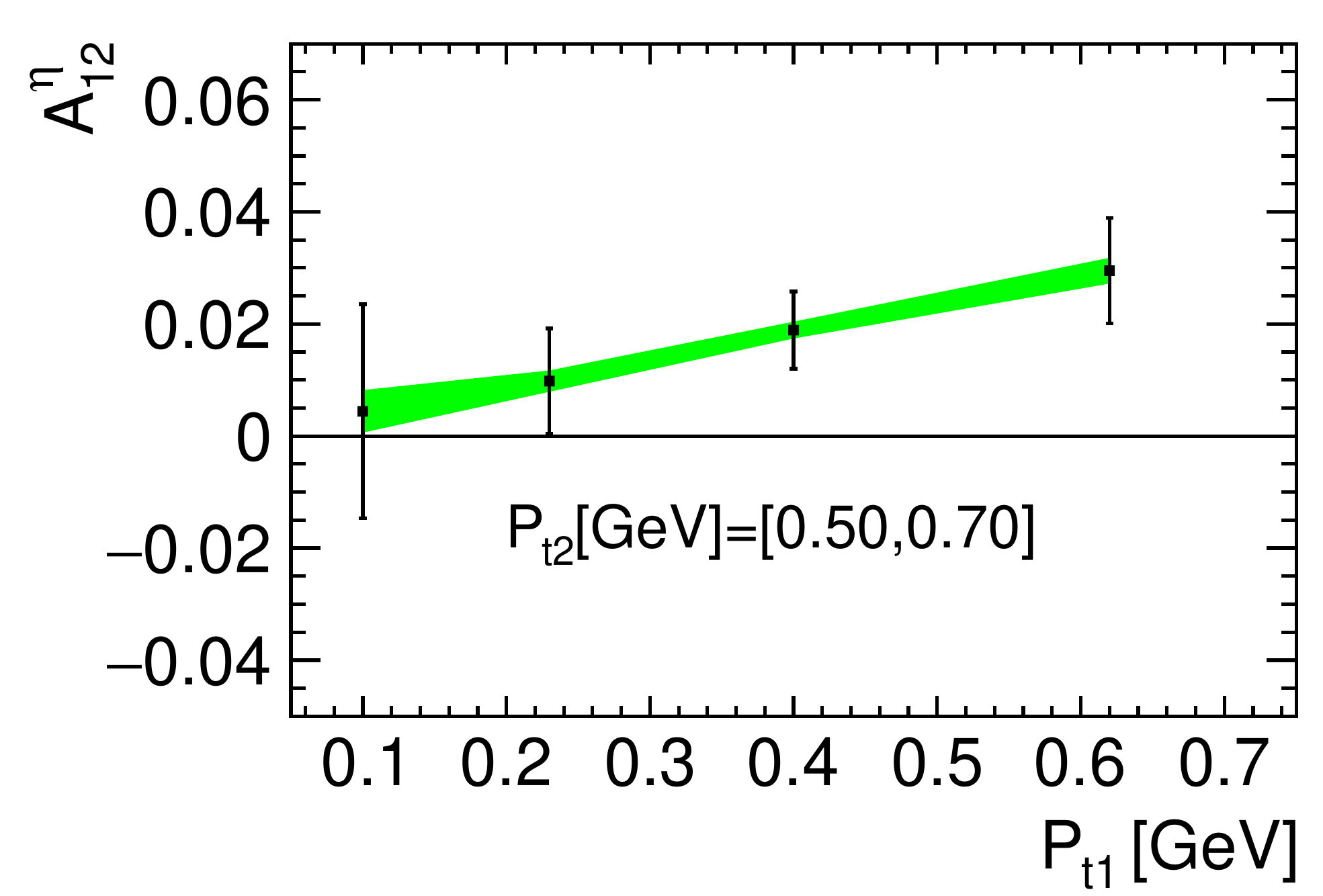}
\includegraphics[clip, trim=4.4cm 0.0cm 0.2cm 0.5cm,height=0.27\textheight,width=0.44\textwidth]{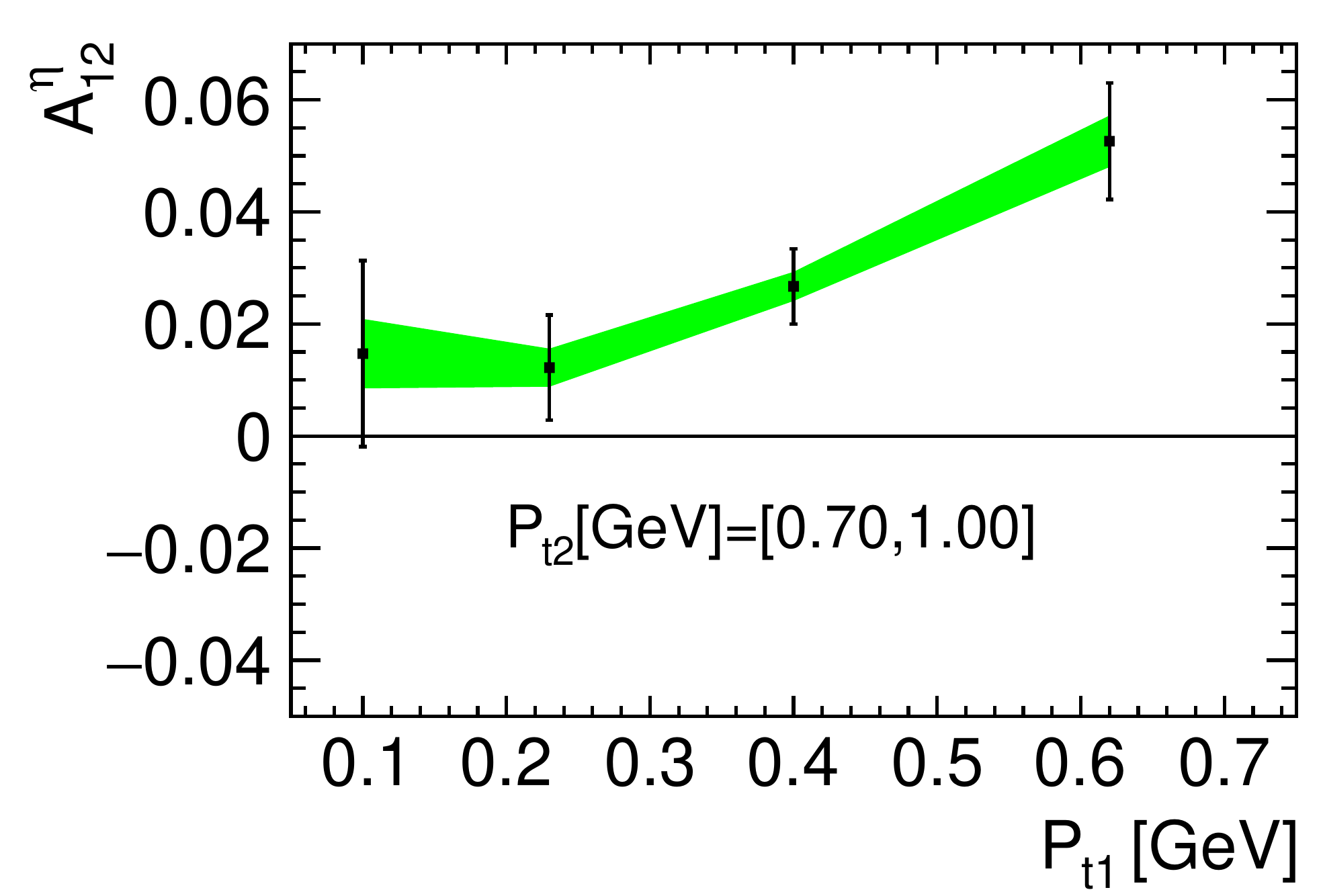}
\caption{Dependence of $A^{\eta}_{12}$ on $P_{t1}$ for four bins in $P_{t2}$ (as labeled), integrating within the overall limits over $z$. Error bars represent statistical uncertainties while the colored bands indicate systematic uncertainties.\label{fig:resEtaVsPt1Pt2}}
\end{figure*}

\begin{figure*} 
\begin{minipage}{0.55\textwidth}
\includegraphics[clip, trim=0cm 2.7cm 0.8cm 0cm,height=0.25\textheight,width=1.0\textwidth]{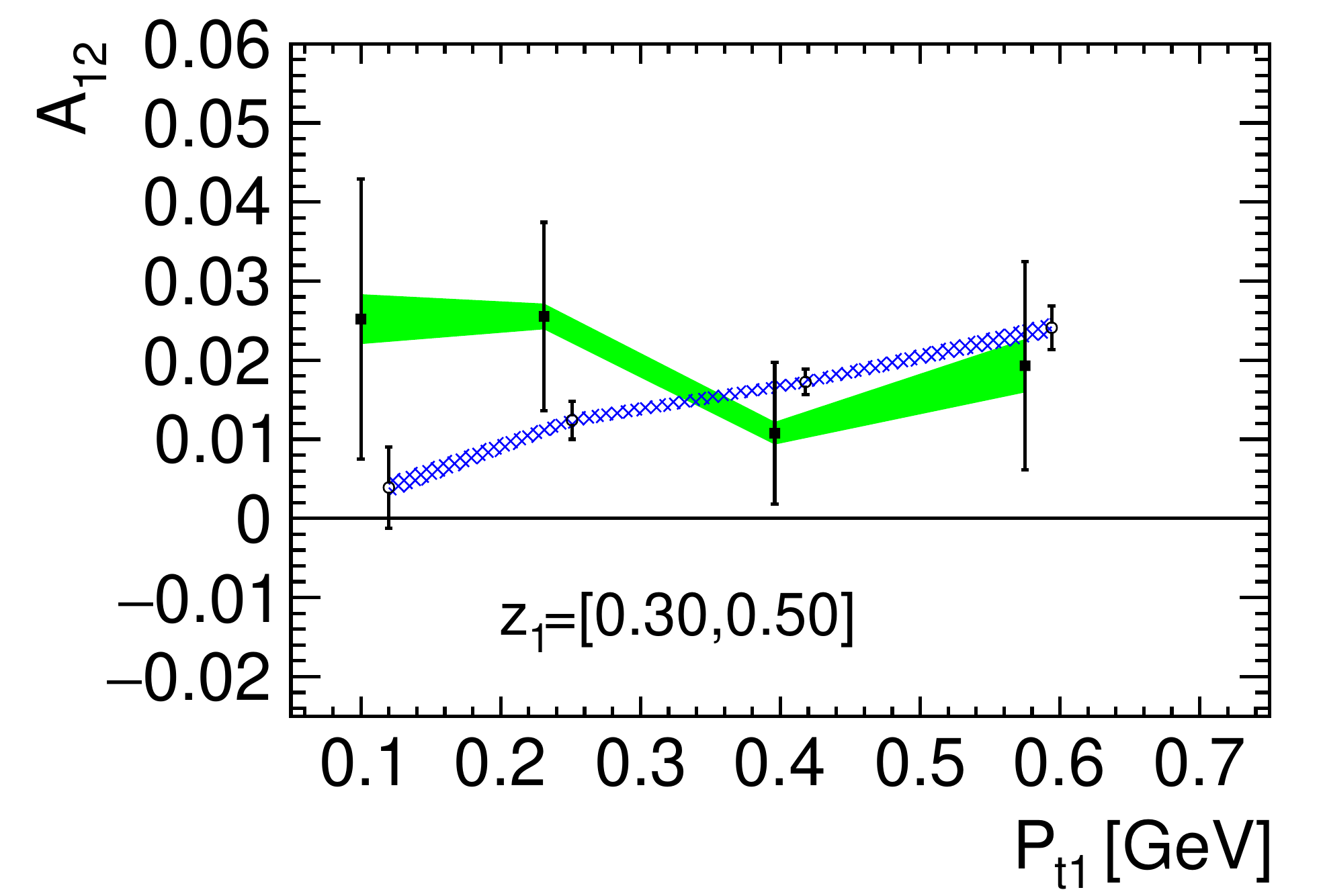}\\
\includegraphics[clip, trim=0cm 0cm 0.8cm 0cm,height=0.29\textheight,width=1.0\textwidth]{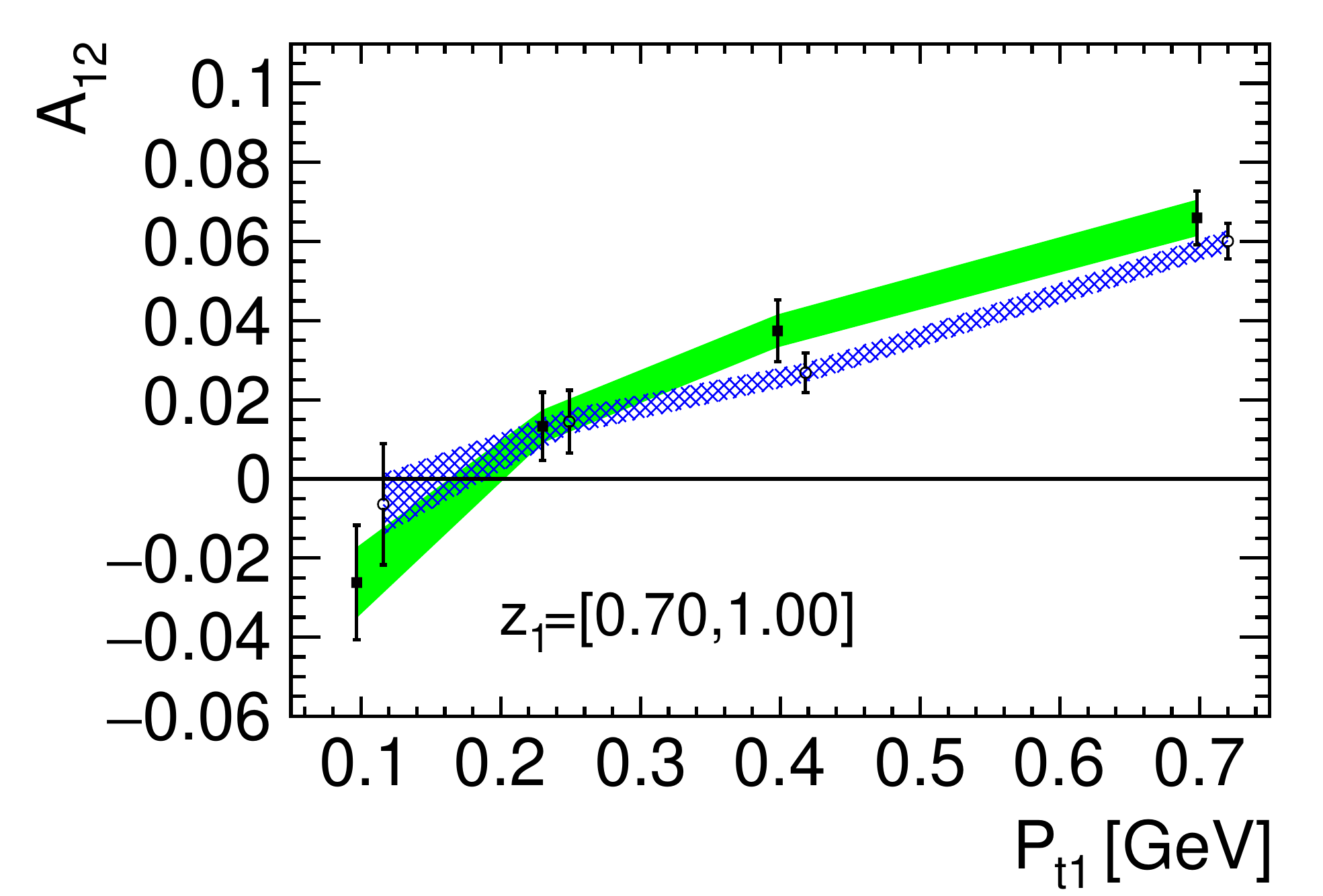}
\end{minipage}
\begin{minipage}{0.44\textwidth}
\includegraphics[clip, trim=4.4cm 0.0cm 0cm 0.305cm,height=0.305\textheight,width=1.0\textwidth]{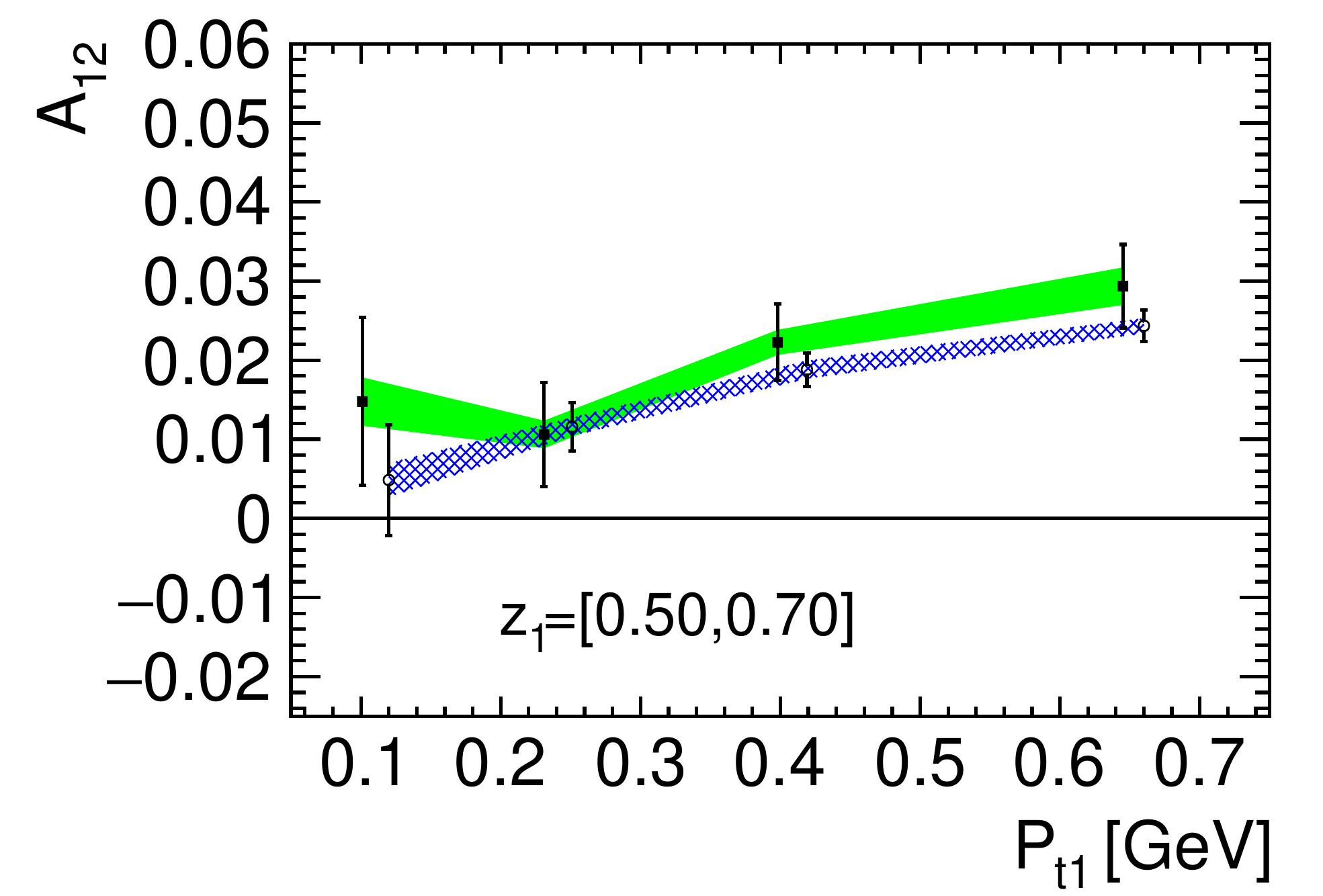}\\
\includegraphics[clip, trim=0cm 2cm 0cm 0cm,height=0.22\textheight,width=1.0\textwidth]{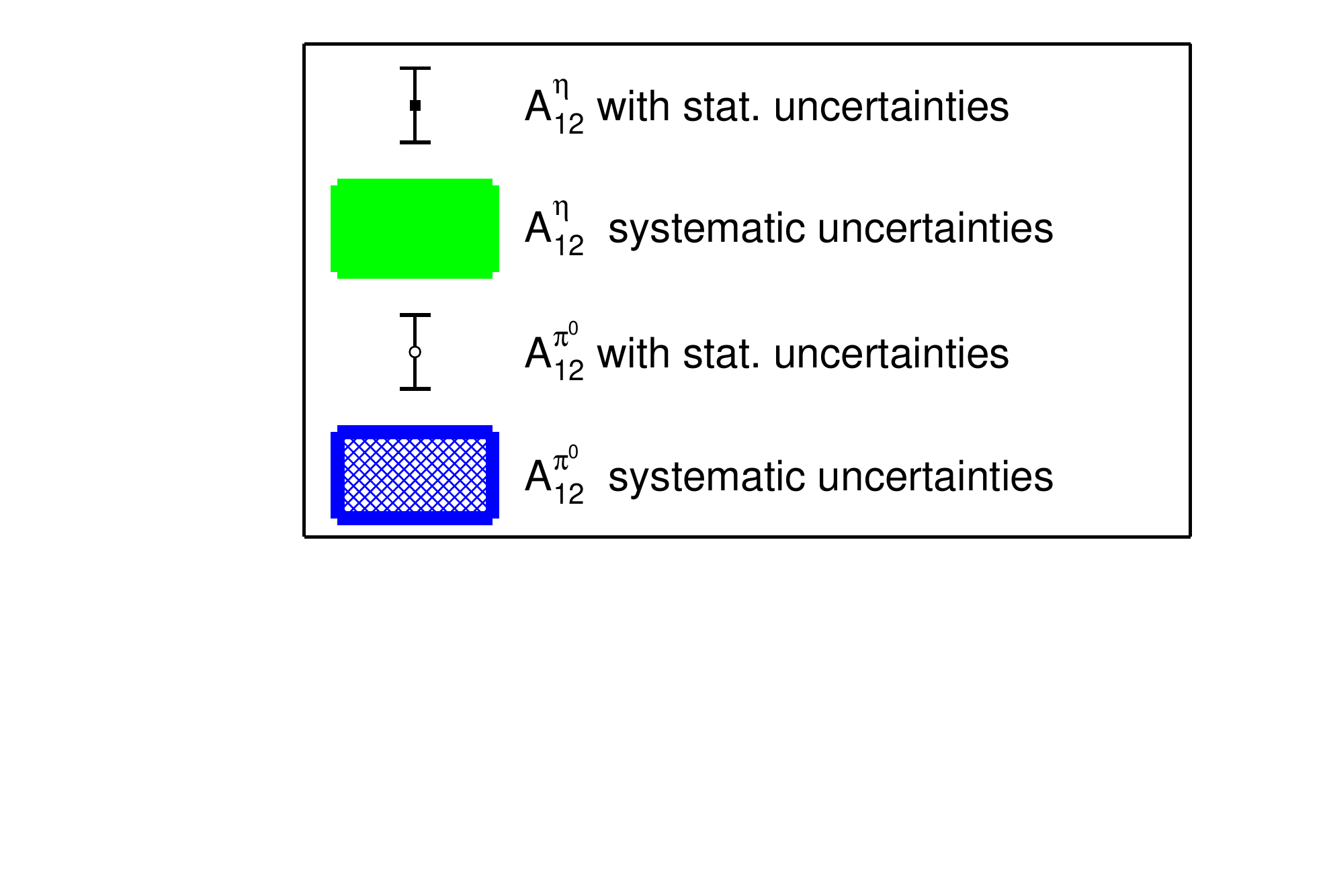}
\end{minipage}
\caption{Comparison of the \(P_{t1}\) dependences of $A^{\pi^0}_{12}$ (open circles) and $A^{\eta}_{12}$ (full squares) for three bins in \(z_{1}\) (as labeled).  A constraint of $z>0.3$ is applied also for $A^{\pi^0}_{12}$ to be consistent with the kinematic constraints used for the \(\eta\) asymmetries.
\label{fig:resEtaPi0ZPt}}
\end{figure*}

In the case of the mixed (\(z_{1},P_{t1}\)) binning, displayed in Fig.~\ref{fig:resEtaVsPt1Pt2}, no definite behavior is visible. While clearly rising with \(P_{t1}\) for the last \(z_1\) bin (\(z_{1} > 0.7\)), the asymmetry is otherwise nearly consistent with a constant, especially as one approaches the lowest \(z_1\) bin. 
Nevertheless, within the much larger uncertainties the \(\eta\) asymmetries are consistent with the $A^{\pi^0}_{12}$ results, which is shown explicitly in Fig.~\ref{fig:resEtaPi0ZPt} for the $(z_1,P_{t1})$ binning, and for which the \(z>0.3\) requirement was also applied to the \(\pi^{0}\) asymmetries. One caveat of this direct comparison is the difference in charm contributions to the \(\pi^{0}\) and \(\eta\), which are about 20--30\% larger for the \(\eta\) sample and cannot be eliminated easily as discussed above. On the other hand, for bins with similar enough charm contributions, a comparison is better motivated. Considering Tables~\ref{tab:sinzcharmratio}-\ref{tab:comptcharmratio}, the best candidates appear to be the first few bins in the $(P_{t1},P_{t2})$ binning, for which the \(\eta\) and \(\pi^{0}\) asymmetries are fully consistent.

\begin{figure} 
    \centering
    \includegraphics[width=0.49\textwidth]{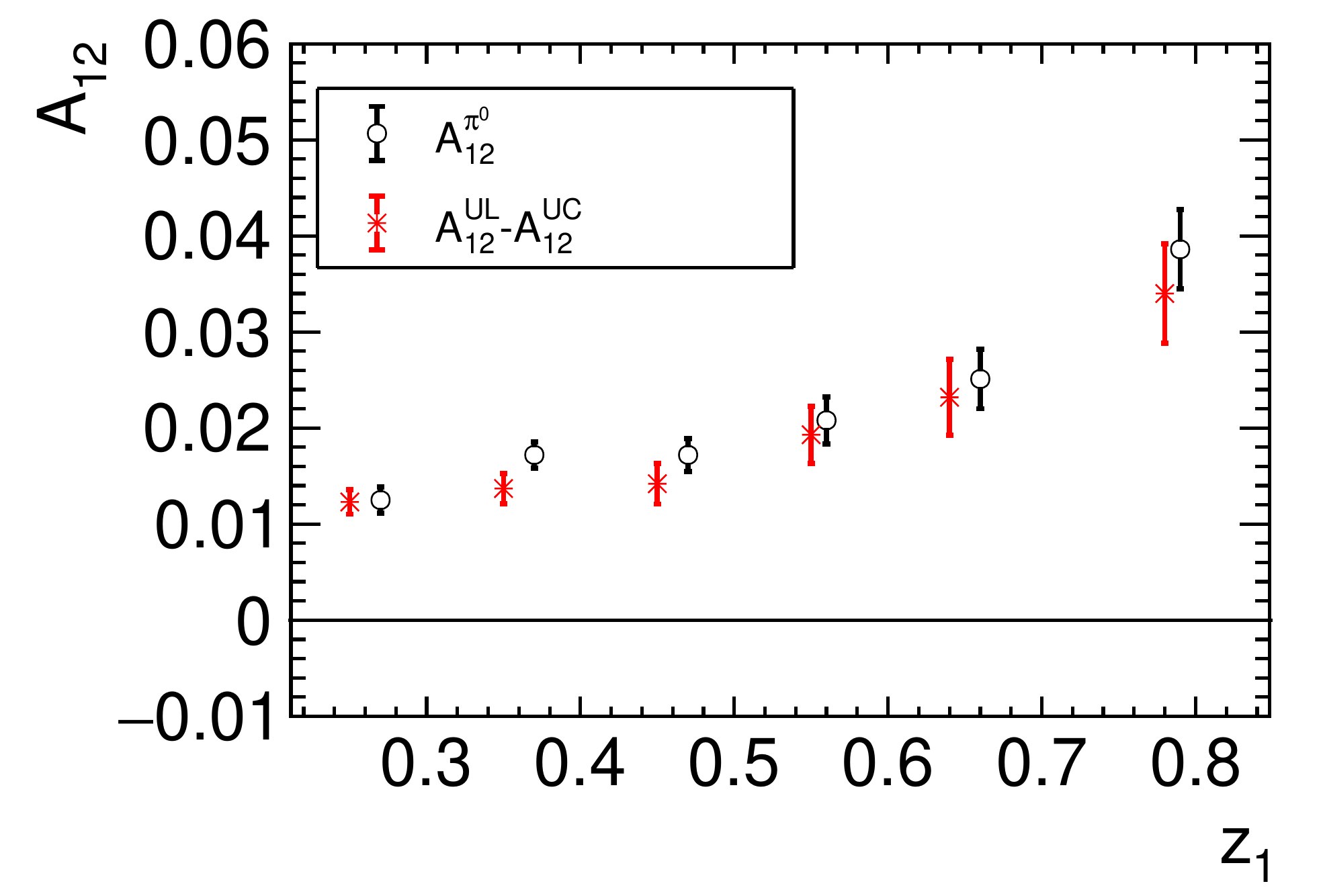}
    \caption{Dependence of $A^{\pi^0}_{12}$ and $A_{12}^{UL}-A_{12}^{UC}$ on $z_{1}$, integrating within the overall limits over \(P_{t}\) and $z_{2}$. The data points of $A_{12}^{\pi^0}$ are offset horizontally by 0.02 for legibility and error bars represent combined statistical and systematic uncertainties.\label{fig:pi0piPMcomparison}}
\end{figure}

Direct extraction of the fragmentation functions for \(\pi^0\) and \(\eta\) from the double ratio results for comparison with those for charged pions requires further assumptions on the charged-pion fragmentation functions, and is hampered by the complexity of the double ratios. This becomes apparent when recalling the rather involved parton-model expressions~\eqref{eqn:allratiosexpress2}-\eqref{eqn:FF5eta} for the various meson combinations. The expression for \( A^{\pi^0}_{12} \) is equal to that of \( A^{UL}_{12} - A^{UC}_{12} \) as a result of the isospin relations \eqref{eqn:FF4} and \eqref{eq:pistrange}.
Figure~\ref{fig:pi0piPMcomparison} displays both \( A^{\pi^0}_{12} \) and the difference between \( A^{UL}_{12}\) and  \(A^{UC}_{12} \), and indeed good agreement is found. The comparison is to be taken with caution as not all potential correlations between the three asymmetries are taken into account.

The non-vanishing asymmetries for double ratios involving \(\pi^0\) and \(\eta\) mesons do not necessarily point to non-vanishing Collins fragmentation functions for these two. It is plausible for non-vanishing asymmetries to arise in the case of vanishing Collins functions for \(\pi^0\) and \(\eta\) due to the presence of the second ratio term in Eqs.~\eqref{eqn:FF5} and \eqref{eqn:FF5eta}, which involves only the charged pions.\footnote{As a reminder, the second term enters because of using charged-pion pairs in the denominator of the double ratios.}
The first ratio term can be rewritten in terms of products of only \(\pi^0\) fragmentation functions (in the case of \( A^{\pi^0}_{12} \)) or of  \(\pi^0\) and \(\eta\) fragmentation functions (in the case of \( A^{\eta}_{12} \)), i.e., the first ratio is governed by neutral-meson fragmentation functions only, while the second term by charged-pion fragmentation functions.
Taking into account that the favored and disfavored pion Collins fragmentation functions are on average of similar magnitude but opposite in sign, thus leading to cancellation effects in the combination relevant for the \(\pi^0\), a scenerio is plausible in which the $\pi^0$ Collins fragmentation is small and the observed signal is due to the term containing the charged-pion fragmentation functions.
This is also consistent with the vanishing \(\pi^0\) Collins asymmetries observed in semi-inclusive DIS~\cite{Airapetian:2010ds}. The non-vanishing results for \( A^{\pi^0}_{12} \) and \( A^{\eta}_{12} \) would then mainly be a reflection of the non-vanishing azimuthal modulation in the denominator of those double ratios.

\section{Summary and Conclusion}
\label{sec:summary}
An analysis of azimuthal asymmetries related to the Collins mechanism has been presented for pairs of back-to-back neutral and charged pions as well as $\eta$ mesons and charged pions. The analysis substantially differs from previous Belle analyses in that results are only presented in the thrust-axis frame without correcting to the $q\bar{q}$ axis, the opening angle of the hadrons to the thrust axis was limited to 0.3 (which effectively corresponds to a $z$-dependent upper limit on $P_t$), and asymmetries were not corrected for charm contributions. Instead, the charm fraction is included and its impact can more properly be treated in future analyses when relevant results on charm azimuthal asymmetries become available, e.g., from Belle II~\cite{Kou:2018nap}. More importantly, this measurement significantly expands the scope of previous Belle measurements by a) including $\pi^0$ and $\eta$ mesons; and b) exploring the transverse-momentum dependence of the azimuthal asymmetries.
Significant asymmetries for all channels are observed. Asymmetries mostly rise, within the given kinematic coverage, with $z$ and $P_t$. The signal for $\eta$ and $\pi^0$ mesons agrees within uncertainties. We show the results for charged-pion pairs agree well with previous Belle measurements~\cite{Abe:2005zx,Seidl:2008xc}.
\begin{acknowledgments}
We thank the KEKB group for the excellent operation of the
accelerator; the KEK cryogenics group for the efficient
operation of the solenoid; and the KEK computer group, and the Pacific Northwest National
Laboratory (PNNL) Environmental Molecular Sciences Laboratory (EMSL)
computing group for strong computing support; and the National
Institute of Informatics, and Science Information NETwork 5 (SINET5) for
valuable network support.  We acknowledge support from
the Ministry of Education, Culture, Sports, Science, and
Technology (MEXT) of Japan, the Japan Society for the 
Promotion of Science (JSPS), and the Tau-Lepton Physics 
Research Center of Nagoya University; 
the Australian Research Council including grants
DP180102629, 
DP170102389, 
DP170102204, 
DP150103061, 
FT130100303; 
Austrian Science Fund (FWF);
the National Natural Science Foundation of China under Contracts
No.~11435013,  
No.~11475187,  
No.~11521505,  
No.~11575017,  
No.~11675166,  
No.~11705209;  
Key Research Program of Frontier Sciences, Chinese Academy of Sciences (CAS), Grant No.~QYZDJ-SSW-SLH011; 
the  CAS Center for Excellence in Particle Physics (CCEPP); 
the Shanghai Pujiang Program under Grant No.~18PJ1401000;  
the Ministry of Education, Youth and Sports of the Czech
Republic under Contract No.~LTT17020;
the Carl Zeiss Foundation, the Deutsche Forschungsgemeinschaft, the
Excellence Cluster Universe, and the VolkswagenStiftung;
the Department of Science and Technology of India; 
the Istituto Nazionale di Fisica Nucleare of Italy; 
National Research Foundation (NRF) of Korea Grants
No.~2015H1A2A1033649, No.~2016R1D1A1B01010135, No.~2016K1A3A7A09005
603, No.~2016R1D1A1B02012900, No.~2018R1A2B3003 643,
No.~2018R1A6A1A06024970, No.~2018R1D1 A1B07047294; Radiation Science Research Institute, Foreign Large-size Research Facility Application Supporting project, the Global Science Experimental Data Hub Center of the Korea Institute of Science and Technology Information and KREONET/GLORIAD;
the Polish Ministry of Science and Higher Education and 
the National Science Center;
the Grant of the Russian Federation Government, Agreement No.~14.W03.31.0026; 
the Slovenian Research Agency;
Ikerbasque, Basque Foundation for Science, Spain;
the Swiss National Science Foundation; 
the Ministry of Education and the Ministry of Science and Technology of Taiwan;
the European Union's Horizon 2020 research and 
innovation programme under grant agreement No 824093;
and the United States Department of Energy and the National Science Foundation.

\end{acknowledgments}
\clearpage
\bibliographystyle{apsrev4-1}

\bibliography{CitationsNewNMCollins}
\clearpage

\begin{appendices}
\renewcommand*{\thesection}{Appendix~\Alph{section}}
\section{Charm fractions}\label{sec:App1}

The fraction of events originating from charm production is given for the various meson combinations and kinematic binning listed in Tables~\ref{tab:sinzcharmratio}-\ref{tab:zptcharmratio}. Here, the charm fraction is defined as the ratio of meson pairs that come out from \(c\bar{c}\) production over those coming out of \(q\bar{q}\) (\(q=u,d,s,c\)) production as determined from Pythia and EvtGen Monte Carlo simulations employing the Belle default tune.  The charm fractions generally are largest at low values of \(z\), reaching fractions as large as 40\%,  and  decrease rapidly with increasing \(z\) to a negligible level in the very last \(z\) bins. A much milder dependence on \(P_{t}\) is observed for all hadron pairs. The fractions are in average larger for pairs involving \(\eta\) mesons compared to those involving only pions.

\begingroup
 \squeezetable
 \begin{table*}[htp]
  \begin{ruledtabular}  
    \begin{tabular}{ccccc} 
  $z_1$ & $\pi^{\pm}\pi^{\pm}$ & $\pi^0\pi^{\pm}$ & $\eta\pi^{\pm}$ & $\pi^0\pi^{\pm}$ $(z>0.3)$ \\
 &   [\%] & [\%] & [\%] &  [\%] \\ \hline\hline
[0.2,0.3]	&	22	&	24	&		&		\\ \hline
[0.3,0.4]	&	18	&	19	&	20	&	16	\\ \hline
[0.4,0.5]	&	16	&	16	&	17	&	14	\\ \hline
[0.5,0.6]	&	15	&	14	&	16	&	11	\\ \hline
[0.6,0.7]	&	10	&	9	&	13	&	7	\\ \hline
[0.7,1.0]	&	5	&	4	&	7	&	3	\\
\end{tabular}
\caption[Charm fraction in $z_1$ bins]{Charm fraction in $z_1$ bins. All numbers are in percent. The minimum $z_i$ for pions is raised to $z_{1,2} > 0.3$ in the last two columns to align with the  $z_i$ constraint for pairs involving $\eta$ mesons.}
\label{tab:sinzcharmratio}
\label{tab:finalulucptbins}
  \end{ruledtabular}
 \end{table*}
\endgroup

\begingroup
 \squeezetable
 \begin{table*}[htp]
  \begin{ruledtabular}  
    \begin{tabular}{ccccc} 
$P_{t1}$ [GeV]  &$\pi^{\pm}\pi^{\pm}$ & $\pi^0\pi^{\pm}$ & $\eta\pi^{\pm}$ & $\pi^0\pi^{\pm}$ $(z>0.3)$  \\
 &   [\%] & [\%] & [\%] &  [\%]\\ \hline\hline
[0,0.15]	&	20	&	21	&	16	&	13	\\ \hline
[0.15,0.30]	&	20	&	21	&	16	&	14	\\ \hline
[0.30,0.50]	&	19	&	19	&	18	&	15	\\ \hline
[0.50,3.0]	&	19	&	18	&	21	&	15	\\ \hline 
\end{tabular}
\caption[Charm fraction in $P_{t1}$ bins]{Charm fraction in $P_{t1}$ bins. All numbers are in percent.}
\label{tab:sinptcharmratio}
  \end{ruledtabular}
 \end{table*}
\endgroup

\begingroup
 \squeezetable
 \begin{table*}[htp]
  \begin{ruledtabular}  
    \begin{tabular}{cccccc} 
    $z_1$& $z_2$ & $\pi^{\pm}\pi^{\pm}$ & $\pi^{\pm}\pi^0$ & $\eta\pi^{\pm}$ & $\pi^0\pi^{\pm}$ $(z>0.3)$ \\
 &  & [\%] & [\%] & [\%] &  [\%] \\ \hline\hline
[0.1,0.2]	&	[0.1,0.2]	&	37	&	42	&	-	&	-	\\ \hline
[0.1,0.2]	&	[0.2,0.3]	&	31	&	35	&	-	&	-	\\ \hline
[0.1,0.2]	&	[0.3,0.5]	&	25	&	29	&	-	&	-	\\ \hline
[0.1,0.2]	&	[0.5,0.7]	&	19	&	22	&	-	&	-	\\ \hline
[0.1,0.2]	&	[0.7,1.0]	&	6	&	8	&	-	&	-	\\ \hline \hline
[0.2,0.3]	&	[0.1,0.2]	&	31	&	33	&	-	&	-	\\ \hline
[0.2,0.3]	&	[0.2,0.3]	&	26	&	27	&	-	&	-	\\ \hline
[0.2,0.3]	&	[0.3,0.5]	&	21	&	22	&	-	&	-	\\ \hline
[0.2,0.3]	&	[0.5,0.7]	&	16	&	17	&	-	&	-	\\ \hline
[0.2,0.3]	&	[0.7,1.0]	&	5	&	6	&	-	&	-	\\ \hline\hline
[0.3,0.5]	&	[0.1,0.2]	&	25	&	25	&	-	&	-	\\ \hline
[0.3,0.5]	&	[0.2,0.3]	&	21	&	21	&	-	&	-	\\ \hline
[0.3,0.5]	&	[0.3,0.5]	&	16	&	16	&	20	&	16	\\ \hline
[0.3,0.5]	&	[0.5,0.7]	&	12	&	12	&	15	&	12	\\ \hline
[0.3,0.5]	&	[0.7,1.0]	&	4	&	4	&	5	&	4	\\ \hline\hline
[0.5,0.7]	&	[0.1,0.2]	&	19	&	18	&	-	&	-	\\ \hline
[0.5,0.7]	&	[0.2,0.3]	&	16	&	15	&	-	&	-	\\ \hline
[0.5,0.7]	&	[0.3,0.5]	&	12	&	11	&	16	&	11	\\ \hline
[0.5,0.7]	&	[0.5,0.7]	&	8	&	8	&	11	&	8	\\ \hline
[0.5,0.7]	&	[0.7,1.0]	&	3	&	3	&	3	&	3	\\ \hline\hline
[0.7,1.0]	&	[0.1,0.2]	&	7	&	5	&	-	&	-	\\ \hline
[0.7,1.0]	&	[0.2,0.3]	&	6	&	5	&	-	&	-	\\ \hline
[0.7,1.0]	&	[0.3,0.5]	&	4	&	3	&	8	&	3	\\ \hline
[0.7,1.0]	&	[0.5,0.7]	&	3	&	2	&	5	&	2	\\ \hline
[0.7,1.0]	&	[0.7,1.0]	&	1	&	1	&	2	&	1	\\
\end{tabular}
\caption[Charm fraction in combined $(z_{1}$,$z_{2})$ bins]{Charm fraction in combined $z_{1}$--$z_{2}$ bins. All numbers are in percent. Empty bins do not fulfill $z_i > 0.3$ required for those columns.}
\label{tab:comzcharmratio}
  \end{ruledtabular}
 \end{table*}
\endgroup

\begingroup
 \squeezetable
 \begin{table*}[htp]
  \begin{ruledtabular}  
    \begin{tabular}{cccccc} 
$P_{t1}$ [GeV] & $P_{t2}$ [GeV] &$\pi^{\pm}\pi^{\pm}$ & $\pi^0\pi^{\pm}$ & $\eta\pi^{\pm}$ & $\pi^0\pi^{\pm}$ $(z>0.3)$  \\
 &  & [\%] & [\%] & [\%] &  [\%]\\ \hline\hline
[0,0.15]	&	[0,0.15]	&	20	&	22	&	14	&	12	\\ \hline
[0,0.15]	&	[0.15,0.30]	&	20	&	22	&	15	&	12	\\ \hline
[0,0.15]	&	[0.30,0.50]	&	19	&	21	&	16	&	14	\\ \hline
[0,0.15]	&	[0.50,3.0]	&	19	&	21	&	18	&	15	\\ \hline\hline
[0.15,0.30]	&	[0,0.15]	&	20	&	22	&	14	&	12	\\ \hline
[0.15,0.30]	&	[0.15,0.30]	&	20	&	21	&	15	&	12	\\ \hline
[0.15,0.30]	&	[0.30,0.50]	&	19	&	21	&	17	&	14	\\ \hline
[0.15,0.30]	&	[0.50,3.0]	&	19	&	21	&	18	&	16	\\ \hline\hline
[0.30,0.50]	&	[0,0.15]	&	19	&	20	&	16	&	13	\\ \hline
[0.30,0.50]	&	[0.15,0.30]	&	19	&	20	&	17	&	13	\\ \hline
[0.30,0.50]	&	[0.30,0.50]	&	18	&	19	&	19	&	15	\\ \hline
[0.30,0.50]	&	[0.50,3.0]	&	18	&	19	&	21	&	17	\\ \hline\hline
[0.50,3.0]	&	[0,0.15]	&	20	&	19	&	19	&	14	\\ \hline
[0.50,3.0]	&	[0.15,0.30]	&	19	&	19	&	20	&	14	\\ \hline
[0.50,3.0]	&	[0.30,0.50]	&	18	&	18	&	21	&	16	\\ \hline
[0.50,3.0]	&	[0.50,3.0]	&	17	&	17	&	24	&	17	\\
\end{tabular}
\caption[Charm fraction in combined $(P_{t1},P_{t2})$ bins]{Charm fraction in $(P_{t1},P_{t2})$ bins. All numbers are in percent.}
\label{tab:comptcharmratio}
  \end{ruledtabular}
 \end{table*}
\endgroup

\begingroup
 \squeezetable
 \begin{table*}[htp]
  \begin{ruledtabular}  
    \begin{tabular}{ccccc} 
$z_1$ & $P_{t1}$ [GeV] &$\pi^{\pm}\pi^{\pm}$ & $\pi^0\pi^{\pm}$ & $\eta \pi^\pm$ \\
 &  [\%] & [\%] & [\%] &  [\%]\\ \hline\hline
[0.2,0.3]	&	[0,0.15]	&	23	&	25    &	-\\ \hline
[0.2,0.3]	&	[0.15,0.30]	&	22	&	24    &	-\\ \hline
[0.2,0.3]	&	[0.30,0.50]	&	22	&	23    &	-\\ \hline
[0.2,0.3]	&	[0.50,3.0]	&	-	&   -	& - \\ \hline\hline	
[0.3,0.5]	&	[0,0.15]	&	16	&	17    &   20	\\ \hline
[0.3,0.5]	&	[0.15,0.30]	&	16	&	17    &	23   \\ \hline
[0.3,0.5]	&	[0.30,0.50]	&	18	&	18    &	27   \\ \hline
[0.3,0.5]	&	[0.50,3.0]	&	21	&	20    &	28   \\ \hline\hline
[0.5,0.7]	&	[0,0.15]	&	11	&	16    &	16   \\ \hline
[0.5,0.7]	&	[0.15,0.30]	&	11	&	16    &   21	\\ \hline
[0.5,0.7]	&	[0.30,0.50]	&	13	&	17    &	24   \\ \hline
[0.5,0.7]	&	[0.50,3.0]	&	16	&	19    &	26   \\ \hline\hline
[0.7,1.0]	&	[0,0.15]	&	3	&	10    &	10   \\ \hline
[0.7,1.0]	&	[0.15,0.30]	&	4	&	10    &	11   \\ \hline
[0.7,1.0]	&	[0.30,0.50]	&	5	&	11    &   15	\\ \hline
[0.7,1.0]	&	[0.50,3.0]	&	6	&	14    &	19   \\
\end{tabular}
\caption[Charm fraction in $(z_{1},P_{t1})$ bins]{Charm fraction in $(z_{1},P_{t1})$ bins. All numbers are in percent.}
\label{tab:zptcharmratio}
  \end{ruledtabular}
 \end{table*}
\endgroup

\clearpage

\section{Tables of results}

In this section, all asymmetry results are tabulated together with the averages in the kinematic variables \(z_1\), \(z_2\), \(P_{t1}\), and \(P_{t2}\), as well as of the quantity \(\sin^{2}\theta/(1+\cos^2\theta)\), which corresponds to a measure of the size of transverse polarization of the quark--anti-quark pair produced. The tabulated values are obtained from the hadron pairs with the same kinematics that are used to bin the data. Then the average of hadron pairs that appear in the double ratio is taken.

\begingroup
 \squeezetable
 \begin{table*}[htp]
  \begin{ruledtabular}  
   \tiny
    \begin{tabular}{ccccccccccc}
$P_{t1}$   [GeV] & $\langle  P_{t1}  \rangle$ [GeV] & $z_1$ &  $\langle  z_1 \rangle$& $P_{t2}$  [GeV] & $\langle  P_{t2}\rangle$  [GeV] & $z_2$ & $\langle  z_2 \rangle$  &$\langle\frac{\sin^2(\theta)}{1+\cos^2(\theta)}\rangle$& $A_{12}^{UL}$ [\%] &  $A_{12}^{UC}$ [\%]   \\ \hline \hline
[0,0.15]	&	0.10	&	[0.2, 1.0]	&	0.32	&	[0,3.0]	&	0.32	&	[0.2, 1.0]	&	0.36	&	0.91	& 1.34  $\pm$ 0.19  $\pm$ 0.07 & 0.67  $\pm$ 0.15  $\pm$ 0.05 \\ \hline						
[0.15,0.3]	&	0.23	&	[0.2, 1.0]	&	0.31	&	[0,3.0]	&	0.32	&	[0.2, 1.0]	&	0.36	&	0.91	& 2.25  $\pm$ 0.10  $\pm$ 0.03  & 1.12  $\pm$ 0.08  $\pm$ 0.03 \\ \hline						
[0.3,0.5]	&	0.38	&	[0.2, 1.0]	&	0.36	&	[0,3.0]	&	0.32	&	[0.2, 1.0]	&	0.36	&	0.91	& 3.18  $\pm$ 0.10  $\pm$ 0.03  & 1.59  $\pm$ 0.08  $\pm$ 0.03 \\ \hline						
[0.5,3.0]	&	0.63	&	[0.2, 1.0]	&	0.51	&	[0,3.0]	&	0.32	&	[0.2, 1.0]	&	0.36	&	0.91	& 5.53  $\pm$ 0.18  $\pm$ 0.07 & 2.76  $\pm$ 0.14  $\pm$ 0.06 \\ \hline						
\hline																							
[0,0.15]	&	0.10	&	[0.2, 1.0]	&	0.32	&	[0,0.15]	&	0.10	&	[0.2, 1.0]	&	0.32	&	0.91	&-0.05  $\pm$ 0.40  $\pm$ 0.13  & -0.02  $\pm$ 0.31  $\pm$ 0.1 \\ \hline						
[0,0.15]	&	0.10	&	[0.2, 1.0]	&	0.32	&	[0.15,0.3]	&	0.23	&	[0.2, 1.0]	&	0.31	&	0.91	&1.06  $\pm$ 0.31  $\pm$ 0.11  & 0.53  $\pm$ 0.25  $\pm$ 0.09 \\ \hline						
[0,0.15]	&	0.10	&	[0.2, 1.0]	&	0.32	&	[0.3,0.5]	&	0.38	&	[0.2, 1.0]	&	0.36	&	0.91	&1.74  $\pm$ 0.30  $\pm$ 0.11   & 0.87  $\pm$ 0.24  $\pm$ 0.09 \\ \hline						
[0,0.15]	&	0.10	&	[0.2, 1.0]	&	0.32	&	[0.5,3.0]	&	0.62	&	[0.2, 1.0]	&	0.51	&	0.91	&2.38  $\pm$ 0.51  $\pm$ 0.20   & 1.19  $\pm$ 0.40  $\pm$ 0.16  \\ \hline						
\hline																							
[0.15,0.3]	&	0.23	&	[0.2, 1.0]	&	0.31	&	[0,0.15]	&	0.10	&	[0.2, 1.0]	&	0.32	&	0.91	&1.50  $\pm$ 0.34  $\pm$ 0.13   & 0.75  $\pm$ 0.27  $\pm$ 0.1  \\ \hline						
[0.15,0.3]	&	0.23	&	[0.2, 1.0]	&	0.32	&	[0.15,0.3]	&	0.23	&	[0.2, 1.0]	&	0.31	&	0.91	&1.74  $\pm$ 0.16  $\pm$ 0.05  & 0.87  $\pm$ 0.12  $\pm$ 0.04 \\ \hline						
[0.15,0.3]	&	0.23	&	[0.2, 1.0]	&	0.31	&	[0.3,0.5]	&	0.38	&	[0.2, 1.0]	&	0.36	&	0.91	&2.12  $\pm$ 0.15  $\pm$ 0.05  & 1.06  $\pm$ 0.12  $\pm$ 0.04 \\ \hline						
[0.15,0.3]	&	0.23	&	[0.2, 1.0]	&	0.31	&	[0.5,3.0]	&	0.62	&	[0.2, 1.0]	&	0.51	&	0.91	&4.63  $\pm$ 0.26  $\pm$ 0.11  & 2.32  $\pm$ 0.21  $\pm$ 0.09 \\ \hline						
\hline																							
[0.3,0.5]	&	0.38	&	[0.2, 1.0]	&	0.36	&	[0,0.15]	&	0.10	&	[0.2, 1.0]	&	0.32	&	0.91	&1.39  $\pm$ 0.29  $\pm$ 0.1   & 0.69  $\pm$ 0.23  $\pm$ 0.08 \\ \hline						
[0.3,0.5]	&	0.38	&	[0.2, 1.0]	&	0.36	&	[0.15,0.3]	&	0.23	&	[0.2, 1.0]	&	0.31	&	0.91	&2.70  $\pm$ 0.15  $\pm$ 0.05   & 1.35  $\pm$ 0.12  $\pm$ 0.04 \\ \hline						
[0.3,0.5]	&	0.38	&	[0.2, 1.0]	&	0.36	&	[0.3,0.5]	&	0.38	&	[0.2, 1.0]	&	0.36	&	0.91	&3.31  $\pm$ 0.15  $\pm$ 0.05  & 1.66  $\pm$ 0.12  $\pm$ 0.04 \\ \hline						
[0.3,0.5]	&	0.38	&	[0.2, 1.0]	&	0.36	&	[0.5,3.0]	&	0.62	&	[0.2, 1.0]	&	0.51	&	0.91	&5.36  $\pm$ 0.26  $\pm$ 0.11  & 2.67  $\pm$ 0.21  $\pm$ 0.08 \\ \hline						
\hline																							
[0.5,3.0]	&	0.63	&	[0.2, 1.0]	&	0.51	&	[0,0.15]	&	0.10	&	[0.2, 1.0]	&	0.32	&	0.91	&2.35  $\pm$ 0.53  $\pm$ 0.21  & 1.18  $\pm$ 0.41  $\pm$ 0.17 \\ \hline						
[0.5,3.0]	&	0.63	&	[0.2, 1.0]	&	0.51	&	[0.15,0.3]	&	0.23	&	[0.2, 1.0]	&	0.31	&	0.91	&4.07  $\pm$ 0.27  $\pm$ 0.11  & 2.04  $\pm$ 0.21  $\pm$ 0.08 \\ \hline						
[0.5,3.0]	&	0.63	&	[0.2, 1.0]	&	0.51	&	[0.3,0.5]	&	0.38	&	[0.2, 1.0]	&	0.36	&	0.91	&5.60  $\pm$ 0.26  $\pm$ 0.11   & 2.80  $\pm$ 0.20  $\pm$ 0.08   \\ \hline						
[0.5,3.0]	&	0.63	&	[0.2, 1.0]	&	0.52	&	[0.5,3.0]	&	0.62	&	[0.2, 1.0]	&	0.51	&	0.91	&9.89  $\pm$ 0.52  $\pm$ 0.24  & 4.95  $\pm$ 0.4  $\pm$ 0.19  \\			
\end{tabular}
\caption[Charged-pion Collins asymmetries $A_{12}^{UC}$ and $A_{12}^{UL}$ binned in $P_t$]{Charged-pion Collins asymmetries $A_{12}^{UC}$ and $A_{12}^{UL}$ binned in $P_t$. Uncertainties are statistical and systematic, respectively. The table contains data binned in $P_{t1}$ while integrated over the other variables, as well as data simultaneously binned in $P_{t1}$ and $P_{t2}$. }
  \end{ruledtabular}
 \end{table*}
\endgroup

\begingroup
 \squeezetable
 \begin{table*}[htp]
  \begin{ruledtabular}
  \tiny
   \begin{tabular}{ccccccccccc}
$z_1$ & $\langle  z_1\rangle$ & $P_{t1}$ & $\langle  P_{t1}\rangle$  [GeV] & $z_2$ & $\langle  z_2 \rangle$ & $P_{t2}$ & $\langle  P_{t2}\rangle$ [GeV]  & $\langle\frac{\sin^2(\theta)}{1+\cos^2(\theta)}\rangle$ & $A_{12}^{UL}$ [\%] &  $A_{12}^{UC}$ [\%]    \\ \hline\hline
[0.2,0.3]	&	0.25	&	[0, 3.0]	&	0.24	&	[0.2,1.0]	&	0.36	&	[0, 3.0]	&	0.32	&	0.90	&2.47  $\pm$ 0.09  $\pm$ 0.03 &1.24  $\pm$ 0.08  $\pm$ 0.03 \\ \hline				
[0.3,0.4]	&	0.35	&	[0, 3.0]	&	0.32	&	[0.2,1.0]	&	0.35	&	[0, 3.0]	&	0.32	&	0.91	&2.75  $\pm$ 0.12  $\pm$ 0.04 &1.38  $\pm$ 0.09  $\pm$ 0.03 \\ \hline				
[0.4,0.5]	&	0.45	&	[0, 3.0]	&	0.39	&	[0.2,1.0]	&	0.35	&	[0, 3.0]	&	0.32	&	0.91	&2.85  $\pm$ 0.16  $\pm$ 0.06 &1.43  $\pm$ 0.12  $\pm$ 0.04 \\ \hline				
[0.5,0.6]	&	0.55	&	[0, 3.0]	&	0.45	&	[0.2,1.0]	&	0.36	&	[0, 3.0]	&	0.32	&	0.91	&3.86  $\pm$ 0.22  $\pm$ 0.08 &1.93  $\pm$ 0.17  $\pm$ 0.06 \\ \hline				
[0.6,0.7]	&	0.64	&	[0, 3.0]	&	0.48	&	[0.2,1.0]	&	0.36	&	[0, 3.0]	&	0.32	&	0.91	&4.64  $\pm$ 0.29  $\pm$ 0.13 &2.32  $\pm$ 0.22  $\pm$ 0.10  \\ \hline				
[0.7,1.0]	&	0.78	&	[0, 3.0]	&	0.44	&	[0.2,1.0]	&	0.36	&	[0, 3.0]	&	0.32	&	0.90	&6.81  $\pm$ 0.36  $\pm$ 0.20  &3.41  $\pm$ 0.27  $\pm$ 0.15 \\ \hline				
\hline																					
[0.1,0.2]	&	0.15	&	[0, 3.0]	&	0.15	&	[0.1,0.2]	&	0.15	&	[0, 3.0]	&	0.15	&	0.91	&0.87  $\pm$ 0.11  $\pm$ 0.04   &	0.44  $\pm$ 0.09  $\pm$ 0.04  \\ \hline			
[0.1,0.2]	&	0.15	&	[0, 3.0]	&	0.15	&	[0.2,0.3]	&	0.25	&	[0, 3.0]	&	0.24	&	0.91	&1.16  $\pm$ 0.13  $\pm$ 0.05   &	0.58  $\pm$ 0.10  $\pm$ 0.04   \\ \hline			
[0.1,0.2]	&	0.15	&	[0, 3.0]	&	0.15	&	[0.3,0.5]	&	0.38	&	[0, 3.0]	&	0.35	&	0.90	&1.44  $\pm$ 0.12  $\pm$ 0.05   &	0.72  $\pm$ 0.10  $\pm$ 0.04   \\ \hline			
[0.1,0.2]	&	0.15	&	[0, 3.0]	&	0.15	&	[0.5,0.7]	&	0.58	&	[0, 3.0]	&	0.46	&	0.90	&2.49  $\pm$ 0.23  $\pm$ 0.10    &	1.25  $\pm$ 0.18  $\pm$ 0.08  \\ \hline			
[0.1,0.2]	&	0.15	&	[0, 3.0]	&	0.15	&	[0.7,1.0]	&	0.78	&	[0, 3.0]	&	0.42	&	0.90	&3.76  $\pm$ 0.52  $\pm$ 0.30    &	1.88  $\pm$ 0.41  $\pm$ 0.24  \\ \hline			
\hline																					
[0.2,0.3]	&	0.25	&	[0, 3.0]	&	0.24	&	[0.1,0.2]	&	0.15	&	[0, 3.0]	&	0.15	&	0.91	&1.20  $\pm$ 0.12  $\pm$ 0.04    &	0.60  $\pm$ 0.10  $\pm$ 0.04    \\ \hline			
[0.2,0.3]	&	0.25	&	[0, 3.0]	&	0.24	&	[0.2,0.3]	&	0.25	&	[0, 3.0]	&	0.24	&	0.91	&1.98  $\pm$ 0.15  $\pm$ 0.05   &	0.99  $\pm$ 0.12  $\pm$ 0.04  \\ \hline			
[0.2,0.3]	&	0.25	&	[0, 3.0]	&	0.24	&	[0.3,0.5]	&	0.38	&	[0, 3.0]	&	0.35	&	0.90	&2.35  $\pm$ 0.14  $\pm$ 0.05   &	1.18  $\pm$ 0.11  $\pm$ 0.04  \\ \hline			
[0.2,0.3]	&	0.25	&	[0, 3.0]	&	0.24	&	[0.5,0.7]	&	0.58	&	[0, 3.0]	&	0.46	&	0.90	&3.97  $\pm$ 0.27  $\pm$ 0.11   &	1.99  $\pm$ 0.21  $\pm$ 0.08  \\ \hline			
[0.2,0.3]	&	0.25	&	[0, 3.0]	&	0.24	&	[0.7,1.0]	&	0.78	&	[0, 3.0]	&	0.43	&	0.90	&4.62  $\pm$ 0.55  $\pm$ 0.28   &	2.31  $\pm$ 0.42  $\pm$ 0.22  \\ \hline			
\hline																					
[0.3,0.5]	&	0.38	&	[0, 3.0]	&	0.35	&	[0.1,0.2]	&	0.15	&	[0, 3.0]	&	0.15	&	0.91	&1.17  $\pm$ 0.12  $\pm$ 0.04   &	0.58  $\pm$ 0.09  $\pm$ 0.04  \\ \hline			
[0.3,0.5]	&	0.38	&	[0, 3.0]	&	0.35	&	[0.2,0.3]	&	0.25	&	[0, 3.0]	&	0.24	&	0.91	&2.03  $\pm$ 0.14  $\pm$ 0.05   &	1.01  $\pm$ 0.11  $\pm$ 0.04  \\ \hline			
[0.3,0.5]	&	0.38	&	[0, 3.0]	&	0.35	&	[0.3,0.5]	&	0.38	&	[0, 3.0]	&	0.35	&	0.91	&2.97  $\pm$ 0.15  $\pm$ 0.05   &	1.48  $\pm$ 0.12  $\pm$ 0.04  \\ \hline			
[0.3,0.5]	&	0.38	&	[0, 3.0]	&	0.35	&	[0.5,0.7]	&	0.58	&	[0, 3.0]	&	0.46	&	0.91	&4.41  $\pm$ 0.28  $\pm$ 0.11   &	2.21  $\pm$ 0.22  $\pm$ 0.08  \\ \hline			
[0.3,0.5]	&	0.38	&	[0, 3.0]	&	0.35	&	[0.7,1.0]	&	0.78	&	[0, 3.0]	&	0.44	&	0.91	&5.95  $\pm$ 0.56  $\pm$ 0.29   &	2.97  $\pm$ 0.43  $\pm$ 0.22  \\ \hline			
\hline																					
[0.5,0.7]	&	0.58	&	[0, 3.0]	&	0.46	&	[0.1,0.2]	&	0.15	&	[0, 3.0]	&	0.15	&	0.91	&2.01  $\pm$ 0.22  $\pm$ 0.09   &	1.01  $\pm$ 0.17  $\pm$ 0.07  \\ \hline			
[0.5,0.7]	&	0.58	&	[0, 3.0]	&	0.46	&	[0.2,0.3]	&	0.25	&	[0, 3.0]	&	0.24	&	0.91	&3.09  $\pm$ 0.26  $\pm$ 0.10    &	1.54  $\pm$ 0.20  $\pm$ 0.08   \\ \hline			
[0.5,0.7]	&	0.58	&	[0, 3.0]	&	0.46	&	[0.3,0.5]	&	0.38	&	[0, 3.0]	&	0.35	&	0.91	&4.33  $\pm$ 0.28  $\pm$ 0.10    &	2.17  $\pm$ 0.22  $\pm$ 0.08  \\ \hline			
[0.5,0.7]	&	0.58	&	[0, 3.0]	&	0.47	&	[0.5,0.7]	&	0.58	&	[0, 3.0]	&	0.47	&	0.91	&6.09  $\pm$ 0.50  $\pm$ 0.22    &	3.06  $\pm$ 0.38  $\pm$ 0.17  \\ \hline			
[0.5,0.7]	&	0.58	&	[0, 3.0]	&	0.47	&	[0.7,1.0]	&	0.78	&	[0, 3.0]	&	0.44	&	0.91	&11.10  $\pm$ 1.05  $\pm$ 0.65   &	5.59  $\pm$ 0.77  $\pm$ 0.49  \\ \hline			
\hline																					
[0.7,1.0]	&	0.78	&	[0, 3.0]	&	0.42	&	[0.1,0.2]	&	0.15	&	[0, 3.0]	&	0.15	&	0.91	&2.57  $\pm$ 0.48  $\pm$ 0.26   &	1.28  $\pm$ 0.36  $\pm$ 0.21  \\ \hline			
[0.7,1.0]	&	0.78	&	[0, 3.0]	&	0.43	&	[0.2,0.3]	&	0.25	&	[0, 3.0]	&	0.24	&	0.91	&5.07  $\pm$ 0.53  $\pm$ 0.29   &	2.55  $\pm$ 0.40  $\pm$ 0.22   \\ \hline			
[0.7,1.0]	&	0.78	&	[0, 3.0]	&	0.44	&	[0.3,0.5]	&	0.38	&	[0, 3.0]	&	0.35	&	0.90	&6.72  $\pm$ 0.57  $\pm$ 0.29   &	3.37  $\pm$ 0.44  $\pm$ 0.23  \\ \hline			
[0.7,1.0]	&	0.78	&	[0, 3.0]	&	0.44	&	[0.5,0.7]	&	0.58	&	[0, 3.0]	&	0.47	&	0.90	&8.76  $\pm$ 0.96  $\pm$ 0.58   &	4.39  $\pm$ 0.71  $\pm$ 0.43  \\ \hline			
[0.7,1.0]	&	0.78	&	[0, 3.0]	&	0.46	&	[0.7,1.0]	&	0.78	&	[0, 3.0]	&	0.46	&	0.90	&25.38  $\pm$ 2.43  $\pm$ 2.40   &	12.81  $\pm$ 1.7  $\pm$ 1.78  \\ 		
   \end{tabular}
   \caption[Charged-pion Collins asymmetries $A_{12}^{UL}$ and $A_{12}^{UC}$ binned in $z$]{Charged-pion Collins asymmetries $A_{12}^{UL}$ and $A_{12}^{UC}$ binned in $z$. Uncertainties are statistical and systematic, respectively. The table contains data binned in $P_{t1}$ while integrated over the other variables, as well as data simultaneously binned in $P_{t1}$ and $P_{t2}$.}
   \label{tab:finaluluczbin}
  \end{ruledtabular}
 \end{table*}
\endgroup

\begingroup
 \squeezetable
 \begin{table*}[htp]
  \begin{ruledtabular}  
    \begin{tabular}{cccccccccc}
$z_1$ & $\langle  z_1\rangle$ & $P_{t1}$ & $\langle  P_{t1}\rangle$  [GeV] & $z_2$ & $\langle  z_2 \rangle$ & $P_{t2}$ & $\langle  P_{t2}\rangle$ [GeV]  & $\langle\frac{\sin^2(\theta)}{1+\cos^2(\theta)}\rangle$& $A_{12}^{\pi^0}$ [\%]  \\ \hline\hline
[0.2,0.3]	&	0.25	&	[0, 3.0]	&	0.24	&	[0.2,1.0]	&	0.36	&	[0, 3.0]	&	0.32	&	0.91	&	1.25  $\pm$ 0.12  $\pm$ 0.06    \\ \hline		
[0.3,0.4]	&	0.35	&	[0, 3.0]	&	0.32	&	[0.2,1.0]	&	0.36	&	[0, 3.0]	&	0.32	&	0.91	&	1.72  $\pm$ 0.13  $\pm$ 0.04    \\ \hline		
[0.4,0.5]	&	0.45	&	[0, 3.0]	&	0.39	&	[0.2,1.0]	&	0.36	&	[0, 3.0]	&	0.32	&	0.91	&	1.72  $\pm$ 0.16  $\pm$ 0.06    \\ \hline		
[0.5,0.6]	&	0.54	&	[0, 3.0]	&	0.45	&	[0.2,1.0]	&	0.36	&	[0, 3.0]	&	0.32	&	0.91	&	2.08  $\pm$ 0.22  $\pm$ 0.11    \\ \hline		
[0.6,0.7]	&	0.64	&	[0, 3.0]	&	0.49	&	[0.2,1.0]	&	0.36	&	[0, 3.0]	&	0.32	&	0.90	&	2.51  $\pm$ 0.28  $\pm$ 0.13    \\ \hline		
[0.7,1.0]	&	0.77	&	[0, 3.0]	&	0.44	&	[0.2,1.0]	&	0.36	&	[0, 3.0]	&	0.32	&	0.90	&	3.86  $\pm$ 0.36  $\pm$ 0.20  	\\ \hline	
\hline																				
[0.1,0.2]	&	0.15	&	[0, 3.0]	&	0.15	&	[0.1,0.2]	&	0.15	&	[0, 3.0]	&	0.15	&	0.91	&0.16  $\pm$ 0.19  $\pm$ 0.04 	\\ \hline		
[0.1,0.2]	&	0.15	&	[0, 3.0]	&	0.15	&	[0.2,0.3]	&	0.25	&	[0, 3.0]	&	0.24	&	0.91	&0.18  $\pm$ 0.22  $\pm$ 0.06 	\\ \hline		
[0.1,0.2]	&	0.15	&	[0, 3.0]	&	0.15	&	[0.3,0.5]	&	0.38	&	[0, 3.0]	&	0.35	&	0.91	&0.47  $\pm$ 0.22  $\pm$ 0.07 	\\ \hline		
[0.1,0.2]	&	0.15	&	[0, 3.0]	&	0.15	&	[0.5,0.7]	&	0.58	&	[0, 3.0]	&	0.46	&	0.91	&1.02  $\pm$ 0.41  $\pm$ 0.11 	\\ \hline		
[0.1,0.2]	&	0.15	&	[0, 3.0]	&	0.15	&	[0.7,1.0]	&	0.78	&	[0, 3.0]	&	0.43	&	0.90	&2.52  $\pm$ 0.92  $\pm$ 0.25 	\\ \hline		
\hline																				
[0.2,0.3]	&	0.25	&	[0, 3.0]	&	0.24	&	[0.1,0.2]	&	0.15	&	[0, 3.0]	&	0.15	&	0.91	&0.74  $\pm$ 0.15  $\pm$ 0.04 	\\ \hline		
[0.2,0.3]	&	0.25	&	[0, 3.0]	&	0.24	&	[0.2,0.3]	&	0.25	&	[0, 3.0]	&	0.24	&	0.91	&1.07  $\pm$ 0.19  $\pm$ 0.05 	\\ \hline		
[0.2,0.3]	&	0.25	&	[0, 3.0]	&	0.24	&	[0.3,0.5]	&	0.38	&	[0, 3.0]	&	0.35	&	0.91	&1.07  $\pm$ 0.19  $\pm$ 0.11 	\\ \hline		
[0.2,0.3]	&	0.25	&	[0, 3.0]	&	0.24	&	[0.5,0.7]	&	0.58	&	[0, 3.0]	&	0.46	&	0.91	&1.85  $\pm$ 0.37  $\pm$ 0.12 	\\ \hline		
[0.2,0.3]	&	0.25	&	[0, 3.0]	&	0.24	&	[0.7,1.0]	&	0.78	&	[0, 3.0]	&	0.43	&	0.90	&4.33  $\pm$ 0.77  $\pm$ 0.32 	\\ \hline		
\hline																				
[0.3,0.5]	&	0.38	&	[0, 3.0]	&	0.35	&	[0.1,0.2]	&	0.15	&	[0, 3.0]	&	0.15	&	0.91	&0.71  $\pm$ 0.13  $\pm$ 0.04 	\\ \hline		
[0.3,0.5]	&	0.38	&	[0, 3.0]	&	0.35	&	[0.2,0.3]	&	0.25	&	[0, 3.0]	&	0.24	&	0.91	&1.40  $\pm$ 0.15  $\pm$ 0.05  	\\ \hline		
[0.3,0.5]	&	0.38	&	[0, 3.0]	&	0.35	&	[0.3,0.5]	&	0.38	&	[0, 3.0]	&	0.35	&	0.91	&1.63  $\pm$ 0.16  $\pm$ 0.06 	\\ \hline		
[0.3,0.5]	&	0.38	&	[0, 3.0]	&	0.35	&	[0.5,0.7]	&	0.58	&	[0, 3.0]	&	0.46	&	0.91	&2.83  $\pm$ 0.30  $\pm$ 0.10   	\\ \hline		
[0.3,0.5]	&	0.38	&	[0, 3.0]	&	0.35	&	[0.7,1.0]	&	0.78	&	[0, 3.0]	&	0.44	&	0.91	&4.07  $\pm$ 0.65  $\pm$ 0.28 	\\ \hline		
\hline																				
[0.5,0.7]	&	0.58	&	[0, 3.0]	&	0.46	&	[0.1,0.2]	&	0.15	&	[0, 3.0]	&	0.15	&	0.91	&1.46  $\pm$ 0.22  $\pm$ 0.09 	\\ \hline		
[0.5,0.7]	&	0.58	&	[0, 3.0]	&	0.46	&	[0.2,0.3]	&	0.25	&	[0, 3.0]	&	0.24	&	0.91	&1.54  $\pm$ 0.26  $\pm$ 0.16 	\\ \hline		
[0.5,0.7]	&	0.58	&	[0, 3.0]	&	0.46	&	[0.3,0.5]	&	0.38	&	[0, 3.0]	&	0.35	&	0.91	&2.49  $\pm$ 0.26  $\pm$ 0.10  	\\ \hline		
[0.5,0.7]	&	0.58	&	[0, 3.0]	&	0.46	&	[0.5,0.7]	&	0.58	&	[0, 3.0]	&	0.46	&	0.91	&3.39  $\pm$ 0.49  $\pm$ 0.24 	\\ \hline		
[0.5,0.7]	&	0.58	&	[0, 3.0]	&	0.47	&	[0.7,1.0]	&	0.78	&	[0, 3.0]	&	0.44	&	0.90	&4.94  $\pm$ 1.06  $\pm$ 0.60  	\\ \hline		
\hline																				
[0.7,1.0]	&	0.77	&	[0, 3.0]	&	0.43	&	[0.1,0.2]	&	0.15	&	[0, 3.0]	&	0.15	&	0.90	&1.56  $\pm$ 0.45  $\pm$ 0.26 	\\ \hline		
[0.7,1.0]	&	0.77	&	[0, 3.0]	&	0.44	&	[0.2,0.3]	&	0.25	&	[0, 3.0]	&	0.24	&	0.90	&1.53  $\pm$ 0.53  $\pm$ 0.28 	\\ \hline		
[0.7,1.0]	&	0.77	&	[0, 3.0]	&	0.44	&	[0.3,0.5]	&	0.38	&	[0, 3.0]	&	0.35	&	0.90	&4.95  $\pm$ 0.57  $\pm$ 0.31 	\\ \hline		
[0.7,1.0]	&	0.77	&	[0, 3.0]	&	0.45	&	[0.5,0.7]	&	0.58	&	[0, 3.0]	&	0.47	&	0.90	&6.17  $\pm$ 1.04  $\pm$ 0.64 	\\ \hline		
[0.7,1.0]	&	0.77	&	[0, 3.0]	&	0.47	&	[0.7,1.0]	&	0.77	&	[0, 3.0]	&	0.46	&	0.90	&18.92 $\pm$ 2.49  $\pm$ 2.64	\\ 	
   \end{tabular}
\caption[Collins asymmetries $A_{12}^{\pi^0}$ binned in $z$]{Collins asymmetries  $A_{12}^{\pi^0}$ binned in $z$. Uncertainties are statistical and systematic, respectively. The table contains data binned in $z_1$ while integrated over the other variables, as well as data simultaneously binned in $z_1$ and $z_2$.}
\label{tab:finalpi0zbin}
  \end{ruledtabular}
 \end{table*}
\endgroup

\begingroup
 \squeezetable
 \begin{table*}[htp]
  \begin{ruledtabular}  
    \tiny
    \begin{tabular}{cccccccccc}
$z_1$ & $\langle  z_1\rangle$ & $P_{t1}$ & $\langle  P_{t1}\rangle$  [GeV] & $z_2$ & $\langle  z_2 \rangle$ & $P_{t2}$ & $\langle  P_{t2}\rangle$ [GeV]  & $\langle\frac{\sin^2(\theta)}{1+\cos^2(\theta)}\rangle$& $A_{12}^{\eta}$ [\%]  \\ \hline\hline
[0.3,0.4]	&	0.35	&	[0, 3.0]	&	0.30	&	[0.3,1.0]	&	0.44	&	[0, 3.0]	&	0.38	&	0.91	&2.52  $\pm$ 0.89  $\pm$ 0.11   \\ \hline			
[0.4,0.5]	&	0.45	&	[0, 3.0]	&	0.38	&	[0.3,1.0]	&	0.44	&	[0, 3.0]	&	0.38	&	0.91	&2.61  $\pm$ 0.65  $\pm$ 0.14   \\ \hline			
[0.5,0.6]	&	0.55	&	[0, 3.0]	&	0.43	&	[0.3,1.0]	&	0.44	&	[0, 3.0]	&	0.38	&	0.91	&2.82  $\pm$ 0.60  $\pm$ 0.20     \\ \hline			
[0.6,0.7]	&	0.64	&	[0, 3.0]	&	0.47	&	[0.3,1.0]	&	0.44	&	[0, 3.0]	&	0.38	&	0.91	&2.63  $\pm$ 0.65  $\pm$ 0.31   \\ \hline			
[0.7,1.0]	&	0.77	&	[0, 3.0]	&	0.43	&	[0.3,1.0]	&	0.44	&	[0, 3.0]	&	0.38	&	0.91	&2.80  $\pm$ 0.64  $\pm$ 0.42    \\ \hline			
\hline																				
[0.3,0.5]	&	0.39	&	[0, 3.0]	&	0.33	&	[0.3,0.5]	&	0.38	&	[0, 3.0]	&	0.35	&	0.91	&	2.15  $\pm$ 0.63  $\pm$ 0.10     \\ \hline		
[0.3,0.5]	&	0.39	&	[0, 3.0]	&	0.33	&	[0.5,0.7]	&	0.58	&	[0, 3.0]	&	0.46	&	0.91	&	3.59  $\pm$ 1.13  $\pm$ 0.19    \\ \hline		
[0.3,0.5]	&	0.39	&	[0, 3.0]	&	0.33	&	[0.7,1.0]	&	0.78	&	[0, 3.0]	&	0.43	&	0.91	&	3.25  $\pm$ 2.38  $\pm$ 0.50  	\\ \hline	
\hline																				
[0.5,0.7]	&	0.58	&	[0, 3.0]	&	0.45	&	[0.3,0.5]	&	0.38	&	[0, 3.0]	&	0.35	&	0.91	&  2.15  $\pm$ 0.51  $\pm$ 0.19  \\ \hline			
[0.5,0.7]	&	0.58	&	[0, 3.0]	&	0.45	&	[0.5,0.7]	&	0.58	&	[0, 3.0]	&	0.46	&	0.91	&  4.91  $\pm$ 0.98  $\pm$ 0.40   \\ \hline			
[0.5,0.7]	&	0.58	&	[0, 3.0]	&	0.45	&	[0.7,1.0]	&	0.78	&	[0, 3.0]	&	0.44	&	0.91	&  3.84  $\pm$ 2.18  $\pm$ 1.10   \\ \hline			
\hline																				
[0.7,1.0]	&	0.77	&	[0, 3.0]	&	0.43	&	[0.3,0.5]	&	0.38	&	[0, 3.0]	&	0.35	&	0.91	& 2.17  $\pm$ 0.73  $\pm$ 0.46  	\\ \hline		
[0.7,1.0]	&	0.77	&	[0, 3.0]	&	0.43	&	[0.5,0.7]	&	0.58	&	[0, 3.0]	&	0.47	&	0.91	& 3.15  $\pm$ 1.38  $\pm$ 0.96     \\ \hline			
[0.7,1.0]	&	0.77	&	[0, 3.0]	&	0.45	&	[0.7,1.0]	&	0.77	&	[0, 3.0]	&	0.46	&	0.90	& 15.42  $\pm$ 3.99  $\pm$ 4.05    \\ 		
\end{tabular}
\caption[Collins asymmetries $A_{12}^{\eta}$  binned in $z$]{Collins asymmetries  $A_{12}^{\eta}$ binned in $z$. Uncertainties are statistical and systematic, respectively. The table contains data binned in $z_1$ while integrated over the other variables, as well as data simultaneously binned in $z_1$ and $z_2$. }
\label{tab:finaletazbin}
  \end{ruledtabular}
 \end{table*}
\endgroup

\begingroup
 \squeezetable
 \begin{table*}[htp]
  \begin{ruledtabular}  
   \tiny
    \begin{tabular}{cccccccccc}
$z_1$ & $\langle  z_1\rangle$ & $P_{t1}$ & $\langle  P_{t1}\rangle$  [GeV] & $z_2$ & $\langle  z_2 \rangle$ & $P_{t2}$ & $\langle  P_{t2}\rangle$ [GeV]  & $\langle\frac{\sin^2(\theta)}{1+\cos^2(\theta)}\rangle$& $A_{12}^{\pi^0}$ [\%]  \\ \hline\hline
[0.3,0.4]	&	0.35	&	[0, 3.0]	&	0.32	&	[0.3,1.0]	&	0.44	&	[0, 3.0]	&	0.38	&	0.91	& 2.19  $\pm$ 0.09  $\pm$ 0.22    	\\ \hline
[0.4,0.5]	&	0.45	&	[0, 3.0]	&	0.39	&	[0.3,1.0]	&	0.44	&	[0, 3.0]	&	0.38	&	0.91	& 2.00  $\pm$ 0.22  $\pm$ 0.08       	\\ \hline
[0.5,0.6]	&	0.54	&	[0, 3.0]	&	0.45	&	[0.3,1.0]	&	0.44	&	[0, 3.0]	&	0.38	&	0.91	& 2.65  $\pm$ 0.29  $\pm$ 0.11    	\\ \hline
[0.6,0.7]	&	0.64	&	[0, 3.0]	&	0.49	&	[0.3,1.0]	&	0.44	&	[0, 3.0]	&	0.38	&	0.90	& 3.02  $\pm$ 0.37  $\pm$ 0.17    	\\ \hline
[0.7,1.0]	&	0.77	&	[0, 3.0]	&	0.45	&	[0.3,1.0]	&	0.44	&	[0, 3.0]	&	0.38	&	0.90	& 5.76  $\pm$ 0.49  $\pm$ 0.28    	\\ \hline
\hline																		
[0.3,0.5]	&	0.38	&	[0, 3.0]	&	0.35	&	[0.3,0.5]	&	0.38	&	[0, 3.0]	&	0.35	&	0.91	& 1.63  $\pm$ 0.16  $\pm$ 0.06  	\\ \hline
[0.3,0.5]	&	0.38	&	[0, 3.0]	&	0.35	&	[0.5,0.7]	&	0.58	&	[0, 3.0]	&	0.46	&	0.91	& 2.83  $\pm$ 0.30  $\pm$ 0.10    	\\ \hline
[0.3,0.5]	&	0.38	&	[0, 3.0]	&	0.35	&	[0.7,1.0]	&	0.78	&	[0, 3.0]	&	0.44	&	0.91	& 4.07  $\pm$ 0.65  $\pm$ 0.28  	\\ \hline
\hline																		
[0.5,0.7]	&	0.58	&	[0, 3.0]	&	0.46	&	[0.3,0.5]	&	0.38	&	[0, 3.0]	&	0.35	&	0.91	&2.49  $\pm$ 0.26  $\pm$ 0.10  	\\ \hline
[0.5,0.7]	&	0.58	&	[0, 3.0]	&	0.46	&	[0.5,0.7]	&	0.58	&	[0, 3.0]	&	0.46	&	0.91	&3.39  $\pm$ 0.49  $\pm$ 0.24 	\\ \hline
[0.5,0.7]	&	0.58	&	[0, 3.0]	&	0.47	&	[0.7,1.0]	&	0.78	&	[0, 3.0]	&	0.44	&	0.90	&4.94  $\pm$ 1.06  $\pm$ 0.60  	\\ \hline
\hline																		
[0.7,1.0]	&	0.77	&	[0, 3.0]	&	0.44	&	[0.3,0.5]	&	0.38	&	[0, 3.0]	&	0.35	&	0.90	&4.95  $\pm$ 0.57  $\pm$ 0.31 	\\ \hline
[0.7,1.0]	&	0.77	&	[0, 3.0]	&	0.45	&	[0.5,0.7]	&	0.58	&	[0, 3.0]	&	0.47	&	0.90	&6.17  $\pm$ 1.04  $\pm$ 0.64 	\\ \hline
[0.7,1.0]	&	0.77	&	[0, 3.0]	&	0.47	&	[0.7,1.0]	&	0.77	&	[0, 3.0]	&	0.46	&	0.90	&18.92  $\pm$ 2.49  $\pm$ 2.64	\\ 
\end{tabular}
\caption[Collins asymmetries $A_{12}^{\pi^0}$ with $z>0.3$ binned in $z$]{Collins asymmetries $A_{12}^{\pi^0}$  with $z>0.3$ binned in $z$. Uncertainties are statistical and systematic, respectively. The table contains data binned in $z_1$ while integrated over the other variables, as well as data simultaneously binned in $z_1$ and $z_2$. }
\label{tab:finaletazbin2}
  \end{ruledtabular}
 \end{table*}
\endgroup

\begingroup
 \squeezetable
 \begin{table*}[htp]
  \begin{ruledtabular}  
    \begin{tabular}{cccccccccc}
$P_{t1}$   [GeV] & $\langle  P_{t1}  \rangle$ [GeV] & $z_1$ &  $\langle  z_1 \rangle$& $P_{t2}$  [GeV] & $\langle  P_{t2}\rangle$  [GeV] & $z_2$ & $\langle  z_2 \rangle$  &$\langle\frac{\sin^2(\theta)}{1+\cos^2(\theta)}\rangle$&  $A_{12}^{\pi^0}$ [\%]  \\ \hline\hline
[0,0.15]	&	0.10	&	[0.2, 1.0]	&	0.31	&	[0,3.0]	&	0.32	&	[0.2, 1.0]	&	0.36	&	0.91	&     0.52  $\pm$ 0.26  $\pm$ 0.10 	\\ \hline
[0.15,0.3]	&	0.23	&	[0.2, 1.0]	&	0.31	&	[0,3.0]	&	0.32	&	[0.2, 1.0]	&	0.36	&	0.91	 &    1.16  $\pm$ 0.12  $\pm$ 0.05 	\\ \hline
[0.3,0.5]	&	0.38	&	[0.2, 1.0]	&	0.36	&	[0,3.0]	&	0.32	&	[0.2, 1.0]	&	0.36	&	0.91	&  1.71  $\pm$ 0.10  $\pm$ 0.08   	\\ \hline
[0.5,3.0]	&	0.62	&	[0.2, 1.0]	&	0.50	&	[0,3.0]	&	0.32	&	[0.2, 1.0]	&	0.36	&	0.91	&     2.95  $\pm$ 0.16  $\pm$ 0.08 	\\ \hline
\hline																		
[0,0.15]	&	0.10	&	[0.2, 1.0]	&	0.31	&	[0,0.15]	&	0.10	&	[0.2, 1.0]	&	0.32	&	0.91	&0.12  $\pm$ 0.61  $\pm$ 0.14  	\\ \hline
[0,0.15]	&	0.10	&	[0.2, 1.0]	&	0.31	&	[0.15,0.3]	&	0.23	&	[0.2, 1.0]	&	0.31	&	0.91	&0.49  $\pm$ 0.42  $\pm$ 0.11  	\\ \hline
[0,0.15]	&	0.10	&	[0.2, 1.0]	&	0.31	&	[0.3,0.5]	&	0.38	&	[0.2, 1.0]	&	0.36	&	0.91	    &0.24  $\pm$ 0.41  $\pm$ 0.17       	\\ \hline
[0,0.15]	&	0.10	&	[0.2, 1.0]	&	0.31	&	[0.5,3.0]	&	0.62	&	[0.2, 1.0]	&	0.51	&	0.91	&1.83  $\pm$ 0.71  $\pm$ 0.19  	\\ \hline
\hline																		
[0.15,0.3]	&	0.23	&	[0.2, 1.0]	&	0.30	&	[0,0.15]	&	0.10	&	[0.2, 1.0]	&	0.32	&	0.91	 &0.99  $\pm$ 0.39  $\pm$ 0.11  	\\ \hline
[0.15,0.3]	&	0.23	&	[0.2, 1.0]	&	0.31	&	[0.15,0.3]	&	0.23	&	[0.2, 1.0]	&	0.31	&	0.91	&   0.92  $\pm$ 0.19  $\pm$ 0.05  	\\ \hline
[0.15,0.3]	&	0.23	&	[0.2, 1.0]	&	0.31	&	[0.3,0.5]	&	0.38	&	[0.2, 1.0]	&	0.36	&	0.91	&1.16  $\pm$ 0.18  $\pm$ 0.08  	\\ \hline
[0.15,0.3]	&	0.23	&	[0.2, 1.0]	&	0.31	&	[0.5,3.0]	&	0.62	&	[0.2, 1.0]	&	0.51	&	0.91	& 1.92  $\pm$ 0.31  $\pm$ 0.13 	\\ \hline
\hline																		
[0.3,0.5]	&	0.38	&	[0.2, 1.0]	&	0.35	&	[0,0.15]	&	0.10	&	[0.2, 1.0]	&	0.32	&	0.91	&0.70  $\pm$ 0.35  $\pm$ 0.12   	\\ \hline
[0.3,0.5]	&	0.38	&	[0.2, 1.0]	&	0.36	&	[0.15,0.3]	&	0.23	&	[0.2, 1.0]	&	0.31	&	0.91	&1.41  $\pm$ 0.17  $\pm$ 0.08  	\\ \hline
[0.3,0.5]	&	0.38	&	[0.2, 1.0]	&	0.36	&	[0.3,0.5]	&	0.38	&	[0.2, 1.0]	&	0.36	&	0.91	&1.93  $\pm$ 0.16  $\pm$ 0.08  	\\ \hline
[0.3,0.5]	&	0.38	&	[0.2, 1.0]	&	0.36	&	[0.5,3.0]	&	0.62	&	[0.2, 1.0]	&	0.51	&	0.91	 &     2.74  $\pm$ 0.28  $\pm$ 0.17  	\\ \hline
\hline																		
[0.5,3.0]	&	0.62	&	[0.2, 1.0]	&	0.50	&	[0,0.15]	&	0.10	&	[0.2, 1.0]	&	0.32	&	0.91	 &1.18  $\pm$ 0.49  $\pm$ 0.22  	\\ \hline
[0.5,3.0]	&	0.62	&	[0.2, 1.0]	&	0.50	&	[0.15,0.3]	&	0.23	&	[0.2, 1.0]	&	0.31	&	0.91	&2.43  $\pm$ 0.25  $\pm$ 0.11  	\\ \hline
[0.5,3.0]	&	0.62	&	[0.2, 1.0]	&	0.50	&	[0.3,0.5]	&	0.38	&	[0.2, 1.0]	&	0.36	&	0.91	&2.72  $\pm$ 0.24  $\pm$ 0.10   	\\ \hline
[0.5,3.0]	&	0.62	&	[0.2, 1.0]	&	0.50	&	[0.5,3.0]	&	0.62	&	[0.2, 1.0]	&	0.51	&	0.91	&5.87  $\pm$ 0.48  $\pm$ 0.24  	\\ 
\end{tabular}
\caption[Collins asymmetries $A_{12}^{\pi^0}$ binned in $P_t$]{Collins asymmetries $A_{12}^{\pi^0}$ binned in $P_t$. Uncertainties are statistical and systematic, respectively. The table contains data binned in $P_{t1}$ while integrated over the other variables, as well as data simultaneously binned in $P_{t1}$ and $P_{t2}$. }
\label{tab:finalpi0ptbin}
  \end{ruledtabular}
 \end{table*}
\endgroup

\begingroup
 \squeezetable
 \begin{table*}[htp]
  \begin{ruledtabular}  
    \begin{tabular}{cccccccccc}
$P_{t1}$   [GeV] & $\langle  P_{t1}  \rangle$ [GeV] & $z_1$ &  $\langle  z_1 \rangle$& $P_{t2}$  [GeV] & $\langle  P_{t2}\rangle$  [GeV] & $z_2$ & $\langle  z_2 \rangle$  &$\langle\frac{\sin^2(\theta)}{1+\cos^2(\theta)}\rangle$& $A_{12}^{\eta}$ [\%]  \\ \hline\hline
[0,0.15]	&	0.10	&	[0.2, 1.0]	&	0.43	&	[0,3.0]	&	0.38	&	[0.2, 1.0]	&	0.44	&	0.91	&	1.29  $\pm$ 1.20  $\pm$ 0.21  	\\ \hline
[0.15,0.3]	&	0.23	&	[0.2, 1.0]	&	0.43	&	[0,3.0]	&	0.38	&	[0.2, 1.0]	&	0.44	&	0.91	&	1.75  $\pm$ 0.64  $\pm$ 0.11 	\\ \hline
[0.3,0.5]	&	0.40	&	[0.2, 1.0]	&	0.44	&	[0,3.0]	&	0.38	&	[0.2, 1.0]	&	0.44	&	0.91	&	1.81  $\pm$ 0.45  $\pm$ 0.10  	\\ \hline
[0.5,3.0]	&	0.63	&	[0.2, 1.0]	&	0.53	&	[0,3.0]	&	0.37	&	[0.2, 1.0]	&	0.44	&	0.91	&	2.91  $\pm$ 0.48  $\pm$ 0.19 	\\ \hline
\hline																			
[0,0.15]	&	0.10	&	[0.2, 1.0]	&	0.43	&	[0,0.15]	&	0.10	&	[0.2, 1.0]	&	0.42	&	0.91	&	-0.63  $\pm$ 3.45  $\pm$ 0.66 	\\ \hline
[0,0.15]	&	0.10	&	[0.2, 1.0]	&	0.43	&	[0.15,0.3]	&	0.23	&	[0.2, 1.0]	&	0.42	&	0.91	&	-3.59  $\pm$ 3.63  $\pm$ 0.85 	\\ \hline
[0,0.15]	&	0.10	&	[0.2, 1.0]	&	0.44	&	[0.3,0.5]	&	0.40	&	[0.2, 1.0]	&	0.42	&	0.91	&	2.66  $\pm$ 1.76  $\pm$ 0.37  	\\ \hline
[0,0.15]	&	0.10	&	[0.2, 1.0]	&	0.44	&	[0.5,3.0]	&	0.62	&	[0.2, 1.0]	&	0.51	&	0.91	&	2.66  $\pm$ 2.07  $\pm$ 0.46  	\\ \hline
\hline																			
[0.15,0.3]	&	0.23	&	[0.2, 1.0]	&	0.43	&	[0,0.15]	&	0.10	 &   	[0.2, 1.0]	&	0.43	&	0.91	&	 -1.73  $\pm$ 2.16  $\pm$ 0.48 	\\ \hline
[0.15,0.3]	&	0.23	&	[0.2, 1.0]	&	0.43	&	[0.15,0.3]	&	0.23	 &   	[0.2, 1.0]	&	0.42	&	0.91	&	 0.34  $\pm$ 1.34  $\pm$ 0.21  	\\ \hline
[0.15,0.3]	&	0.23	&	[0.2, 1.0]	&	0.43	&	[0.3,0.5]	&	0.40	 &   	[0.2, 1.0]	&	0.42	&	0.91	&	 1.52  $\pm$ 0.97  $\pm$ 0.16  	\\ \hline
[0.15,0.3]	&	0.23	&	[0.2, 1.0]	&	0.43	&	[0.5,3.0]	&	0.62	 &   	[0.2, 1.0]	&	0.51	&	0.91	&	4.77  $\pm$ 1.28  $\pm$ 0.27  	\\ \hline
\hline																			
[0.3,0.5]	&	0.40	&	[0.2, 1.0]	&	0.43	&	[0,0.15]	&	0.10	&	[0.2, 1.0]	&	0.43	&	0.91	&	0.44  $\pm$ 1.91  $\pm$ 0.38  	\\ \hline
[0.3,0.5]	&	0.40	&	[0.2, 1.0]	&	0.44	&	[0.15,0.3]	&	0.23	&	[0.2, 1.0]	&	0.42	&	0.91	&	0.98  $\pm$ 0.94  $\pm$ 0.19  	\\ \hline
[0.3,0.5]	&	0.40	&	[0.2, 1.0]	&	0.44	&	[0.3,0.5]	&	0.40	&	[0.2, 1.0]	&	0.42	&	0.91	&	1.89  $\pm$ 0.69  $\pm$ 0.15  	\\ \hline
[0.3,0.5]	&	0.40	&	[0.2, 1.0]	&	0.44	&	[0.5,3.0]	&	0.62	&	[0.2, 1.0]	&	0.51	&	0.91	&	2.95  $\pm$ 0.94  $\pm$ 0.23  	\\ \hline
\hline																			
[0.5,3.0]	&	0.63	&	[0.2, 1.0]	&	0.53	&	[0,0.15]	&	0.10	    &	[0.2, 1.0]	&	0.44	&	0.91	&	1.47  $\pm$ 1.66  $\pm$ 0.62  	\\ \hline
[0.5,3.0]	&	0.63	&	[0.2, 1.0]	&	0.53	&	[0.15,0.3]	&	0.23	&	[0.2, 1.0]	&	0.42	&	0.91	&	1.22  $\pm$ 0.94  $\pm$ 0.34  	\\ \hline
[0.5,3.0]	&	0.63	&	[0.2, 1.0]	&	0.53	&	[0.3,0.5]	&	0.40	&	[0.2, 1.0]	&	0.42	&	0.91	&	2.67  $\pm$ 0.67  $\pm$ 0.26  	\\ \hline
[0.5,3.0]	&	0.62	&	[0.2, 1.0]	&	0.53	&	[0.5,3.0]	&	0.62	&	[0.2, 1.0]	&	0.51	&	0.91	&	5.26  $\pm$ 1.04  $\pm$ 0.46  	\\ 
\end{tabular}
\caption[Collins asymmetries $A_{12}^{\eta}$ binned in $P_t$]{Collins asymmetries $A_{12}^{\eta}$ binned in $P_t$. Uncertainties are statistical and systematic, respectively. The table contains data binned in $P_{t1}$ while integrated over the other variables, as well as data simultaneously binned in $P_{t1}$ and $P_{t2}$. }
\label{tab:finaletaptbin}
  \end{ruledtabular}
 \end{table*}
\endgroup

\begingroup
 \squeezetable
 \begin{table*}[htp]
  \begin{ruledtabular}  
    \begin{tabular}{cccccccccc}
$P_{t1}$   [GeV] & $\langle  P_{t1}  \rangle$ [GeV] & $z_1$ &  $\langle  z_1 \rangle$& $P_{t2}$  [GeV] & $\langle  P_{t2}\rangle$  [GeV] & $z_2$ & $\langle  z_2 \rangle$  &$\langle\frac{\sin^2(\theta)}{1+\cos^2(\theta)}\rangle$&  $A_{12}^{\pi^0}$ [\%]  \\ \hline\hline
[0,0.15]	&	0.10	&	[0.2, 1.0]	&	0.42	&	[0,3.0]	&	0.38	&	[0.2, 1.0]	&	0.44	&	0.91	&	0.38  $\pm$ 0.54  $\pm$ 0.20  	\\ \hline	
[0.15,0.3]	&	0.23	&	[0.2, 1.0]	&	0.41	&	[0,3.0]	&	0.38	&	[0.2, 1.0]	&	0.44	&	0.91	&	1.59  $\pm$ 0.24  $\pm$ 0.09    \\ \hline		
[0.3,0.5]	&	0.40	&	[0.2, 1.0]	&	0.41	&	[0,3.0]	&	0.38	&	[0.2, 1.0]	&	0.44	&	0.91	&	2.15  $\pm$ 0.17  $\pm$ 0.10     \\ \hline		
[0.5,3.0]	&	0.62	&	[0.2, 1.0]	&	0.50	&	[0,3.0]	&	0.37	&	[0.2, 1.0]	&	0.44	&	0.91	&	3.60  $\pm$ 0.22  $\pm$ 0.09     \\ \hline		
\hline																				
[0,0.15]	&	0.10	&	[0.2, 1.0]	&	0.41	&	[0,0.15]	&	0.10	&	[0.2, 1.0]	&	0.43	&	0.91	&-1.57  $\pm$ 1.88  $\pm$ 0.64   	\\ \hline		
[0,0.15]	&	0.10	&	[0.2, 1.0]	&	0.42	&	[0.15,0.3]	&	0.23	&	[0.2, 1.0]	&	0.42	&	0.91	&0.35  $\pm$ 1.07  $\pm$ 0.37    	\\ \hline		
[0,0.15]	&	0.10	&	[0.2, 1.0]	&	0.42	&	[0.3,0.5]	&	0.40	&	[0.2, 1.0]	&	0.42	&	0.91	&0.28  $\pm$ 0.80  $\pm$ 0.29     	\\ \hline		
[0,0.15]	&	0.10	&	[0.2, 1.0]	&	0.42	&	[0.5,3.0]	&	0.62	&	[0.2, 1.0]	&	0.51	&	0.91	&1.10  $\pm$ 1.09  $\pm$ 0.35         	\\ \hline		
\hline																				
[0.15,0.3]	&	0.23	&	[0.2, 1.0]	&	0.41	&	[0,0.15]	&	0.10	&	[0.2, 1.0]	&	0.43	&	0.91	&1.22  $\pm$ 1.13  $\pm$ 0.33    	\\ \hline		
[0.15,0.3]	&	0.23	&	[0.2, 1.0]	&	0.41	&	[0.15,0.3]	&	0.23	&	[0.2, 1.0]	&	0.42	&	0.91	&0.61  $\pm$ 0.46  $\pm$ 0.18        	\\ \hline		
[0.15,0.3]	&	0.23	&	[0.2, 1.0]	&	0.41	&	[0.3,0.5]	&	0.40	&	[0.2, 1.0]	&	0.42	&	0.91	&1.77  $\pm$ 0.36  $\pm$ 0.10     	\\ \hline		
[0.15,0.3]	&	0.23	&	[0.2, 1.0]	&	0.41	&	[0.5,3.0]	&	0.63	&	[0.2, 1.0]	&	0.51	&	0.91	&2.42  $\pm$ 0.47  $\pm$ 0.16    	\\ \hline		
\hline																				
[0.3,0.5]	&	0.40	&	[0.2, 1.0]	&	0.41	&	[0,0.15]	&	0.10	&	[0.2, 1.0]	&	0.43	&	0.91	&1.55  $\pm$ 0.77  $\pm$ 0.28    	\\ \hline		
[0.3,0.5]	&	0.40	&	[0.2, 1.0]	&	0.41	&	[0.15,0.3]	&	0.23	&	[0.2, 1.0]	&	0.42	&	0.91	&1.68  $\pm$ 0.34  $\pm$ 0.12    	\\ \hline		
[0.3,0.5]	&	0.40	&	[0.2, 1.0]	&	0.41	&	[0.3,0.5]	&	0.40	&	[0.2, 1.0]	&	0.42	&	0.91	&2.01  $\pm$ 0.24  $\pm$ 0.11    	\\ \hline		
[0.3,0.5]	&	0.40	&	[0.2, 1.0]	&	0.42	&	[0.5,3.0]	&	0.62	&	[0.2, 1.0]	&	0.51	&	0.91	&3.13  $\pm$ 0.33  $\pm$ 0.15    	\\ \hline		
\hline																				
[0.5,3.0]	&	0.62	&	[0.2, 1.0]	&	0.50	&	[0,0.15]	&	0.10	&	[0.2, 1.0]	&	0.43	&	0.91	&2.43  $\pm$ 0.86  $\pm$ 0.33    	\\ \hline		
[0.5,3.0]	&	0.62	&	[0.2, 1.0]	&	0.50	&	[0.15,0.3]	&	0.23	&	[0.2, 1.0]	&	0.42	&	0.91	&2.70  $\pm$ 0.43  $\pm$ 0.17     	\\ \hline		
[0.5,3.0]	&	0.62	&	[0.2, 1.0]	&	0.50	&	[0.3,0.5]	&	0.40	&	[0.2, 1.0]	&	0.42	&	0.91	&3.03  $\pm$ 0.30  $\pm$ 0.12     	\\ \hline		
[0.5,3.0]	&	0.62	&	[0.2, 1.0]	&	0.50	&	[0.5,3.0]	&	0.62	&	[0.2, 1.0]	&	0.51	&	0.91	&5.78  $\pm$ 0.47  $\pm$ 0.24    	\\ 	
\end{tabular}
\caption[Collins asymmetries $A_{12}^{\pi^0}$ with $z>0.3$ binned in $P_t$]{Collins asymmetries $A_{12}^{\pi^0}$ with $z>0.3$ binned in $P_t$. Uncertainties are statistical and systematic, respectively. The table contains data binned in $P_{t1}$ while integrated over the other variables, as well as data simultaneously binned in $P_{t1}$ and $P_{t2}$. }\label{tab:finaletaptbin2}
  \end{ruledtabular}
 \end{table*}
\endgroup

\begingroup
 \squeezetable
 \begin{table*}[htp]
  \begin{ruledtabular}  
    \begin{tabular}{cccccccccc}
$z_1$& $\langle  z_{1}  \rangle$ & $z_2$ & $\langle  z_{2}\rangle$& $P_{t1}$ [GeV]   & $\langle  P_{t1} \rangle$ [GeV]  & $P_{t2}$  [GeV]  &  $\langle P_{t2}\rangle$  [GeV]  &$\langle\frac{\sin^2(\theta)}{1+\cos^2(\theta)}\rangle$& $A_{12}^{UC}$ [\%]   \\ \hline\hline
[0.2,0.3]	&	0.24	&	[0.2,1.0]	&	0.36	&	[0,0.15]	&	0.10	&	[0,3.0]	&	0.32	&	0.91	& 0.55  $\pm$ 0.11  $\pm$ 0.05 \\ \hline			
[0.2,0.3]	&	0.24	&	[0.2,1.0]	&	0.36	&	[0.15,0.30]	&	0.23	&	[0,3.0]	&	0.32	&	0.91	& 0.96  $\pm$ 0.07  $\pm$ 0.03 \\ \hline			
[0.2,0.3]	&	0.26	&	[0.2,1.0]	&	0.36	&	[0.30,0.50]	&	0.35	&	[0,3.0]	&	0.32	&	0.90	& 1.43  $\pm$ 0.10  $\pm$ 0.04  \\ \hline			
[0.2,0.3]	&	0.00	&	[0.2,1.0]	&		&	[0.50,3.0]	&		&	[0,3.0]	&		&		&			\\ \hline
\hline																				
[0.3,0.5]	&	0.37	&	[0.2,1.0]	&	0.36	&	[0,0.15]	&	0.10	&	[0,3.0]	&	0.32	&	0.91	& 0.49  $\pm$ 0.17  $\pm$ 0.07  	\\ \hline		
[0.3,0.5]	&	0.37	&	[0.2,1.0]	&	0.36	&	[0.15,0.30]	&	0.23	&	[0,3.0]	&	0.32	&	0.91	& 0.98  $\pm$ 0.10  $\pm$ 0.04   	\\ \hline		
[0.3,0.5]	&	0.38	&	[0.2,1.0]	&	0.36	&	[0.30,0.50]	&	0.40	&	[0,3.0]	&	0.32	&	0.91	& 1.41  $\pm$ 0.08  $\pm$ 0.03  	\\ \hline		
[0.3,0.5]	&	0.42	&	[0.2,1.0]	&	0.36	&	[0.50,3.0]	&	0.58	&	[0,3.0]	&	0.32	&	0.91	& 1.97  $\pm$ 0.16  $\pm$ 0.07  	\\ \hline		
\hline																				
[0.5,0.7]	&	0.56	&	[0.2,1.0]	&	0.35	&	[0,0.15]	&	0.10	&	[0,3.0]	&	0.31	&	0.91	& 0.23  $\pm$ 0.30  $\pm$ 0.12   	\\ \hline		
[0.5,0.7]	&	0.56	&	[0.2,1.0]	&	0.35	&	[0.15,0.30]	&	0.23	&	[0,3.0]	&	0.31	&	0.91	& 1.04  $\pm$ 0.20  $\pm$ 0.09   	\\ \hline		
[0.5,0.7]	&	0.56	&	[0.2,1.0]	&	0.35	&	[0.30,0.50]	&	0.40	&	[0,3.0]	&	0.31	&	0.91	& 1.71  $\pm$ 0.15  $\pm$ 0.07  	\\ \hline		
[0.5,0.7]	&	0.58	&	[0.2,1.0]	&	0.36	&	[0.50,3.0]	&	0.67	&	[0,3.0]	&	0.32	&	0.91	& 2.47  $\pm$ 0.16  $\pm$ 0.08  	\\ \hline		
\hline																				
[0.7,1.0]	&	0.81	&	[0.2,1.0]	&	0.35	&	[0,0.15]	&	0.09	&	[0,3.0]	&	0.33	&	0.91	& 0.34  $\pm$ 0.48  $\pm$ 0.28  	\\ \hline		
[0.7,1.0]	&	0.77	&	[0.2,1.0]	&	0.35	&	[0.15,0.30]	&	0.23	&	[0,3.0]	&	0.32	&	0.91	& 1.07  $\pm$ 0.37  $\pm$ 0.22  	\\ \hline		
[0.7,1.0]	&	0.75	&	[0.2,1.0]	&	0.35	&	[0.30,0.50]	&	0.40	&	[0,3.0]	&	0.32	&	0.91	& 2.24  $\pm$ 0.33  $\pm$ 0.2   	\\ \hline		
[0.7,1.0]	&	0.75	&	[0.2,1.0]	&	0.36	&	[0.50,3.0]	&	0.72	&	[0,3.0]	&	0.31	&	0.90	& 5.14  $\pm$ 0.33  $\pm$ 0.24  	\\ 		
\end{tabular}
\caption[Collins asymmetries $A_{12}^{UC}$ binned in $(z_{1},P_{t1})$]{Collins asymmetries $A_{12}^{UC}$ binned in $(z_{1},P_{t1})$. Uncertainties are statistical and systematic, respectively. The fourth row is empty because the hadron pairs have to possess low $z$ and high $P_t$ kinematics simultaneously, which is unlikely to happen.}\label{tab:finalucptbins}
  \end{ruledtabular}
 \end{table*}
\endgroup

\begingroup
 \squeezetable
 \begin{table*}[htp]
  \begin{ruledtabular}  
    \begin{tabular}{cccccccccc}
$z_1$& $\langle  z_{1}  \rangle$ & $z_2$ & $\langle  z_{2}\rangle$& $P_{t1}$  [GeV]   & $\langle  P_{t1} \rangle$ [GeV]  & $P_{t2}$  [GeV]  &  $\langle P_{t2}\rangle$  [GeV]  &$\langle\frac{\sin^2(\theta)}{1+\cos^2(\theta)}\rangle$& $A_{12}^{\pi^0}$ [\%]   \\ \hline\hline
[0.2,0.3]	&	0.24	&	[0.2,1.0]	&	0.36	&	[0,0.15]	&	0.10	&	[0,3.0]	&	0.32	&	0.91	& 0.50  $\pm$ 0.33  $\pm$ 0.11   	\\ \hline		
[0.2,0.3]	&	0.24	&	[0.2,1.0]	&	0.36	&	[0.15,0.30]	&	0.23	&	[0,3.0]	&	0.32	&	0.91	& 1.08  $\pm$ 0.15  $\pm$ 0.05        \\ \hline			
[0.2,0.3]	&	0.26	&	[0.2,1.0]	&	0.36	&	[0.30,0.50]	&	0.35	&	[0,3.0]	&	0.32	&	0.91	& 1.53  $\pm$ 0.18  $\pm$ 0.11        \\ \hline			
[0.2,0.3]	&		&	[0.2,1.0]	&		&	[0.50,3.0]	&		&	[0,3.0]	&		&		&			\\ \hline
 \hline																				
[0.3,0.5]	&	0.37	&	[0.2,1.0]	&	0.36	&	[0,0.15]	&	0.10	&	[0,3.0]	&	0.32	&	0.91	&0.39  $\pm$ 0.52  $\pm$ 0.13   \\ \hline			
[0.3,0.5]	&	0.37	&	[0.2,1.0]	&	0.36	&	[0.15,0.30]	&	0.23	&	[0,3.0]	&	0.32	&	0.91	&1.24  $\pm$ 0.24  $\pm$ 0.09   \\ \hline			
[0.3,0.5]	&	0.37	&	[0.2,1.0]	&	0.36	&	[0.30,0.50]	&	0.40	&	[0,3.0]	&	0.32	&	0.91	&1.72  $\pm$ 0.16  $\pm$ 0.07   \\ \hline			
[0.3,0.5]	&	0.42	&	[0.2,1.0]	&	0.36	&	[0.50,3.0]	&	0.57	&	[0,3.0]	&	0.32	&	0.91	&2.41  $\pm$ 0.27  $\pm$ 0.12   \\ \hline			
 \hline                                                     
[0.5,0.7]	&	0.49	&	[0.2,1.0]	&	0.36	&	[0,0.15]	&	0.10	&	[0,3.0]	&	0.32	&	0.91	&0.48  $\pm$ 0.70  $\pm$ 0.21    \\ \hline			
[0.5,0.7]	&	0.49	&	[0.2,1.0]	&	0.36	&	[0.15,0.30]	&	0.23	&	[0,3.0]	&	0.32	&	0.91	&1.16  $\pm$ 0.31  $\pm$ 0.12    \\ \hline			
[0.5,0.7]	&	0.49	&	[0.2,1.0]	&	0.36	&	[0.30,0.50]	&	0.40	&	[0,3.0]	&	0.32	&	0.91	&1.88  $\pm$ 0.21  $\pm$ 0.11    \\ \hline			
[0.5,0.7]	&	0.53	&	[0.2,1.0]	&	0.36	&	[0.50,3.0]	&	0.64	&	[0,3.0]	&	0.32	&	0.91	&2.43  $\pm$ 0.20  $\pm$ 0.10   \\ \hline			
 \hline                         						
[0.7,1.0]	&	0.72	&	[0.2,1.0]	&	0.36	&	[0,0.15]	&	0.10	&	[0,3.0]	&	0.32	&	0.90	&-0.64  $\pm$ 1.53  $\pm$ 0.77   \\ \hline			
[0.7,1.0]	&	0.69	&	[0.2,1.0]	&	0.36	&	[0.15,0.30]	&	0.23	&	[0,3.0]	&	0.32	&	0.90	&1.45  $\pm$ 0.79  $\pm$ 0.32    \\ \hline			
[0.7,1.0]	&	0.68	&	[0.2,1.0]	&	0.36	&	[0.30,0.50]	&	0.40	&	[0,3.0]	&	0.32	&	0.90	&2.68  $\pm$ 0.50  $\pm$ 0.24    \\ \hline          			
[0.7,1.0]	&	0.68	&	[0.2,1.0]	&	0.36	&	[0.50,3.0]	&	0.70	&	[0,3.0]	&	0.32	&	0.90	& 6.01  $\pm$ 0.45  $\pm$ 0.27   \\ 
\end{tabular}
\caption[Collins asymmetries $A_{12}^{\pi^0}$ binned in $(z_{1},P_{t1})$]{Collins asymmetries $A_{12}^{\pi^0}$ binned in $(z_{1},P_{t1})$. Uncertainties are statistical and systematic, respectively. Again, the fourth row is empty due to kinematic constraints.}
\label{tab:finalpi0ptbins}
  \end{ruledtabular}
 \end{table*}
\endgroup

\begingroup
 \squeezetable
 \begin{table*}[htp]
  \begin{ruledtabular}  
    \begin{tabular}{cccccccccc}
$z_1$& $\langle  z_{1}  \rangle$ & $z_2$ & $\langle  z_{2}\rangle$& $P_{t1}$ [GeV] & $\langle  P_{t1} \rangle$ [GeV] & $P_{t2}$ [GeV] &  $\langle P_{t2}\rangle$ [GeV]  &$\langle\frac{\sin^2(\theta)}{1+\cos^2(\theta)}\rangle$& $A_{12}^{\eta}$ [\%]   \\ \hline\hline
[0.3,0.5]	&	0.38	&	[0.3,1.0]	&	0.44	&	[0,0.15]	&	0.10	&	[0,3.0]	&	0.38	&	0.91	&  2.52  $\pm$ 1.77  $\pm$ 0.32     \\ \hline			
[0.3,0.5]	&	0.38	&	[0.3,1.0]	&	0.44	&	[0.15,0.30]	&	0.23	&	[0,3.0]	&	0.38	&	0.91	&  2.55  $\pm$ 1.19  $\pm$ 0.16     \\ \hline			
[0.3,0.5]	&	0.38	&	[0.3,1.0]	&	0.44	&	[0.30,0.50]	&	0.40	&	[0,3.0]	&	0.38	&	0.91	&  1.08  $\pm$ 0.90  $\pm$ 0.14      \\ \hline			
[0.3,0.5]	&	0.43	&	[0.3,1.0]	&	0.44	&	[0.50,3.0]	&	0.58	&	[0,3.0]	&	0.37	&	0.91	&  1.93  $\pm$ 1.31  $\pm$ 0.34     \\ \hline			
\hline																				
[0.5,0.7]	&	0.50	&	[0.3,1.0]	&	0.44	&	[0,0.15]	&	0.10	&	[0,3.0]	&	0.38	&	0.91	&  1.48  $\pm$ 1.06  $\pm$ 0.31     \\ \hline			
[0.5,0.7]	&	0.50	&	[0.3,1.0]	&	0.44	&	[0.15,0.30]	&	0.23	&	[0,3.0]	&	0.38	&	0.91	&  1.06  $\pm$ 0.66  $\pm$ 0.17     \\ \hline			
[0.5,0.7]	&	0.51	&	[0.3,1.0]	&	0.44	&	[0.30,0.50]	&	0.40	&	[0,3.0]	&	0.38	&	0.91	&  2.23  $\pm$ 0.48  $\pm$ 0.16     \\ \hline			
[0.5,0.7]	&	0.55	&	[0.3,1.0]	&	0.44	&	[0.50,3.0]	&	0.65	&	[0,3.0]	&	0.37	&	0.91	&  2.94  $\pm$ 0.53  $\pm$ 0.24     \\ \hline			
\hline																				
[0.7,1.0]	&	0.73	&	[0.3,1.0]	&	0.44	&	[0,0.15]	&	0.10	&	[0,3.0]	&	0.39	&	0.91	&  -2.62  $\pm$ 1.45  $\pm$ 0.89    \\ \hline			
[0.7,1.0]	&	0.70	&	[0.3,1.0]	&	0.44	&	[0.15,0.30]	&	0.23	&	[0,3.0]	&	0.38	&	0.91	&  1.33  $\pm$ 0.86  $\pm$ 0.42     \\ \hline			
[0.7,1.0]	&	0.69	&	[0.3,1.0]	&	0.44	&	[0.30,0.50]	&	0.40	&	[0,3.0]	&	0.38	&	0.91	&  3.74  $\pm$ 0.78  $\pm$ 0.42     \\ \hline			
[0.7,1.0]	&	0.69	&	[0.3,1.0]	&	0.44	&	[0.50,3.0]	&	0.70	&	[0,3.0]	&	0.38	&	0.91	&  6.60  $\pm$ 0.68  $\pm$ 0.46    	\\ 		
\end{tabular}
\caption[Collins asymmetries $A_{12}^{\eta}$ binned in $(z_{1},P_{t1})$]{Collins asymmetries $A_{12}^{\eta}$ binned in $(z_{1},P_{t1})$. Uncertainties are statistical and systematic, respectively.}
\label{tab:finaletaptbins}
  \end{ruledtabular}
 \end{table*}
\endgroup

\begingroup
 \squeezetable
 \begin{table*}[htp]
  \begin{ruledtabular}  
    \begin{tabular}{cccccccccc}
$z_1$& $\langle  z_{1}  \rangle$ & $z_2$ & $\langle  z_{2}\rangle$& $P_{t1}$ [GeV] & $\langle  P_{t1} \rangle$ [GeV] & $P_{t2}$ [GeV] &  $\langle P_{t2}\rangle$ [GeV] &$\langle\frac{\sin^2(\theta)}{1+\cos^2(\theta)}\rangle$& $A_{12}^{\pi^0}$ [\%]   \\ \hline\hline
[0.3,0.5]	&	0.37	&	[0.3,1.0]	&	0.44	&	[0,0.15]	&	0.10	&	[0,3.0]	&	0.38	&	0.91	& -0.08  $\pm$ 0.50  $\pm$ 0.16     \\ \hline			
[0.3,0.5]	&	0.37	&	[0.3,1.0]	&	0.44	&	[0.15,0.30]	&	0.23	&	[0,3.0]	&	0.38	&	0.91	& 1.40  $\pm$ 0.26  $\pm$ 0.10       \\ \hline			
[0.3,0.5]	&	0.37	&	[0.3,1.0]	&	0.44	&	[0.30,0.50]	&	0.40	&	[0,3.0]	&	0.38	&	0.91	& 1.98  $\pm$ 0.19  $\pm$ 0.09     \\ \hline			
[0.3,0.5]	&	0.42	&	[0.3,1.0]	&	0.44	&	[0.50,3.0]	&	0.58	&	[0,3.0]	&	0.38	&	0.91	& 2.55  $\pm$ 0.42  $\pm$ 0.28     \\ \hline			
\hline																				
[0.5,0.7]	&	0.49	&	[0.3,1.0]	&	0.44	&	[0,0.15]	&	0.10	&	[0,3.0]	&	0.38	&	0.91	& 0.44  $\pm$ 0.66  $\pm$ 0.23     \\ \hline			
[0.5,0.7]	&	0.49	&	[0.3,1.0]	&	0.44	&	[0.15,0.30]	&	0.23	&	[0,3.0]	&	0.38	&	0.91	& 1.40  $\pm$ 0.39  $\pm$ 0.13      \\ \hline			
[0.5,0.7]	&	0.49	&	[0.3,1.0]	&	0.44	&	[0.30,0.50]	&	0.40	&	[0,3.0]	&	0.38	&	0.91	& 2.24  $\pm$ 0.27  $\pm$ 0.13     \\ \hline			
[0.5,0.7]	&	0.53	&	[0.3,1.0]	&	0.44	&	[0.50,3.0]	&	0.64	&	[0,3.0]	&	0.38	&	0.91	& 3.03  $\pm$ 0.26  $\pm$ 0.11     \\ \hline			
\hline																				
[0.7,1.0]	&	0.72	&	[0.3,1.0]	&	0.44	&	[0,0.15]	&	0.10	&	[0,3.0]	&	0.39	&	0.91	& 0.22  $\pm$ 1.39  $\pm$ 0.63     \\ \hline			
[0.7,1.0]	&	0.69	&	[0.3,1.0]	&	0.44	&	[0.15,0.30]	&	0.23	&	[0,3.0]	&	0.39	&	0.90	& 1.31  $\pm$ 1.06  $\pm$ 0.38     \\ \hline			
[0.7,1.0]	&	0.68	&	[0.3,1.0]	&	0.44	&	[0.30,0.50]	&	0.40	&	[0,3.0]	&	0.38	&	0.91	& 3.67  $\pm$ 0.73  $\pm$ 0.35     \\ \hline			
[0.7,1.0]	&	0.68	&	[0.3,1.0]	&	0.45	&	[0.50,3.0]	&	0.70	&	[0,3.0]	&	0.38	&	0.90	& 7.90  $\pm$ 0.64  $\pm$ 0.36   	\\ 
\end{tabular}
\caption[Collins asymmetries $A_{12}^{\pi^0}$ with $z>0.3$ binned in $(z_{1},P_{t1})$]{Collins asymmetries $A_{12}^{\pi^0}$ with $z>0.3$ binned in $(z_{1},P_{t1})$. Uncertainties are statistical and systematic, respectively.}
  \end{ruledtabular}
 \end{table*}
\endgroup

\end{appendices}
\end{document}